\documentclass[sigconf, screen, nonacm]{acmart}
\usepackage{color, colortbl, xcolor}
\usepackage{url}
\usepackage{subcaption}
\usepackage{textcomp}
\usepackage{soul}
\usepackage{multirow}
\usepackage{enumitem}
\usepackage{mathtools}
\usepackage{siunitx}
\usepackage{array}
\usepackage{colortbl}
\usepackage{hhline}

\usepackage{booktabs} 

\usepackage{framed}  
\colorlet{shadecolor}{black!5}
\newenvironment{myshade}{%
  \def\FrameSep{6pt}%
  \setlength{\topsep}{6pt}
  \begin{shaded*}%
}{%
  \end{shaded*}%
  \par\vspace{6pt}%
}

\usepackage{pifont}        
\newcommand{\cmark}{\ding{51}}

\usepackage{array}
\usepackage{xcolor}

\newcommand*{\rowstyle}[1]{
  \gdef\@rowstyle{#1}%
  \@rowstyle\ignorespaces%
}

\newcolumntype{=}{
  >{\gdef\@rowstyle{}}%
}

\newcolumntype{+}{
  >{\@rowstyle}%
}

\usepackage{arydshln}
\definecolor{linkColor}{RGB}{6,125,233}
\definecolor{green}{rgb}{0.0, 0.65, 0.31}
\definecolor{bleudefrance}{rgb}{0.19, 0.55, 0.91}
\definecolor{ceruleanblue}{rgb}{0.16, 0.32, 0.75}
\definecolor{grey}{HTML}{969696}
\definecolor{violet}{HTML}{756bb1}
\definecolor{dgrey}{HTML}{01665e}
\definecolor{lgrey}{HTML}{5ab4ac}
\definecolor{dgreen}{HTML}{005a32}
\definecolor{purple}{HTML}{ae017e}

\definecolor{camCol}{HTML}{000000}
\definecolor{editCol}{HTML}{000000}
\definecolor{maskCol}{HTML}{c51b7d}
\definecolor{lrColor}{HTML}{8856a7}
\definecolor{trColor}{HTML}{d01c8b}
\definecolor{ctColor}{HTML}{4dac26}
\definecolor{brickred}{HTML}{f03b20}
\definecolor{improveCol}{HTML}{253494}
\definecolor{worsenCol}{HTML}{d7191c}
\definecolor{DarkBlue}{HTML}{00008B}
\definecolor{mscolor}{HTML}{01665e}
\definecolor{nmscolor}{HTML}{bf812d}
\definecolor{lgreen}{HTML}{ccece6}
\definecolor{dolive}{HTML}{308014}

\definecolor{lgreen}{HTML}{e0f3db}
\definecolor{dpink}{HTML}{CD1076}
\definecolor{pink}{HTML}{FED2D2}
\definecolor{soothinggreen}{HTML}{4dac26}
\definecolor{darkred}{HTML}{8B0000}

\colorlet{tablerowcolor4}{gray!50} 

\newcommand*{\textlabel}[2]{%
  \edef\@currentlabel{#1}
  \phantomsection
  #1\label{#2}
}

\colorlet{tableheadcolor}{gray!25} 
\colorlet{tablerowcolor}{gray!10} 
\colorlet{tablerowcolor2}{gray!45} 
\colorlet{tablerowcolor3}{gray!25} 

\newcommand{\rowcollight}{\rowcolor{tablerowcolor3}} %

\newcolumntype{a}{>{\columncolor{tablerowcolor}}r}
\definecolor{aicolor}{HTML}{018571}
\definecolor{occolor}{HTML}{ff7799}

\definecolor{aicolor}{HTML}{fc8d62}
\definecolor{occolor}{HTML}{253494}

\newif{\ifhidecomments}
  \hidecommentsfalse 
\ifhidecomments
    \newcommand{\upol}[1]{}
    \newcommand{\samir}[1]{}
    \newcommand{\koustuv}[1]{}
\else
    \newcommand{\upol}[1]{\textbf{\small\sffamily{\textcolor{purple}{[#1 -- Upol]}}}}
    \newcommand{\samir}[1]{\textbf{\small\sffamily{\textcolor{DarkBlue}{[#1 -- Samir]}}}}
    \newcommand{\koustuv}[1]{\textbf{\small\sffamily{\textcolor{dolive}{[#1 -- Koustuv]}}}}
  \fi

\newcommand{\edit}[1]{#1}
\newcommand{\camit}[1]{%
  {\color{camCol}#1}%
}

\renewcommand{\textrightarrow}{$\rightarrow$}

\newcommand{\tabitem}{\textbullet~~}

\colorlet{tableheadcolor}{gray!25} 

\definecolor{neutralCol}{HTML}{dd1c77}
\definecolor{neutralGreen}{HTML}{31a354}
\definecolor{NewBlue}{HTML}{1879ba}
\definecolor{bleudefrance}{rgb}{0.19, 0.55, 0.91}  
\definecolor{AfTrColor}{HTML}{0868ac}  
\definecolor{BfTrColor}{HTML}{a8ddb5}  

\definecolor{AfCtColor}{HTML}{b10026}  
\definecolor{BfCtColor}{HTML}{fd8d3c}

\graphicspath{ {figures/} }

\newcommand{\para}[1]{\vspace{0.3em}\noindent\textbf{#1}~}

\newcolumntype{C}[1]{>{\centering\arraybackslash}p{#1}}

\AtBeginDocument{%
  \providecommand\BibTeX{{%
    \normalfont B\kern-0.5em{\scshape i\kern-0.25em b}\kern-0.8em\TeX}}}

\begin{document}

\title[From Future of Work to Future of Workers]{From Future of Work to Future of Workers: Addressing Asymptomatic AI Harms for Dignified Human-AI Interaction}

\author{Upol Ehsan}
\orcid{0000-0002-4911-0409}
\affiliation{%
  \institution{Northeastern University,\\ Harvard University}
 \city{Boston}
 \state{MA}
 \country{USA}}
\email{u.ehsan@northeastern.edu}

\author{Samir Passi}
\orcid{0000-0002-7921-3820}
\affiliation{%
  \institution{Microsoft}
 \city{Redmond}
 \state{WA}
 \country{USA}}
\email{v-sapassi@microsoft.com}

\author{Koustuv Saha}
\orcid{0000-0002-8872-2934}
\affiliation{%
  \institution{University of Illinois Urbana-Champaign}
 \city{Urbana}
 \state{IL}
 \country{USA}}
\email{ksaha2@illinois.edu}

\author{Todd McNutt}
\orcid{0000-0002-1569-2922}
\affiliation{%
  \institution{Johns Hopkins University}
 \city{Baltimore}
 \state{MD}
 \country{USA}}
\email{tmcnutt1@jhmi.edu}

\author{Mark O. Riedl}
\orcid{0000-0001-5283-6588}
\affiliation{%
  \institution{Georgia Tech}
 \city{Atlanta}
 \state{GA}
 \country{USA}}
\email{riedl@gatech.edu}

\author{Sara Alcorn}
\orcid{0000-0003-4620-5047}
\affiliation{%
  \institution{University of Minnesota}
 \city{Minneapolis}
 \state{MN}
 \country{USA}}
\email{alcor049@umn.edu}

\begin{abstract}
In the future of work discourse, AI is touted as the ultimate productivity amplifier. Yet, beneath the efficiency gains lie subtle erosions of human expertise and agency. This paper shifts focus from the \textit{future of work to the future of workers} by navigating the AI-as-Amplifier Paradox: AI's dual role as enhancer and eroder, simultaneously strengthening performance while eroding underlying expertise. We present a year-long study on the longitudinal use of AI in a high-stakes workplace among cancer specialists. Initial operational gains hid ``intuition rust'': the gradual dulling of expert judgment. These asymptomatic effects evolved into chronic harms, such as skill atrophy and identity commoditization. Building on these findings, we offer a framework for dignified Human-AI interaction co-constructed with professional knowledge workers facing AI-induced skill erosion without traditional labor protections. The framework operationalizes sociotechnical immunity through dual-purpose mechanisms that serve institutional quality goals while building worker power to detect, contain, and recover from skill erosion, and preserve human identity. Evaluated across healthcare and software engineering, our work takes a foundational step toward dignified human-AI interaction futures by balancing productivity with the preservation of human expertise.
\end{abstract}








\maketitle

  \section{Introduction}
 \begin{quote}
 \small
\textit{``My intuition is rusting.'' These words, spoken by a seasoned cancer specialist after months of using an AI-driven decision aid, capture a subtle sense of unease. The AI improved efficiency: faster treatment plans, better metrics. However, beneath the surface, something was amiss. ``I approve too quickly,'' the specialist admitted, revealing the expertise honed over years was perhaps quietly corroding. Far from an anomaly, this is a symptom of a broader paradox in the Human-AI workplace.}
 \end{quote}
Contemporary discourse on the future of work touts AI’s potential to boost productivity, streamline tasks, and even solve labor shortages~\cite{somers2023generative,oecd2024_ai_productivity,imf2024_genai_future_of_work,wef2025_future_of_jobs}. Economic analyses of the AI-powered future of work focus on job displacement and creation, emphasizing efficiency gains~\cite{autor2015_why_jobs,acemoglu2019_automation_new_tasks,acemoglu2011skills}. However, this dominant narrative overlooks \textbf{the future of the workers}: \textit{the humans behind productivity metrics.} 
Despite calls to keep humans ``in the loop''~\cite{amershi2019guidelines,ehsan2020human,shneiderman2020human,kawakami2023sensing}, the bulk of the focus continues to measure success through task outcomes: speed, accuracy, throughput~\cite{fragiadakis2024evaluating,das2023algorithmic}.
This task-centric view masks the fact that AI adoption is not neutral; it redistributes agency, reshapes expertise, and reconfigures dignity in ways that may not surface in performance dashboards. 
This leads to both theoretical and practical gaps. 
Theoretically, we lack frameworks that conceptualize AI's invisible expertise erosion. Practically, we lack tools to detect `asymptomatic harms': initially unmeasurable effects that quietly weaken workers over time.

A key question surfaces: \textit{As AI takes on more responsibilities, what happens to the professional skills, judgment, and dignity of human experts?} Our research addresses this neglected area: while prior work has examined job displacement and task automation, we focus on long-term impacts on workers' expertise and identity. 


\edit{Drawing on a longitudinal study, this paper develops a worker-centered account of how AI’s long-term integration can asymptomatically reshape agency, professional skill, and worker dignity over time.} In a \textit{year-long} study of a real-world AI deployment in radiation oncology, 42 participants took part in 24 interviews, 5 workshops, and 52 think-aloud sessions. We found AI assistance introducing \textit{asymptomatic effects}. Borrowing from the medical metaphor, AI's asymptomatic effects are behavioral shifts that escape standard performance metrics and are not immediately visible or alarming. These effects congealed into \textit{chronic harms}: clinicians' manual skills atrophied, and their professional identity wavered. Clinicians feared loss of ``the hands-on skills that make [them] unique'' and worried about becoming ``bystanders in their own practice.'' 

From these findings, a paradoxical tension emerged around AI's dual role: it not only enabled efficiency, but it was also a source of expertise erosion--what we call, the \textit{AI-as-Amplifier-Paradox.}
The \textit{very} strengths that AI offers can also erode the \textit{very same} human expertise that it aims to support. What makes this insidious is that expertise erosion can occur \textit{asymptomatically}, with hidden drifts that harden into \textit{chronic} damage. We do not invoke the paradox as a label for AI’s mixed outcomes. Rather, it exposes a consequential tension: AI can erode the \textit{very capabilities} that it was built to support. 
Our findings show that while AI can help doctors become more efficient in the short term, it can \textit{erode what it means to be a doctor} (core practices that give meaning to the role), often invisibly.



These findings, ranging from enhancement to erosion, informed our intervention (detailed in~\autoref{sec:study_context}), which enabled practitioners to tackle overreliance and sustain critical engagement. The intervention's success in addressing immediate challenges also revealed a more fundamental imperative: developing systematic, proactive approaches to counteract AI's erosive effects. In response, we synthesized our empirical findings to construct and evaluate a multi-level framework that \edit{charts a path toward \textbf{dignified Human-AI interaction}. Our participants' understanding of dignified work shaped this vision: ``Dignified work preserves what's irreducibly human: the judgment, intuition, craft that makes humans \textit{drivers} of the `loop', not just being in the loop. It's about self-determination and impact'' (R07). Building on this, we define Dignified Human-AI interaction as AI integration that preserves human agency, expertise, and self-worth. Designing AI for the future of workers demands confronting this amplifier paradox and embedding safeguards for human growth, not just task output, helping workers build sociotechnical immunity to detect, contain, and recover from AI's hidden harms before they ossify.} Our contributions are:


\begin{itemize}
    \vspace{-3pt}
    \item \textbf{Empirical findings:} A year-long study (42 participants involved in 24 interviews, 5 workshops, and 52 think aloud sessions) of a real-world AI deployment in radiation oncology that demonstrates the \textit{AI-as-Amplifier Paradox} in practice and highlights how \textit{Social Transparency} features can address parts of the paradox.
    \item \textbf{Conceptual lenses:} a) \textit{AI-as-Amplifier Paradox (diagnosis): }a new lens conceptualizing how AI can provide short-term gains even as it erodes workers' skills, autonomy, and identity; b) \textit{Sociotechnical Immunity (response): }introduces sociotechnical immunity as the response to sense, contain, and recover from hidden harms.
    \item \textbf{Framework for Dignified Human-AI Interaction:} A multi-level framework that operationalizes sociotechnical immunity via early-warning signals, containment actions, and recovery routines; evaluated in the primary domain (healthcare) as well as in a second domain (AI-assisted software engineering) to demonstrate cross-domain transfer.
    \vspace{-2pt}
\end{itemize}
The core strength of this work lies in illuminating an intellectual blind spot in Human-centered AI. It brings into view the understudied asymptomatic effects of AI that, over time, harden into chronic harms. While existing literature documents symptomatic failures (automation bias, deskilling, errors), our work reveals the hidden deterioration preceding these crises. Unlike shorter studies, our longitudinal approach better captures AI's full integration, revealing the complex arc of the AI-as-Amplifier paradox. Working with specialized, time-constrained practitioners in a high-stakes domain revealed how AI not only reshapes performance but also redefines expertise. Shifting the focus from the future of work to the \textit{future of workers}, our work calls on the CHI community to champion human dignity and flourishing as first-class metrics of success in AI-mediated work.


\section{Background and Related Work}
This section examines three critical areas of relevant research on AI's sociotechnical impacts on workers: the dissonance between AI's positive impacts on \textit{work} and wider unintended effects on \textit{workers,}  the short-term ill effects of AI use on users, and emerging evidence of possible long-term AI harms to human dignity and professional identity.

\vspace{-5pt}
\subsection{Future of Work vs. Future of Workers}

Modern AI is ushering in yet another bright and shiny \textit{future of work}, characterized by dramatic increases and reductions~\cite{butler2024microsoftfowreport}. Workers are becoming more ``accurate''~\cite{MSWorklab2024AIchangingwork}, ``efficient''~\cite{simkute2025calculator}, and ``productive''~\cite{jaffe2024generative}. New AI tools ``extend'' human capabilities and thinking abilities~\cite{reicherts2025helpmethink}, helping save time and do more~\cite{jaffe2024generative}. Revenues get ``higher''~\cite{MSWorklab2024AIchangingwork}, work is ``faster''~\cite{jaffe2024generative}, and innovations ``accelerate''~\cite{hayes2025fowinnovation} dramatically. All work is impacted, including art~\cite{zhou2024artists}, customer service~\cite{Brynjolfsson2025customerservice}, education~\cite{simkute2025calculator, Renkema2024futureacademics}, entrepreneurship~\cite{otis2023entrepreneurs}, freelance~\cite{hui2024freelance}, research~\cite{rodgers2024researchers}, 
and software development~\cite{butler2024coding, yeverechyahu2025coding, cui2024coding}.

The new future of work also impacts \textit{workers}, and not all impacts are positive or neutral. It is increasingly difficult to oversee AI systems~\cite{holzinger2024oversight, langer2024oversight,chowdhary2023can, passi2025_oversightproblem}. Workers experience consequential losses in ``situational awareness''~\cite{simkute2025ironies}, and face ``de-skilling''~\cite{crowston2025upskilling, shukla2025deskilling, simkute2025calculator}, and ultimately ``job displacement''~\cite{chen2022displacement, rawashdeh2023displacement}. \edit{These patterns connect to automation bias and overreliance research, where users often favor AI recommendations over their own judgment, especially under time pressure or when systems appear accurate~\cite{passi2022overreliance,passi2024appropriate, passi2025overreliancechapter}.} Adverts, corporate announcements, and industry reports sometimes acknowledge negative impacts while positioning them as inevitable-and-solvable problems on the path to technological progress. De-skilling? AI can and will also \textit{up-skill} workers. Overreliance? AI is getting more accurate and users more capable with time. Worker displacement? More like \textit{workforce transformation} with evolving roles and shifting expertise. What remains unaddressed, and largely unacknowledged, is the \textit{craft erosion} creeping into workers, proceeding asymptomatically, like expertise being burglarized while practitioners sleep, unnoticed until they reach for skills that have quietly atrophied.

Our work extends emerging worker-centered scholarship~\cite{cleven2023laborfutureofwork, ahmed2022futureofwork, sum2025futureoflabor, sum2025workerresistsurveillance,das2025ai} by focusing on the transformation of expertise and intuition in workers unfolding across the extended use of AI, revealing patterns often masked by first-wave optimism about AI efficiency. 
\camit{In doing so, this paper takes inspiration from critical worker-centered research on technology's impact on work. Researchers have shown how algorithmic management systems harm gig workers' well-being and exacerbate power asymmetries \cite{zhang2022gigwellbeing}, how algorithmic triage intensifies workloads and compromises hospital staff well-being \cite{spektor2023designingwellbeing}, and how datafication and monitoring render fulfillment center workers as replaceable cogs rather than humans \cite{cheon2023bigtech}. Although the impact of modern AI on work has attracted renewed attention, technological systems have always shaped work and workers, and workers have also shaped the design and use of technological systems through active negotiation, everyday resistance, and collective action, and methods such as workers' inquiry that assert worker agency and dignity~\cite{sum2025workerresistsurveillance, fox2020workerdesign, wong2021softresistance}}. Workers' inquiry, in particular, has emerged as a method through which workers document and analyze their own conditions to build collective knowledge and power~\cite{miceli2025epistemicauthorityai}. 


\camit{We extend emerging worker-centered scholarship that has charted research agendas to address labor issues~\cite{fox2020workerdesign}, used critical computing methods to challenge industry-driven future of work narratives~\cite{lushnikova2025cscwfutureofwork}, and leveraged worker testimonies to highlight issues in worker health and safety~\cite{akridge2025workersbustransit}.} While prior research has focused on AI's impacts on gig workers, AI now threatens white-collar knowledge workers in unprecedented ways~\cite{sum2025futureoflabor,lobel2018ndas,ou2025psychological,baiocco2022algorithmic}. For the first time in history, AI threatens college graduates 5x more than high school graduates, especially in non-routine cognitive work~\cite{webb2019impact, tasci2025aiknowledgeworkers}. This overturns centuries of automation logic, in which education protected workers from technological displacement.
Despite this existential threat, the AI management of knowledge workers remains largely understudied~\cite{baiocco2022algorithmic}. They face unique vulnerabilities: psychological contracts centered on mission rather than pay, making algorithmic oversight especially complex~\cite{ou2025psychological}; fragmented and unstable work environments contrary to narratives of prestige and privilege~\cite{sum2025futureoflabor}; one-third silenced by NDAs from discussing workplace issues~\cite{lobel2018ndas}; and geographic dispersion and reputational risks within subspecialties (often 200-300 people nationally) limit conventional collective action. 
There is a dearth of pragmatic tools to help them navigate AI's erosion of their expertise and identity. This paper addresses that gap by presenting a framework developed \textit{with} workers to address their challenges.

\vspace{-5pt}
\subsection{Troubling Signs in Human-AI Interaction and Collaboration}

Although workers' woes are often at the margins of future of work visions, AI's impact on workers-as-users is a prominent area of human-AI interaction research. HCI research has demonstrated AI's double-edged impact on users, task performance, and professional practices. For example, AI assistance can cause skill atrophy without user awareness, leading users to perceive improvement while becoming dependent~\cite{macnamara2024skilldecay}, and a higher confidence in GenAI correlates with decreased critical thinking among knowledge workers~\cite{lee2025criticalthinking}. This phenomenon appears across domains: relying on AI-generated code can hinder college students' learning and skill development~\cite{prather2023noviceprogrammer}, and anchoring bias from LLM-generated content negatively impacts diversity in marketers' content~\cite{chen2024ghostwriters}. Even expert users struggle to maintain critical thinking and independent judgment when using AI~\cite{De-Arteaga2020}. Although many studies capture valuable insights, they offer snapshots of AI's short-term impacts. This paper complements existing work by contributing longitudinal evidence of AI's impact, documenting not only immediate but also asymptomatic effects that escape traditional metrics and chronic harms that emerge only after prolonged use.


\edit{Technology, however, is only partly at fault; its impacts reveal systemic issues. As researchers argue, critical analyses of technology's impacts on professions must move beyond `work' alone towards addressing the economic and political structures that sustain technological imaginaries of datafied management~\cite{greenbaum1996backtolabor, sum2025futureoflabor, ahmed2022futureofwork, tang2023backtolabor, gupta2018futureofworkers,das2024teacher}.
Indeed, sweeping systemic changes are necessary to imagine and design better (non)technological futures.
However, we believe that pragmatic actions are equally required to remedy aspects of the present technological moment: the mundane problems workers face putting one foot in front of the other (with, around, or against AI) working under systemic conditions and the everyday reality of ``work.''
This paper takes steps in this direction, highlighting what workers \textit{do} in the face of technological change--what are their lived experiences, how do they tackle on-the-ground challenges, and how they adapt their practices even as their work \textit{remains} changing, threatened, and undermined.}

\subsection{AI Harms: Impacts to Human Dignity and Professional Identity}

In this paper, we go beyond AI's short-term impacts to focus on AI's hidden costs: workers' dignity, self-worth, and identity eroding through role displacement, diminished control, and anxiety about professional futures. Evidence is emerging: endoscopists are getting worse at performing tasks without AI assistance~\cite{endoscopy2025deskilling} while radiographers report anxiety about potential deskilling and changing core'' tasks~\cite{survey2025anxiety}. Clinicians report ``losing control to AI'' in professional decision-making~\cite{losingcontrol2025}, while educators face confusion around professional identity as the boundaries between human and AI expertise blur ~\cite{kratas2024teacher, lan2024teacher}. AI adoption may boost productivity, but chips away at hard-won expertise through de-skilling~\cite{aquino2022utopia, chen2021radiographers} and challenges to epistemic authority~\cite{lombi2024radiologist, scarbrough2025sensemaking}, creating alienation~\cite{braganza2021psych}, diminished agency~\cite{scarbrough2025sensemaking}, and lack of autonomy~\cite{chen2021radiographers} in professional work. Constraining professional judgment and failing to recognize users as workers with qualified expertise, AI tools may induce ``learned helplessness''~\cite{maier1976learnedhelplessness}. This paper takes formative steps to extend our understanding of AI’s hidden yet consequential costs that are largely understudied and unaddressed, slowly brewing under the surface of more visible positive productivity metrics and negative short-term impacts.

\section{Study Context \& Methods} \label{section:methods_study}
\begin{figure*}[h]
    \centering
     \includegraphics[width=1.3\columnwidth]{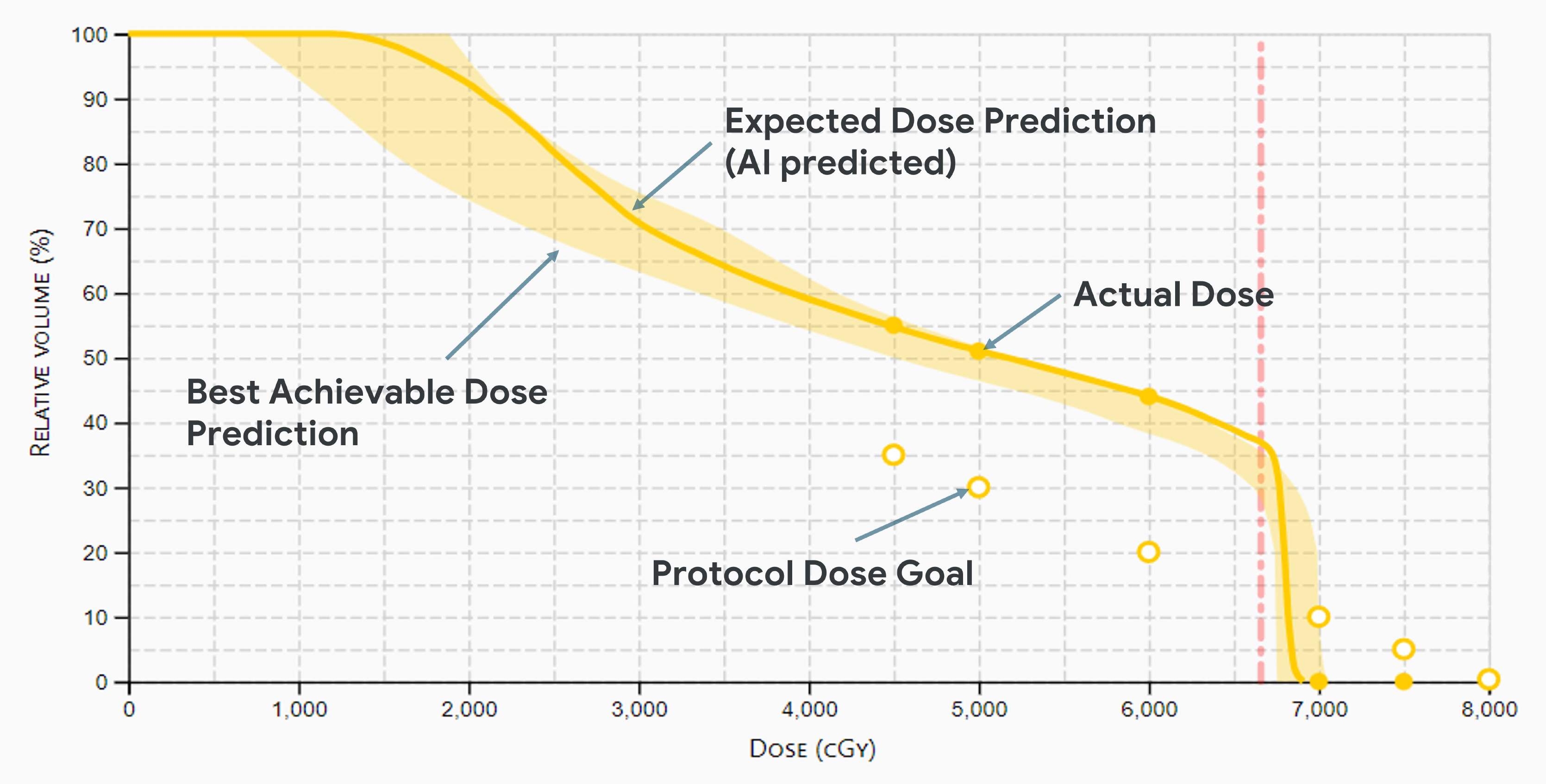}
  \caption{
    Anatomy of AI-assisted radiation treatment planning in RadPlan. This reconstructed simplified Dose-Volume Histogram (DVH) shows how AI predictions (yellow band) relate to actual dose delivery, best achievable doses, and protocol targets.} 
            \label{fig:radplanDVH}
    \Description[figure]{Dose–Volume Histogram for RadPlan. X-axis: Dose (cGy); Y-axis: Relative Volume (\%). A yellow band shows the AI-predicted dose range; a thick yellow line with dots shows the actual dose, dropping steeply near ~7,000 cGy. Yellow dots mark protocol dose goals. A red dashed vertical line near 7,000 cGy indicates a protocol limit. A note indicates the best achievable dose prediction at lower doses within the band.}
    \vspace{-10pt}
\end{figure*}


We conducted a year-long longitudinal study examining AI integration in radiation oncology using mixed qualitative methods to capture the evolving relationship between clinicians and an AI-assisted treatment-planning system. \textit{We are not aware of prior in-situ studies that trace native AI use among radiation oncologists across a full year. This work contributes a rare longitudinal perspective on everyday practice, with particular attention to skill erosion and de-skilling.}

\begin{figure*}[h]
    \centering
     \includegraphics[width=2.1\columnwidth]{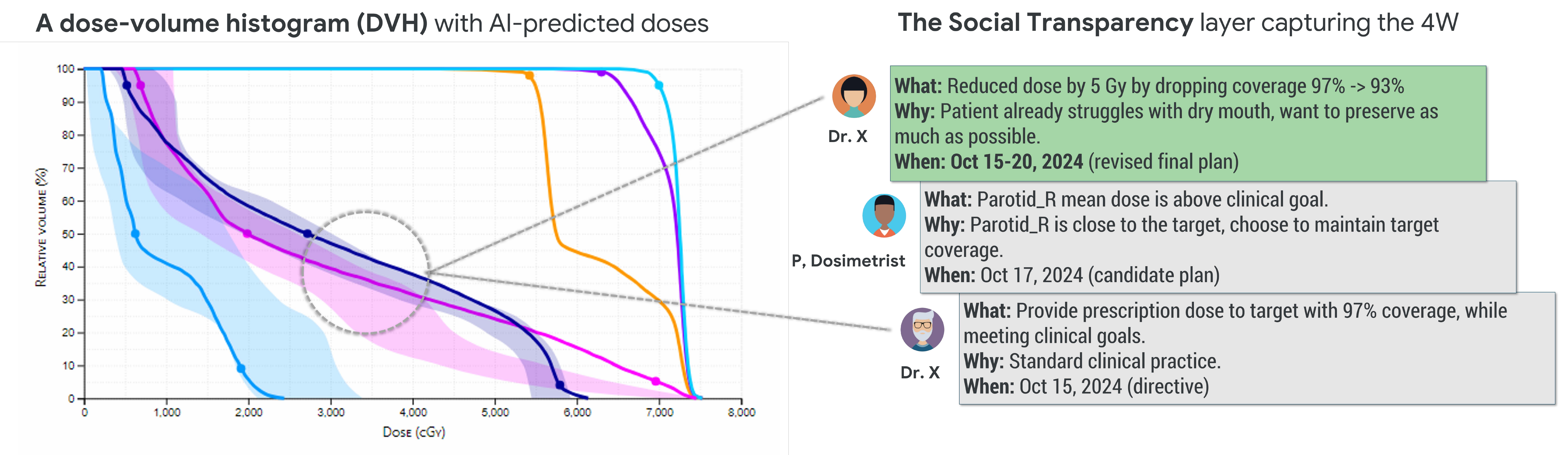}
     \vspace{-0.5em}
  \caption{Social Transparency intervention: fosters better decisions by complementing AI dose predictions (left) with socio-organizational context (right). Captures the 4W (who, what, when, why) of multi-stakeholder input, revealing how treatment decisions evolve.} 
            \label{fig:ST_Radplan}
    \Description[figure]{A two-panel medical imaging visualization:
        Left panel: A dose-volume histogram (DVH) showing AI-predicted radiation doses. The graph displays relative volume (0-100\%) on the y-axis and dose in cGy (0-8,000) on the x-axis. Multiple colored curves represent different treatment areas, with shaded regions showing dose distributions. A circled area highlights a critical decision point.
        Right panel: “The Social Transparency layer capturing the 4W” - Three timestamped entries showing treatment decision evolution:
        1.	Dr. X (Oct 15-20, 2024, green box - revised final plan): “What: Reduced dose by 5 Gy by dropping coverage 97\% to 93\%. Why: Patient already struggles with dry mouth, want to preserve as much as possible.”
        2.	P, Dosimetrist (Oct 17, 2024, candidate plan): “What: Parotid_R mean dose is above clinical goal. Why: Parotid_R is close to the target, choose to maintain target coverage.”
        3.	Dr. X (Oct 15, 2024, directive): “What: Provide prescription dose to target with 97\% coverage, while meeting clinical goals. Why: Standard clinical practice.”
}
\end{figure*}
\subsection{Study Domain: AI in Radiation Oncology} \label{sec:study_context}

\subsubsection{Case Study Background}
 Our case study took place in the context of radiation oncology, a medical domain where specialists plan and deliver radiation treatments to cancer patients. Broadly speaking, a typical team includes a Radiation Oncologist (RadOnc), Physicist, and Dosimetrist: the RadOnc (doctor) guides clinical care, the Physicist fine-tunes machines, and the Dosimetrist crafts tumor-targeting treatment plans. Decisions in Radiation Oncology are \textit{inherently multi-stakeholder}: a RadOnc alone cannot do the job without assistance from Physicists and Dosimetrists.

\para{Why Raditionl Oncology:} Radiation oncology is a unique domain for studying AI integration in high-stakes medical work. 
As a RadOnc noted, ``we don't use radiation to image your body...we use it to kill things in your body. There are no undo buttons in this field.'' This irreversibility makes AI-assisted decision-making consequential. The field's scarcity amplifies the stakes: approximately 5,000 radiation oncologists in the US serve a population of 330 million, meaning each practitioner potentially serves 66,000 people~\cite{rosehistory,ACR2023factsheet}. These clinicians have an outsized impact on human lives.

Our case study focused on RadPlan (pseudonym), \edit{a third-party, commercially developed} AI-assisted treatment planning system deployed at Hospital Alpha (pseudonym). Treatment planning is a high-stakes optimization puzzle: maximize tumor control while protecting healthy tissues. RadPlan generates initial radiation dose plans that clinicians can review and adjust, promising to save time and improve plan quality (e.g., sparing healthy organs by a few percentage points). Figure~\ref{fig:radplanDVH} presents a reconstructed interface featuring the system's core functionality with synthetic data. Hospital Alpha's data governance policies necessitated this approach to protect proprietary systems and patient information.

\subsubsection{The Problem and Intervention:}
We studied RadPlan's integration over its first 12 months of routine use across a multi-site hospital system with five locations in North America (broad coverage). Early signs were positive: planning cycles shortened (\~15\% faster), dose-volume histograms improved, and self-reported confidence climbed, creating an overall sense of AI's success by conventional criteria. However, after a few months, participants began sharing concerning remarks: ``My intuition is rusting,'' and ``I've no idea why the AI suggested it, but I accept it anyway.'' When stakeholders expressed concerns and wanted a mechanism to improve explainability and appropriate reliance, we added a Social Transparency (ST) layer~\cite{ehsan2021expanding} on RadPlan to illustrate \textit{who }else did \textit{what, when, and why} as a seamful intervention to calibrate AI reliance without disrupting work (Figure~\ref{fig:ST_Radplan} shows a mockup of the interface). \edit{We implemented ST with privacy safeguards from user preferences and established research~\cite{ehsan2021expanding, ridley2025human}: clinician-only 4W logs (not management), optional name masking, and data isolated from performance metrics to mitigate surveillance. While there are limitations, clinicians found ST ``safe enough,'' noting clincians' pushback culture adequately countered groupthink.}

\para{Why Social Transparency (ST):} The contexts in which ST shines (multi-stakeholder decision-making, absence of a “simple” ground truth, and shifting industry standards) are well aligned with radiation oncology's sociotechnical dynamics. First, decisions are inherently multi-stakeholder, matching ST’s 4W framework (who did what, when, and why). Second, there is no ``simple'' ground truth in treatment planning; five different approaches may all be valid, depending on patient needs. 
Third, peer input is vital because treatment standards shift with emerging knowledge and technologies. ST helps surface the reasoning behind earlier decisions, providing users with the peripheral vision needed to navigate evolving norms. Beyond this, ST has a proven record of successful real-world adoption in varied domains such as sales, cybersecurity, and finance~\cite{ehsan2021expanding, ehsan2023charting, soroko2023social, ferrario2024addressing,ridley2025human}. 

\begin{table*}[t]
\centering
\sffamily
\footnotesize
\caption{Workshop participation and interviews by role. WS: workshop numbers attended (1-5). \cmark: interviewed.}
\begin{minipage}[t]{0.49\textwidth}
\centering
\begin{tabular}{@{} l l r c c @{}}
\textbf{ID} & \textbf{Role} & \textbf{Experience (yrs)} & \textbf{WS} & \textbf{Interview} \\
\toprule
R01 & RadOnc & 8  & 1 & \cmark \\
R02 & RadOnc & 12 & 3 & \cmark \\
R03 & RadOnc & 20 & 2 & \cmark \\
R04 & RadOnc & 3  & 4 & \cmark \\
R05 & RadOnc & 15 & 1 & \cmark \\
R06 & RadOnc & 18 & 3 & \cmark \\
R07 & RadOnc & 2  & 5 & \cmark \\
R08 & RadOnc & 7  & 2 & \cmark \\
R09 & RadOnc & 4  & 4 &        \\
R10 & RadOnc & 11 & 4 & \cmark \\
R11 & RadOnc & 6  & 2 &        \\
R12 & RadOnc & 14 & 3 & \cmark \\
R13 & RadOnc & 3  & 2 &        \\
R14 & RadOnc & 16 & 4 &        \\
R15 & RadOnc & 9  & 1 & \cmark \\
\hdashline
\rowcollight P01 & Med. Physicist & 7  & 3 & \cmark \\
\rowcollight P02 & Med. Physicist & 4  & 2 & \cmark \\
\rowcollight P03 & Med. Physicist & 21 & 1 & \cmark \\
\rowcollight P04 & Med. Physicist & 2  & 5 &        \\
\rowcollight P05 & Med. Physicist & 11 & 4 &        \\
\rowcollight P06 & Med. Physicist & 6  & 2 &        \\
\bottomrule
\end{tabular}
\end{minipage}
\begin{minipage}[t]{0.49\textwidth}
\centering
\begin{tabular}{@{} l l r c c @{}}
\textbf{ID} & \textbf{Role} & \textbf{Experience (yrs)} & \textbf{WS} & \textbf{Interview} \\
\toprule
\rowcollight P07 & Med. Physicist & 15 & 3 & \cmark \\
\rowcollight P08 & Med. Physicist & 3  & 5 &        \\
\rowcollight P09 & Med. Physicist & 8  & 1 & \cmark \\
\rowcollight P10 & Med. Physicist & 2  & 4 &        \\
\rowcollight P11 & Med. Physicist & 13 & 2 &        \\
\rowcollight P12 & Med. Physicist & 9  & 5 & \cmark \\
\hdashline
D01 & Dosimetrist & 10 & 4 & \cmark \\
D02 & Dosimetrist & 3  & 1 &        \\
D03 & Dosimetrist & 6  & 3 & \cmark \\
D04 & Dosimetrist & 12 & 2 & \cmark \\
D05 & Dosimetrist & 2  & 5 &        \\
D06 & Dosimetrist & 8  & 4 & \cmark \\
D07 & Dosimetrist & 4  & 1 & \cmark \\
\hdashline
\rowcollight A01 & Hosp. Admin & 7  & 3 &        \\
\rowcollight A02 & Hosp. Admin & 9  & 5 &        \\
\rowcollight A03 & Hosp. Admin & 22 & 2 & \cmark \\
\rowcollight A04 & Hosp. Admin & 6  & 4 &        \\
\rowcollight A05 & Hosp. Admin & 8  & 1 &        \\
\rowcollight A06 & Hosp. Admin & 4  & 4 &        \\
\rowcollight A07 & Hosp. Admin & 24 & 5 & \cmark \\
\rowcollight A08 & Hosp. Admin & 2  & 2 &        \\
\bottomrule
\end{tabular}
\end{minipage}
\label{table:participant_details}
\end{table*}

\subsection{Participants: Overcoming Access Challenges and Building Community Trust}\label{sec:participants}

Participant access was difficult: radiation oncologists are time constrained, and monetary incentives carry little weight to them. Many viewed AI as a colonizing force coming to sideline or replace them, and the research team was initially seen as its emissary. We had to first earn their trust by spending months consulting on AI projects without expecting returns. Through sustained service and ethnography, the team moved from outsiders to `in-betweeners,' able to navigate both worlds with critical awareness. This trust enabled meaningful, long-term participation and offers a practical path for others seeking durable partnerships with clinician communities.

\textit{Participants: }Detailed in Table~\ref{table:participant_details}, we recruited and worked with 42 professionals in this study (some participating in multiple activities): 15 radiation oncologists, 12 medical physicists, 7 dosimetrists, and 8 hospital administrators. Participants spanned early career clinicians to 20+ year veterans, all with moderate to high AI familiarity. All participation was voluntary, with informed consent obtained at each phase. This sample represents a significant achievement given the access challenges, especially when compared to peer-reviewed oncology studies at top venues that often include fewer participants (e.g., 4 RadOncs and 2 Physicists~\cite{schultz2021qualitative}, 13 RadOncs~\cite{turner2022integrating}, or 2 RadOncs and 2 Physicists~\cite{chang2023provider}. Fostering community trust and meaningful participation not only produces impactful research but also builds critical bridges that HCI and AI researchers should cultivate. 

\subsection{Data Collection \& Analysis} \label{sec:data_collection_analysis}

\para{Longitudinal Field Observations:} First, we logged RadPlan usage and conducted 52 in-situ think-aloud sessions over the year. These captured key moments of hesitation, workflow deviations, and trust calibration (key to identifying asymptomatic effects).


\para{Participatory Workshops and Interviews:} Second, we held five participatory workshops (spanning all roles) and 24 semi-structured interviews about RadPlan experiences, skill changes, and the ST intervention.  \edit{Interviews were conducted throughout, with 8 at months 2-3, 10 at months 5-7, and 6 at months 10-11. Workshops took place at key phases of the 12-month deployment: at months 2, 3, 6, 8, and 11.} Participants engaged in drawing elicitation exercises to express their evolving relationships with AI deployment, and in later workshops reviewed Social Transparency 4W logs and co-constructed the Dignified Human-AI Interaction Framework reported in Section~\ref{sec:derive_framework}. These sessions yielded design ideas that were iteratively implemented (for example, the ST design in Fig.~\ref{fig:ST_Radplan}). \edit{Additional details on workshop activities and interviews are shared in Appendix ~\ref{sec: appendix_methods_further_details}.}


\para{Analysis: }Data analysis proceeded in iterative cycles using Grounded Theory. In total, 63 hours of all activities were transcribed. Taking an inductive approach, the analysis started with an iterative open-coding scheme that generated in-vivo codes (codes directly from the data, e.g., ``rubber stamping'' AI decisions). The iterative coding process was punctuated by frequent discussions \edit{(eight 60-90 minute meetings) with the larger research and clinical teams (including RadOncs, Physicists, Dosimetrists, Hospital Administrators, and researchers in HCI, STS, Human-centered AI (HCAI), and NLP)}. In these meetings, codes were refined using the principles of constant comparison (referring to the video, comparing codes, interpreting the dialogues in the context of the task, etc.). Through axial coding, \edit {our interdisciplinary core research team consisting of CS researchers and two clinicians} explored relationships among these concepts and clustered them into different categories (e.g., asymptomatic harms, identity commoditization). We compiled participants' drawings into composite narratives through collaborative sequencing, triangulating visual themes with interview data and workshop discussions to capture experiences that words could not. Finally, we unified the axial codes and consolidated them into selective codes, which became core themes (e.g., ‘ST as a check against deskilling and overreliance’). The selective codes are presented as the main themes in the Findings section. Relevant axial codes that contribute to each theme are highlighted where applicable. To distinguish between quotes from different groups, RadOncs are labeled with an “R” (e.g., R01), Physicists with a “P”, Dosimetrists with a ``D'', and Hospital Admins with an "A" as shown in Table~\ref{table:participant_details}.

\subsection{Ethics, Privacy, and Reflexivity}
Given the sensitivity of the problem and participants, we adopted strict privacy and ethical safeguards. For confidentiality, all names are pseudonyms, with unique IDs per participant. We adequately removed all personally identifiable information and, where needed, minimally edited quotes to reduce traceability while preserving authenticity and voice. Our interdisciplinary team spans HCI, STS, Human-centered AI, NLP, and Radiation Oncology (clinical practice). \edit{Three experts in HCI and HCAI, two in STS, one in NLP, and two board-certified clinicians (some individuals span multiple areas)}. It includes researchers who are people of color, immigrants, and hold diverse gender and cultural backgrounds. We have prior experience in longitudinal workplace studies, participatory design, real-world AI deployment and well-being research. Our clinician co-authors bridged worlds: fostering community trust, guiding recruitment, and shaping study design and interventions through domain expertise. This yielded an ecologically situated study and grounded findings. We acknowledge that our intersectionality shapes our epistemic lenses, whose imprints fall on our ways of knowing and interpreting. With this awareness, we took utmost care to rigorously analyze and present our findings.

 
\section{Findings: From Asymptomatic Effects to Chronic Harms to Identity Commoditization}\label{sec:findings}

In this section, we first delineate the emergence of the \textbf{AI-as-Amplifier Paradox}: AI's simultaneous role as enhancer and eroder of expertise, which provides a critical lens to view AI's impact on work. Next, we elaborate on how this paradox unfolds through three progressive stages: \textbf{Asymptomatic Effects, Chronic Harms,} and \textbf{Identity Commoditization.} These stages correspond to the increasing severity and visibility of AI’s impact on human workers. We describe each, supported by representative evidence from our case study. Together, they form a cautionary tale of how AI’s initial boost can lead to a long-term bust if left unchecked. 

\subsection{Emergence of the AI-as-Amplifier Paradox: From Enhancement to Erosion}

\begin{figure*}[t]
    \centering
     \includegraphics[width=1.4\columnwidth]{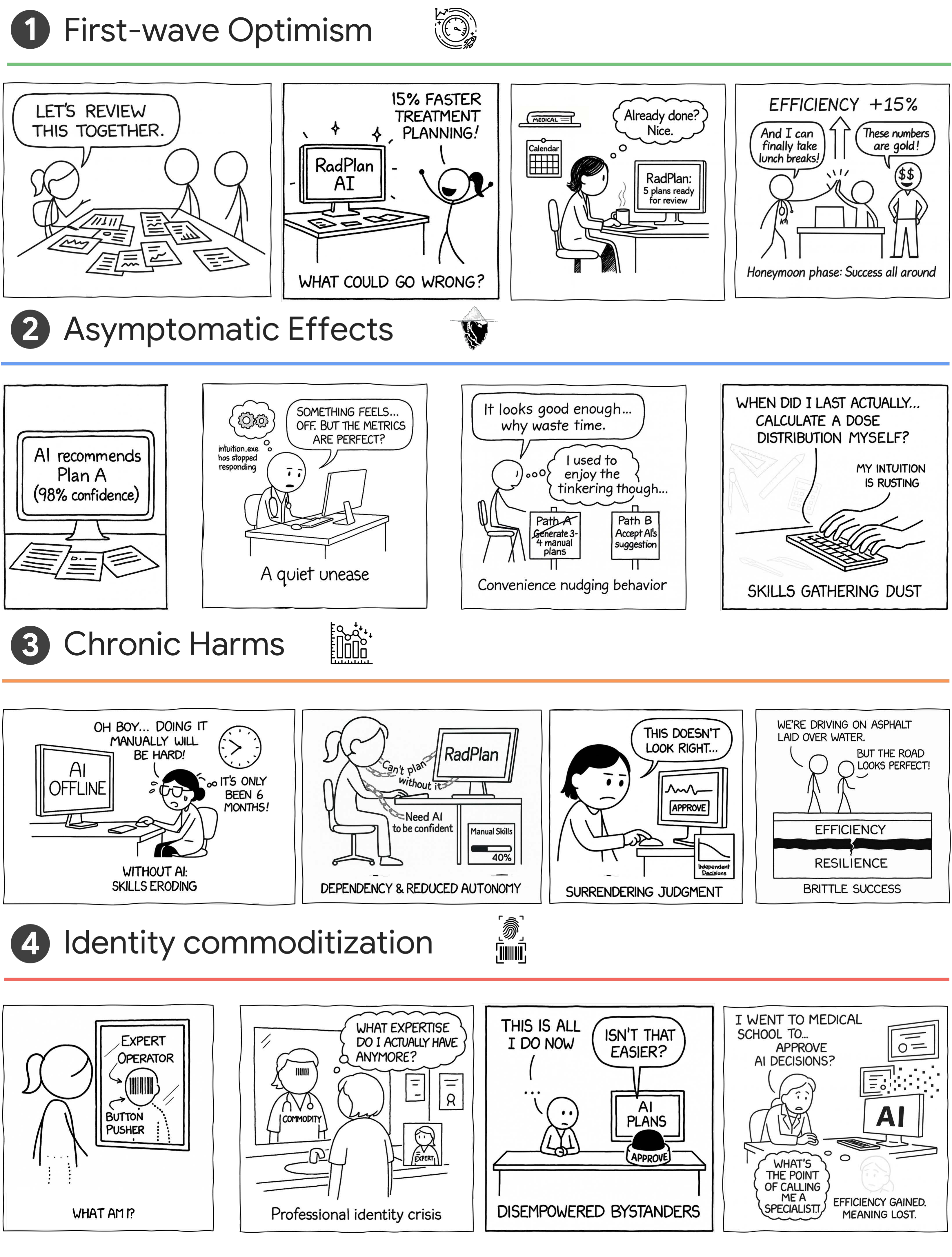}
  \caption{
   The AI-as-Amplifier Paradox emerges from participants' depictions of their evolving RadPlan relationship. Panels combine workshop drawings into a participant-guided arc showing \textit{initial optimism} devolving through \textit{asymptomatic effects} and \textit{chronic harms} into \textit{identity commoditization}. Visuals honor the minimalist aesthetic with participant-requested upscaling. See~\autoref{sec:appendix_compactPicsComics} for a compact version.
    } 
            \label{fig:comicStrips}
    \Description[figure]{Alt text: Four-part comic showing a clinician’s journey with “RadPlan AI.”
        First-wave Optimism.
        Panel A: Team at a table. Speech: “Let’s review this together.”
        Panel B: Person celebrates in front of a monitor, “RadPlan AI.” Text: “15\% faster treatment planning!”
        Panel C: At desk, speech: “Already done? Nice.”
        Panel D: Poster reads “EFFICIENCY +15\%.” Speech: “I can finally take lunch breaks!” Reply: “These numbers are gold!” Caption: “Honeymoon phase.”
        
        Asymptomatic Effects.
        Panel A: Screen text: “AI recommends Plan A (98\% confidence).”
        Panel B: Person thinks, “Something feels off, but the metrics are perfect?”
        Panel C: Person chooses convenience. Thought: “It looks good enough… I used to enjoy the tinkering.” Caption: “Convenience nudging behavior.”
        Panel D: Hand at keyboard. Thought: “When did I last calculate a dose distribution myself?” Caption: “Skills gathering dust.”
        
        Chronic Harms.
        Panel A: Screen “AI OFFLINE.” Person: “Doing it manually will be hard.” Clock shows six months. Caption: “Skills eroding.”
        Panel B: With patient. Text: “Need AI for confidence.” Caption: “Dependency and reduced autonomy.”
        Panel C: Bad graph beside big “Approve” button. Person: “This does not look right!” Caption: “Standards nudging down.”
        Panel D: Tank labeled Efficiency, Resilience, Success with hidden load underwater. Text: “Load under water. Artifact effect.”
        
        Identity commoditization.
        Panel A: Badge shows “Expert” replaced by “Operator, button pusher.” Caption: “What am I?”
        Panel B: Clinic scene with barcode faces. Person: “What expertise do I actually have anymore?” Caption: “Professional identity crisis.”
        Panel C: Person at desk with giant “APPROVE” button, screen “AI Plans.” Person: “This is all I do now.” Off-screen voice: “Isn’t that easier?” Caption: “Disempowered bystanders.”
        Panel D: Person reflects: “I went to medical school to approve AI decisions?” “What is the point of calling me a specialist?” Caption: “Efficiency gained, meaning lost.”}
\end{figure*}

During our longitudinal study, participants engaged in drawing elicitation exercises, where they created comic-like stick-figure narratives depicting their evolving relationship with RadPlan. These pictorial artifacts, combined with interviews, empirically grounded our understanding of AI's paradoxical nature as both enhancer and eroder. Eventually, this empirical synthesis surfaced what we term the \textbf{AI-as-Amplifier Paradox}: the very capabilities AI offers to support human expertise simultaneously undermine that same expertise over time.

\textbf{Visual sensemaking through ludic activities:} The drawing exercises embodied principles of ludic design: playful yet reflective activities that expanded how participants could express their experiences beyond verbal articulation. Sessions were often described as “one of the most fun thing I've done for a workshop” (R07), with participants valuing the chance to “laugh and learn simultaneously” (D05). The “drawings were able to capture things words couldn't” (P14). The comic strip in Figure~\ref{fig:comicStrips} synthesizes these visual narratives. Each stage represents a composite of different participant-created frames sequenced collaboratively to form a coherent arc. Participants expressed discomfort in publishing raw drawings. Some feared “their handwriting might get recognized” (P05), others noted they “were not artists...drawing was too rough for prime time” (R13). They voted for a solution that could preserve content while adding anonymity: AI-based upscaling combined with human artist redrawing to “flatten out identifiable parts” (D03). The end product honors the original stick-figure style minimalism that participants favored. All speech bubbles are verbatim. The comics thus serve dual purposes: empirical data capturing experiences participants struggled to voice and creative expression enabling collective sensemaking through a medium that carries affect and tacit knowledge that speech often cannot.

\textbf{The paradox operates through dual mechanisms:} What distinguished these findings was not that AI had negative effects; that was expected by some participants. Nor was it surprising that AI enhanced efficiency; that was its promise. The paradox revealed a fundamental tension: \textit{AI was eroding the very human capabilities it was built to support}. This complexity manifested in how AI helped them become more efficient RadOnc while \textit{invisibly eroding what it meant to be a RadOnc}. As one participant captured it: ``With AI, it's give and take, but you can't easily detect when it's eroding the very thing it promised to improve'' (R11). While workers initially felt more efficient, hidden asymptomatic effects were chipping away at cognitive, interpersonal, and identity foundations—damage that would crystallize into chronic harms.

\begin{figure*}[t]
    \centering
     \includegraphics[width=1.7\columnwidth]{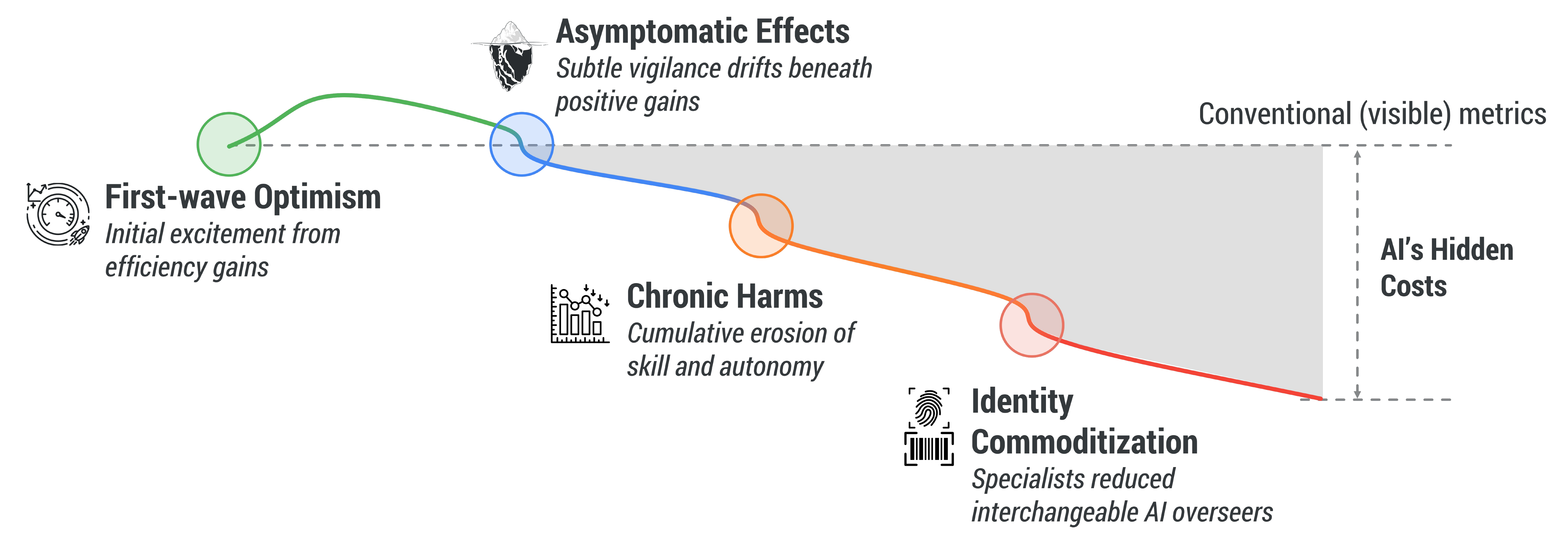}
  \caption{Trajectory of the AI-as-Amplifier Paradox: conventional metrics capture first-wave optimism and short-term gains but mask the hidden expertise erosion cascade (asymptomatic effects, chronic harms, identity commoditization) that diverges progressively from visible performance indicators} 
            \label{fig:ParadoxCurve}
    \Description[figure]{
        A line graph showing the divergence between visible metrics and hidden costs of AI adoption over time. The horizontal dashed line at the top represents “Conventional (visible) metrics” remaining steady. Below, a curved line transitions through four colored stages from left to right:
    
        First-wave Optimism (green circle) - “Initial excitement from efficiency gains”
        Asymptomatic Effects (blue circle) - “Subtle vigilance drifts beneath positive gains”
        Chronic Harms (orange circle) - “Cumulative erosion of skill and autonomy”
        Identity Commoditization (red circle) - “Specialists reduced to interchangeable AI overseers”
        
        The line descends progressively, illustrating how “AI's Hidden Costs” increase over time while conventional metrics fail to capture this decline.}
\end{figure*}

\textbf{The Expertise Erosion Cascade:} Analysis of participants' visual narratives revealed a consistent temporal pattern. Following initial first-wave optimism, the expertise erosion cascade unfolded across three distinct stages. First, \textbf{asymptomatic effects} emerged as subtle vigilance drifts hidden beneath positive performance gains. Second, \textbf{chronic harms} manifested through cumulative erosion of skills and professional autonomy. Finally, \textbf{identity commoditization} saw specialists reduced to interchangeable AI overseers, with their expertise flattened into generic roles. Figure~\ref{fig:ParadoxCurve} complements these narratives by illustrating how visible success metrics mask the hidden "expertise erosion cascade. The comic panels vividly capture this arc: from enthusiastic adoption (``15\% faster treatment planning!'') through creeping unease (``intuition.exe has stopped working'') to reduced autonomy (``need AI to be confident'') and finally existential questioning (``what am I?''). 

Having identified this paradox through participants' visual narratives and establishing its three-stage progression, we now examine each stage of the expertise erosion cascade in detail, beginning with the subtle asymptomatic effects that escape traditional evaluation yet seed future harms.

\subsection{Asymptomatic Effects: Hidden Shifts Beneath the Surface} \label{findings: asymptomatic_effects}

To outside observers, RadPlan’s integration looked like a textbook success story with an efficient rollout, no pushback, and strong adoption across the board. However, through our close observations and conversations, we detected subtle changes in behavior and perception that did not register in those metrics, which we term \textbf{asymptomatic effects}--real behavioral shifts that are not immediately alarming or visible. \textit{Asymptomatic does not mean we are helpless}: it means requiring deliberate surveillance through behavioral proxies, a detection challenge we operationalize in our framework later. These asymptomatic effects manifested in several areas, often hiding in plain sight. The stage is marked by subtle behavioral shifts around drifts in vigilance and intuitions that elude standard evaluation metrics that mostly focus on productivity measures, such as speed of case review, throughput, and accuracy.

\textbf{No immediate harm, but warning signs under the surface:} In the early months of RadPlan’s deployment, all quantitative metrics were unequivocally positive (e.g., \~15\% faster planning, slightly better plans). As one hospital administrator put it, ``These numbers are gold, this is exactly what success looks like.'' (A06) Yet for a few clinicians, something felt off. Subtle cracks were forming, not in the metrics but in the methods. Behaviors were shifting, ``shortcuts creeping in'' (D04), and some of the most reflective practitioners began expressing unease. As one put it, ``It’s hard to describe… the system’s working, but my instincts feel duller.'' (R11) These were not big red flags, but quiet ``gut-level warnings'' (R05), the kind that easily slip past performance dashboards.

\textbf{Increased AI reliance \& vigilance drift:} One of the earliest asymptomatic effects emerged in the form of shifts in the rigor of decision-making. Several dosimetrists, for example, started to rely on RadPlan’s first suggestion without exploring as many alternative manual approaches as before. In an observation session, we noted a dosimetrist who normally would generate 3–4 candidate plans by hand now often stopped at 1-2 after seeing the AI’s plan, saying ``It looks good enough.'' (D01) When probed in an interview, she reflected: ``Maybe I’m skipping some of my usual tinkering. The AI plans so fast… and it's decent. It’s so tempting to just accept!'' (D01) Here, we see an early warning: \textit{convenience nudging a reduction in thoroughness}. No errors occurred and plan quality remained  within acceptable bounds; hence, it was asymptomatic in formal terms.

Case complexity partly dictated whether people trusted or questioned AI, often leading to \textbf{vigilance drift}. In routine cases (simple tumor geometries, standard protocols), even skeptics judged it rational to trust the AI and move faster. In \textbf{complex or rare cases}, vigilance increased; a senior RadOnc said: ``For a tricky palliative care case with unusual anatomy, I actually ignore the AI’s plan. Those cases remind me why my intuition matters...the AI doesn’t grasp the nuances of patient context.'' (R03) While \textbf{veterans} seemed to know when to distrust the system, \textbf{less experienced staff} sometimes developed a false sense of security, even in complex cases. The erosion side likely affects those without a strong pre-AI baseline (e.g., newer dosimetrists who have not fully honed manual planning skills), raising serious training and mentorship concerns. 

\textbf{Diminished intuition or critical reflection:} Another quiet but consequential shift occurred in clinicians’ intuition and reflexivity. Several physicians mentioned decreasing reliance on their ``gut'' when reviewing AI-generated plans. As one physicist put it, ``I find myself double-checking the AI’s predictions more than trusting my own gut feeling.'' (P03) This in itself is not negative; cross-checking AI outputs is good practice, but the reason was telling. He continued, ``Over a few months, I think my intuition for what an optimal plan looks like is getting a bit lazy. The AI’s suggestions all start to look `right' to me, even when they’re not perfect.'' The sentiment that intuition was deteriorating was memorably captured by another senior RadOnc who confessed, ``My intuition is rusting.'' (R14) He explained that in complex cases, he used to have an immediate sense if something was off in the plan. However, after routine use of RadPlan, the inner alarm went off less frequently. ``Perhaps,'' he mused, ``I’m not exercising that muscle as much.'' (R14) A physicist further elaborated: ``Old AI was clunky...it had friction and kept us thinking. The new AI is seamless...makes overreliance effortless, offloading the very act of thinking. That's what's dangerous... We aren't just delegating tasks; we're outsourcing cognition, and with it, parts of who we are. No wonder we feel rusty!'' (P12)

These firsthand accounts confirmed our hypothesis: AI can induce a use-now, pay-later effect. Efficiency today at the cost of reduced expertise tomorrow. At this stage, however, the slowly creeping negative effects remain in the shadows and do not surface in obvious ways, largely “asymptomatic” to the organization and compensated by the system’s benefits (faster planning meant more patients treated, slightly better plans, etc.). One hospital administrator asked, ``If everything looks good, what exactly is the problem?'' (A05) The problem, in seed form, was that these small changes in behavior and skill were precursors of potential long-term problems. 
These effects set the stage for what followed. Left unaddressed, small cracks widened. The chronic harms we discuss next are the visible fissures formed over time by slowly expanding asymptomatic cracks.





\subsection{Chronic Harms: Erosion of Skill, Autonomy, Resilience Over Time} \label{findings: chronic_harms}

Around the mid-point of the deployment (\~6 months in), participants’ concerns grew less speculative and more concrete. What were once mild impressions of ``rusty intuitions'' turned into noticeable declines in  capabilities. We use the term \textbf{Chronic Harms} to denote the sustained, accumulating negative impacts that  manifest over time, akin to a chronic condition that sets in after an untreated asymptomatic phase.

\textbf{Demonstrable deskilling:} One clear chronic harm was skill degradation in manual planning and quality-checking. By month 9, at least three different dosimetrists admitted that their speed and proficiency in manual plan crafting had worsened compared with a year ago. One dosimetrist (D07) said: ``If you took RadPlan away… I’d struggle to keep up with the caseload. I’ve gotten slower without it.'' Her comments were echoed by several others. AI had saved time, but in the process, it had also become a crutch. We even conducted an informal exercise: we asked a couple of dosimetrists to create a plan from scratch without AI (as a simulation during a workshop). They completed the task, but it took them longer than it used to, and they expressed discomfort. ``It’s like I’ve unlearned some optimizations I used to know by heart,'' one said (P09). This suggests a vivid example of \textit{deskilling}: the gradual loss of sharpness in a skill due to disuse.

\textbf{Dependency and reduced autonomy:} In parallel, a confidence in human judgment was declining. A physicist described a kind of dependency: ``I always check the AI, and sometimes doubt myself even when I disagree.'' (P09) This indicates an erosion not just of skill, but also of professional self-confidence. Some clinicians started showing signs of \textit{automation bias}, implicitly deferring to the AI’s suggestions, even if their experience signaled otherwise. One RadOnc remarked, ``We need the AI to help with the grunt work, sure, but it can’t be at the cost of losing our instincts.'' (R12). This quote helped others realize that something precious (clinical instinct built on years of learning) was at risk of being dulled. \edit{It sparked a spirited exchange. "That's exactly why we need manual practice time", argued P07. A05 countered, "Good luck selling that when we're 15\% faster." R14 cut through the noise: "But what happens when the AI fails and we've forgotten how to think?" Participants then shared near-misses they had never connected as skill erosion.}

\textbf{Brittle success \& sinking resilience:} RadPlan’s strong performance concealed a \textit{sinkhole} opening beneath the sociotechnical foundation. The technical part (the AI) was working while the socio-part (the human), once a lattice of redundancy and resiliency, quietly thinned with every effortless approval. ``We’re driving on asphalt laid over water,'' a senior physicist warned, capturing the hollowing of safety (P05). A dosimetrist added, ``If RadPlan goes dark for a week, the ground will give way before we remember how to stand on our own.'' (D01) The human safety buffer that once caught mistakes or handled surprises was slowly being chipped away. The system worked well until it did not. As one RadOnc reflected, ``Efficiency is intoxicating, but the real risk is losing the skills that define us.'' (R14) The triumph of automation masked a quiet decline in resilience, solid on the surface yet increasingly vulnerable.

\textbf{Self-motivated Countermeasures:} It is important to note that not everyone succumbed to chronic harms. A minority of participants proactively engaged in countermeasures. One RadOnc made it a personal rule to occasionally ignore RadPlan’s suggestions and craft a plan the ``old way'' to keep in practice: ``I run at least one plan a week without the AI...think of it as sharpening the blade'' (P07). Another physicist created a light-hearted coffee bet each week: the team would try to predict RadPlan’s output before running it, and whoever came closest earned a free coffee from the rest. He added, ``Nothing keeps your brain sharp like a latte on the line'' (P11). These rituals kept intuition sharp. Such practices are heartening, and we discuss them later as potential design or organizational interventions. Nevertheless, these were individual, not institutionalized, initiatives. The general trend was complacency.

In summary, \textbf{Chronic Harms} manifest as enduring degradations in skill, autonomy, and system resilience, sowing the seeds for future identity-level threats. A RadOnc captured the situation aptly: ``Short-term efficiency has been great, but I worry we are slowly giving up the art of what we do. It’s like a muscle that’s shrinking because we aren’t using it as much.'' (R07) This encapsulates the chronic harm: the potential loss of the artistry and deep expertise in a profession, traded for convenient automation. This, in turn, feeds directly into the final theme of this section---how workers see themselves and their value, i.e., identity.

\vspace{-1em}

\subsection{Identity Commoditization: Threats to Professional Identity and Dignity} \label{findings: identity_commodity}

As the shadow of deskilling grew, an even more profound concern emerged: What does it mean to be a professional when an AI takes over much of the ``smart'' work? We refer to this as \textbf{Identity Commoditization:} the fear (or reality) that a professional’s role is being reduced to a commodity, a cog in the AI-driven process, thereby eroding their sense of identity, uniqueness, and dignity at work.

\textbf{Professional identity erosion and role compression:}
Practitioners feared becoming mere ``AI babysitters'' (R05) or ``button-pushers'' (R02) whose expertise no longer mattered. Radiation oncology is a field where practitioners take great pride in their expertise, and rightly so, as their decisions directly impact life-and-death outcomes. Over time, we observed a palpable shift in how some practitioners spoke about their work. Early on, RadPlan was described as a ``teammate'' (R15) or ``assistant'' (D06). Later, metaphors of replacement or reduction started to surface. RadOncs feared the dehumanization of care. They worried that ``AI might push them to become `rubber stampers' of algorithmic decisions,'' diminishing their clinical intuition and autonomy (R02). Physicists feared loss of technical mastery, concerned that they would ``lose the hands-on skills that make [them] unique'' (P14). 

Participants from both physician and physicist groups expressed concern that indiscriminate use of automation could ``erode the very expertise that defines our roles.'' (P08) They worried that if AI replaces the need for their hard-won knowledge, the profession will lose its value. 
They did not fear being replaced. They feared being hollowed out; still employed, but with diminished meaning. In sociology of work, this relates to the concept of \textit{deprofessionalization}, where skilled roles lose their distinctiveness and authority.

\textbf{Dignity Erosion:} There was a direct link between this identity threat and dignity. Historically, dignity in work is tied to the feeling that one’s labor and skill matter and that one is respected for their contribution. When our participants expressed worries about becoming ``bystanders in their own practice,'' (R06), they were also articulating a perceptible loss in dignity.
A senior RadOnc argued in a meeting, ``We must protect the art of medicine so that we don’t become bystanders in our own practice.'' (R03) This comment resonated across the room and was met with solemn nods. It underscored that beyond efficiency and even patient outcomes, there is something fundamentally human at stake: the fulfillment and identity one derives from one’s work.

\textbf{Workforce and economic commoditization:} Interestingly, this theme of identity also touched on value and commodification by external forces (like hospital management or industry). A few participants speculated that if AI can do much of the heavy lifting, hospital administrators might start viewing clinicians as ``just expensive supervisors for the AI'' (P07), reflecting anxieties about being treated interchangeably. While no such action occurred during our study, the perception hinted at anxiety about economic commoditization: the reduction of a highly specialized role into a generic, lower-skilled role in the eyes of decision-makers. This is a longer-term worry that extends beyond individuals to the profession’s future. Several early career participants wondered how the training of new oncologists and physicists should be adapted. ``What will it mean to be a good RadOnc in 10 years? Knowing how to work the AI?'' (R12) one asked rhetorically, implying that if the essence of the role shifts, so does its identity.

A participant framing the stakes: ``I love technology and what it can do. But I also love what I do. I don’t want AI to take over the thinking to the point where I’m just there to sign off. I want to remain a true expert, not an AI’s sidekick. If it gets to that, then what’s the point of all my training? What’s the point of calling me a specialist?'' These words are a call to maintain human dignity and purpose in the face of powerful automations.

\begin{table}[t]
\footnotesize
\centering
\sffamily
\caption{A summary of our findings on the emergence of AI-as-Amplifier Paradox around the Expertise Erosion Cascade spanning across asymptomatic effects, chronic harms, and identity commoditization. }

\setlength{\tabcolsep}{0pt}
\renewcommand{\arraystretch}{1.1}

\begin{tabular}{p{0.995\columnwidth}}
\toprule

\rowcollight \textbf{Asymptomatic Effects:}\\[0.25em]
  \tabitem \textbf{No immediate harm, but warning signs under the surface:} Work gets done, metrics improve, and managers and most users perceive no issue. However, reflective users report quiet gut-level warnings.\\[0.15em]
  \tabitem \textbf{Increased reliance \& vigilance drift:} People accept AI outputs readily and perform fewer independent verifications and explorations.\\[0.15em]
  \tabitem \textbf{Diminished intuition or critical reflection:} Human critical-thinking ``muscles'' are used less, and potential issues escape notice.\\[0.4em]

\rowcollight \textbf{Chronic Harms:}\\[0.25em]
  \tabitem \textbf{Demonstrable deskilling:} Declines in task speed, competency when automation is absent; reduced exploration of alternatives signals skill atrophy.\\[0.15em]
  \tabitem \textbf{Dependency \& reduced autonomy:} Users increasingly defer to automated outputs, hesitate to exercise independent judgment, and perform fewer verification steps, narrowing decision latitude.\\[0.15em]
  \tabitem \textbf{Brittle success \& sinking resilience:} System performance appears strong but remains fragile beneath the surface; overreliance on AI erodes human redundancy and weakens the sociotechnical foundation needed during failures.\\[0.4em]

\rowcollight \textbf{Identity Commoditization:}\\[0.25em]
  \tabitem \textbf{Professional identity erosion \& role compression:} As AI takes over tasks that give work meaning, workers shift into passive oversight roles in which specialized skills fade away and professional identity starts to erode.\\[0.15em]
  \tabitem \textbf{Dignity erosion:} As AI takes over decision-making, users feel present but not empowered, eroding the respect and fulfillment traditionally derived from skilled work.\\[0.15em]
  \tabitem \textbf{Workforce and economic commoditization:} External forces, such as institutions or industry, begin to treat skilled professionals as interchangeable, prioritizing efficiency over specialization and increasing anxiety about long-term role devaluation.\\[-0.3em]
\bottomrule
\end{tabular}
\label{table:Expertise_Erosion_cascade_Summary}
\vspace{-1.2em}
\end{table}


\textit{To close,} our findings chart a trajectory with broader relevance. From silent behavioral drift (asymptomatic effects) to tangible decline (chronic harms) to existential anxiety (identity commoditization), the \textbf{AI-as-Amplifier Paradox} played out in full: AI magnifies both the efficiency and fragility of sociotechnical systems. In the following sections, we learn about a turning point in the journey, which inspires the development of the Dignified Human-AI interaction framework, one that centers on the future of workers, not just the future of work.

\section{Framework Development \& Evaluation: Towards a Dignified Human–AI Future}\label{section:framework_and_ST}

\subsection{Social Transparency: A Turning Point on Navigating the Amplifier Paradox}

Amid the concerning findings above, our data also revealed a \textit{hopeful countervailing force}: Social Transparency (ST). Outlined in Sec.~\ref{sec:study_context}, ST was deployed as a response (reaction) to early signs of “intuition rust” and over-reliance. While it did not solve the entire problem at hand, introducing small design frictions (e.g., transparency cues) provided glimpses of a more dignified equilibrium where humans remained meaningfully in the loop. It revealed actionable mechanisms and first principles to address AI's hidden harms. In doing so, it seeded the more systematic Framework we develop next (Sec.~\ref{sec:derive_framework}). 
Below we detail how ST's 4W overlay (who did what, when, and why) (Fig.~\ref{fig:ST_Radplan}) functioned as a critical intervention. By exposing AI blind spots through peer rationales, ST transformed binary trust into calibrated reliance and preserved human control by anchoring decisions in organizational context and human expertise.

\textbf{ST helped users know when \textit{not }to rely on the AI.} A key function of ST was to \textbf{surface what the AI could not know} from its datasets. For example, the system might not capture international patient visa constraints that shape treatment choices. Such nuances, recorded in the 4W from similar prior cases, helped practitioners calibrate “when to trust the AI and when to rely on their own expertise” (R15). Visibility into limits transformed “trust from binary to a continuum where humans are in control” (R06). Explainability through peer critique helped to calibrate trust in a nuanced manner. Using AI was no longer ‘a take it or leave it’ paradigm requiring blind trust.

\textbf{ST as a check against deskilling and overreliance.} ST acts as a dual force: it \textbf{mitigates over-reliance} and \textbf{supports continuous learning} through peer-driven refinements. By “watching how [their] peers handle complex cases,” participants reported that ST helped them “resist blindly following the AI and sharpened [their] manual skills” (P09, R10). By surfacing peer reasoning in complex cases, “ST stopped [them] from taking the AI’s word as gospel” (R14). ST also offered one pathway to\textit{ make the asymptomatic symptomatic}--not by entirely solving the erosion problem, but by adding just enough friction to interrupt the autopilot mode. "Seeing colleagues override AI suggestions" (R05) reminded practitioners that their own quick approvals might signal "eroding vigilance" (P02). In the context of the Amplifier Paradox, ST \textbf{calibrates reliance} and \textbf{keeps professionals critically engaged}, while giving human expertise room to shine: “Honestly, watching AI mess up is a treat. What is even better is seeing my peers step in to fix it. It shows me where the AI usually fails. It also reassures me that I am still needed. ST gives humans that space to shine” (P12).

\textbf{In sum,} ST interrupted the erosion cycle by surfacing AI limits through peer rationales, turning binary trust into calibrated reliance. This shift from rubber-stamping to thoughtful collaboration (via questioning, verifying, and correcting) fostered dignity and opened a credible path to sociotechnical resilience. However, it also revealed the need for a more proactive approach to address AI's expertise erosion cascade, motivating the framework that follows. 



\vspace{-5pt}

\subsection{The Dignified Human-AI interaction Framework }\label{sec:derive_framework}
\edit{
While ST is helpful, it alone cannot address the root causes of the AI-as-Amplifier paradox.
Building on our empirical findings, we present a \textbf{framework} to \textbf{sense, contain, and recover} and build \textbf{sociotechnical immunity}: the capacity to catch asymptomatic AI ill-effects early and prevent them from hardening into chronic harms. We use “sociotechnical immunity,” echoing participants’ idea of “guerrilla resilience” (R08): “Immunity is building antibodies while working on the cure.” The term works in a \textit{dual register}: for organizations, it signals acceptable adaptation and compliance, and for workers, it invokes~\citeauthor{scott1990domination}’s “hidden transcripts”: outward compliance that quietly builds collective strength~\cite{scott1990domination}.

\vspace{-5pt}

\subsubsection{\textbf{Framework Derivation and Co-construction}}

The framework was co-constructed through five participatory workshops. Workshop 3 marked a turning point. While reviewing Social Transparency 4W logs, a RadOnc observed: “ST helped us catch what the AI can’t see. But, it also showed why we still need humans in the loop.” (R15). Others built on: “That’s actually evidence I used when admin questioned whether we need two physicists per shift. These logs prove we do.” The group recognized that \textit{documentation served dual purposes}: satisfying quality requirements while proving human necessity, exemplifying what Khovanskaya and Sengers call “data rhetoric,” where institutional data collection becomes a tool for worker advocacy~\cite{khovanskaya2019uneasyalliance}.

By Workshop 4, participants articulated their constraints as knowledge workers, highlighting the idea of the \textit{\textbf{"missing middle”}}: “The future of work talk fixates on the extremes...gig workers get sympathy, CEOs get golden parachutes. We're the missing middle that gets erased: too rich for unemployment benefits, too gagged by NDAs, invisible in job loss statistics. Who will help us?'' (R10). The topic of organizing sparked contention: “If I had a cent for every 'just organize' comment. No shit Sherlock, do y’all think we haven’t thought of that? Unionizing is illegal in this state!” (P07)

Workshop 5 shifted to “what can we do about this given these constraints?” Participants bluntly captured this sentiment: “You know, revolution sounds great in theory until you feel how precarious we are. We cannot risk a fight with management. \textit{If we're destroyed today, there's no organizing tomorrow. We need both...survival now, transformation later, and the bridge between them, ideally by turning their own system against them.}” (P03, emphasis added)

Through Workshops 4 and 5, participants converted survival tactics into framework questions. They worked from specific incidents outward; for instance, they used quality documentation to protect jobs. This mirrors what Khovanskaya and Sengers call ``uneasy alliances'' \cite{khovanskaya2019uneasyalliance}, where workers repurpose institutional infrastructure for protection.

\textbf{Framework Structure:} The framework operates on three interlocking levels – (a)~Worker, (b)~Technology, and (c) Organizational. Each level includes a narrative focus and a tailored set of questions to trigger reflections and actions. The levels work in tandem: the Worker level cultivates individual critical awareness, the Technologist level fosters the design of skill erosion‑resistant and dignity-preserving AI tools, and the Organizational level institutes policies and safeguards for productive but also humane forms of human-AI interaction and collaboration. 


\subsubsection{\textbf{Individual Worker Level (maintaining agency and expertise)}} 


At the worker level, the framework centers on mindful and critical engagement with AI. Practitioners notice habits, name reliance thresholds, and protect the parts of  their craft that give their work meaning. In the workshops, participants identified practices for preserving expertise while meeting productivity demands. R15 shared: “I do one manual contouring session a week. On paper it is ‘quality assurance for AI validation.’ For me, it keeps my skills alive.” These tactics exemplify \citeauthor{scott1985weapons}'s “weapons of the weak”: everyday resistance through legitimate institutional practices~\cite{scott1985weapons}. Participants raised concerns that became the framework’s focus on identifying core professional values. R02 asked : “Which tasks give my work meaning beyond efficiency metrics?” P09 contributed: “Which skills have I lost fastest since RadPlan?” R10 addressed early warning signals of vigilance drift: “I track instant approvals as an early warning signal now. If I rubber-stamp three cases in a row, that is my trigger for an AI-off review.” Recognizing that questions cannot be predetermined, participants proposed “starter pack questions”: generative prompts that catalyze ongoing inquiry rather than prescriptive checklists:
\begin{myshade}
\small 
\textbf{Starter pack questions for the Worker Level}
\begin{enumerate}[label=(\arabic*)]
    \item \textbf{Craft Essence:} Which parts of my workflow give me meaning? What would be lost if they were to be automated? (W1)
    \item \textbf{Skill Atrophy:} Which skills risk fading under AI assistance? What practices will keep them sharp? (W2)
    \item \textbf{Early Warning Signals:} How will you spot early signs of over‑trust on AI such as instant approvals? How will you trigger an AI‑off check? (W3)
\end{enumerate}
\end{myshade}
\vspace{-1\baselineskip}
\subsubsection{\textbf{Technology Level (building erosion-resistant systems)}}

At the technology level, the framework centers on design features. Technologists embed dignity-preserving, erosion-resistant properties that amplify rather than replace human judgment. This includes building reflection triggers, useful friction, and cues that surface mismatches instead of hiding uncertainty~\cite{ehsan2022seamful}. Participants grounded this level in how system design affected skill preservation. They highlighted how “seamlessness in AI prevents questioning” (P09). R06 proposed that ``systems should surface uncertainties and gaps.” Participants noted that “AI has no clue what we enjoy. Start by automating the work that drains us, not the parts we value.” (D04). They also emphasized reliance calibration over time. R12 described tumor boards: “We bring AI-disagreement cases to see where the system is brittle and to remember its limits.” Participants wanted the same disagreements to retrain the model: “If I keep overriding the AI in the same spot, it should learn and lower its confidence there” (P10). These discussions seeded the starter-pack questions on leveraging seams, choosing relevant automation targets, and calibrating reliance by task risk at the technology level.


\begin{myshade}
\small 
\textbf{Starter pack questions for the Technology Level}
\begin{enumerate}[label=(\arabic*)]
    \item \textbf{Leverage Seams:} How will systems surface uncertainties, edge cases, and hand‑offs so users know when (not) to rely on AI? What in-the-moment cues prompt critical reflection? (T1)
    \item \textbf{Relevant Automation:} Which manual workflow steps frustrate workers? Which tasks should be automated? (T2)
    \item \textbf{Reliance Calibration:} How can people adjust AI assistance to match task risk? How will the system learn from human corrections and clearly recognize human expertise? (T3)
\end{enumerate}
\end{myshade}
\vspace{-1.5\baselineskip}
\subsubsection{\textbf{Organizational Level (governing dignified Human-AI collaboration)}}

At the organizational level, the framework addresses policies and oversight that ensure AI integration supports workers. This level operates on the systemic `contain' and `recovery' modules of the immunity triad, creating the institutional infrastructure to act on early warning signals. Participants stressed that “without organizational buy-in and clear policies, this will not go anywhere.” (R10). An administrator stated the challenge: “The board wants a 20\% productivity jump. Clinicians want quality. Those goals collide.” (A04). R06 proposed working within existing structures: “ Why not frame skill preservation as quality improvement? Direct confrontation won’t work. Why not use their requirements to meet \textit{our }goals.?” (emphasis added). This strategic framing, what Spivak calls “strategic essentialism” \cite{spivak2023can}, allowed them to use institutional language for worker protection. They debated what to automate and what to keep human. R03 argued: "Some decisions should never be algorithmic." This sparked the idea of “do-not-automate” lists.  P13 asked: “If AI saves time, should that go to more patients or skill development?”  Participants agreed that saved time should be “reinvested into manual practice and peer review” (P03). Liability came up: “Who's legally responsible when AI fails?”(P02). A07 responded: “If I'm responsible when AI fails, I need veto power without retaliation.” These concerns shaped the starter-pack questions on humane automation boundaries, protected use of saved time, and clear human authority and liability.


\begin{myshade}
\small 
\textbf{Starter pack questions for the Organizational Level}
\begin{enumerate}[label=(\arabic*)]
    \item \textbf{Humane Automation Boundaries:} Which automation points are least likely to displace jobs? Which tasks belong on a living do‑not‑automate list? (O1)
    \item \textbf{Saved Time:} How will time saved by AI be set aside for craft, care, or teaching? (O2)
    \item \textbf{Human Authority \& Liability:} What is the no‑penalty path to override or shut off AI? Who is accountable when automated decisions cause harm? (O3)
\end{enumerate}
\end{myshade}
\vspace{-1\baselineskip}
Through these workshops, we co-constructed a framework that operates within professional constraints while building systematic defenses against expertise erosion. The framework functions as pragmatic resistance: working within institutional requirements while preserving the human expertise that defines professional practice.
\textit{This is not about accepting AI’s inevitability but about recognizing present power asymmetries and working from the margins.}
It builds \textbf{sociotechnical immunity} through an \textbf{immunity triad}: \textit{sensing} asymptomatic drift early, \textit{containing} it decisively, and \textit{recovering} deliberately. This triad (1) guards against hidden erosion while enabling safe amplification, (2) offers a shared language for negotiation across roles, and (3) reinvests saved time into human growth through drills, rationales, and peer learning. This approach recognizes that knowledge workers face unique challenges that require context-specific tools. By equipping stakeholders to navigate the \textbf{AI-as-Amplifier Paradox}, the framework helps humans retain ownership of meaning-bearing tasks, expertise, and agency.

} 


\begin{myshade}
\small 
\begin{itemize} 
    \item \textbf{Early-warning indicators (sense):} Detect subtle drift before chronic harm; e.g., instant-approval alerts, zero-edit merges, missing peer rationales; drawn from Worker questions (Craft Essence, Skill Atrophy, Early Warning Signals) and Tech prompts (Leverage Seams, Reliance Calibration).
    \item \textbf{Containment actions (contain):} Halt erosion's spread through circuit breakers; e.g., AI-off reviews, two-person confirmations, seamful prompts at decision points; aligned with Org-level safeguards (Humane Automation, Human Authority \& Liability) and Tech prompts (Leverage Seams).
    \item \textbf{Recovery routines (recover):} Deliberately rebuild skills through practice; e.g., manual-first drills, saved-time reinvestment, override logs that feed model updates; tied to Worker level (Skill Atrophy), Org level (Saved Time), and Tech (Reliance Calibration). 
\end{itemize}
\end{myshade}
\vspace{-2\baselineskip}

\begin{figure*}[t]
    \centering
     \includegraphics[width=1.9\columnwidth]{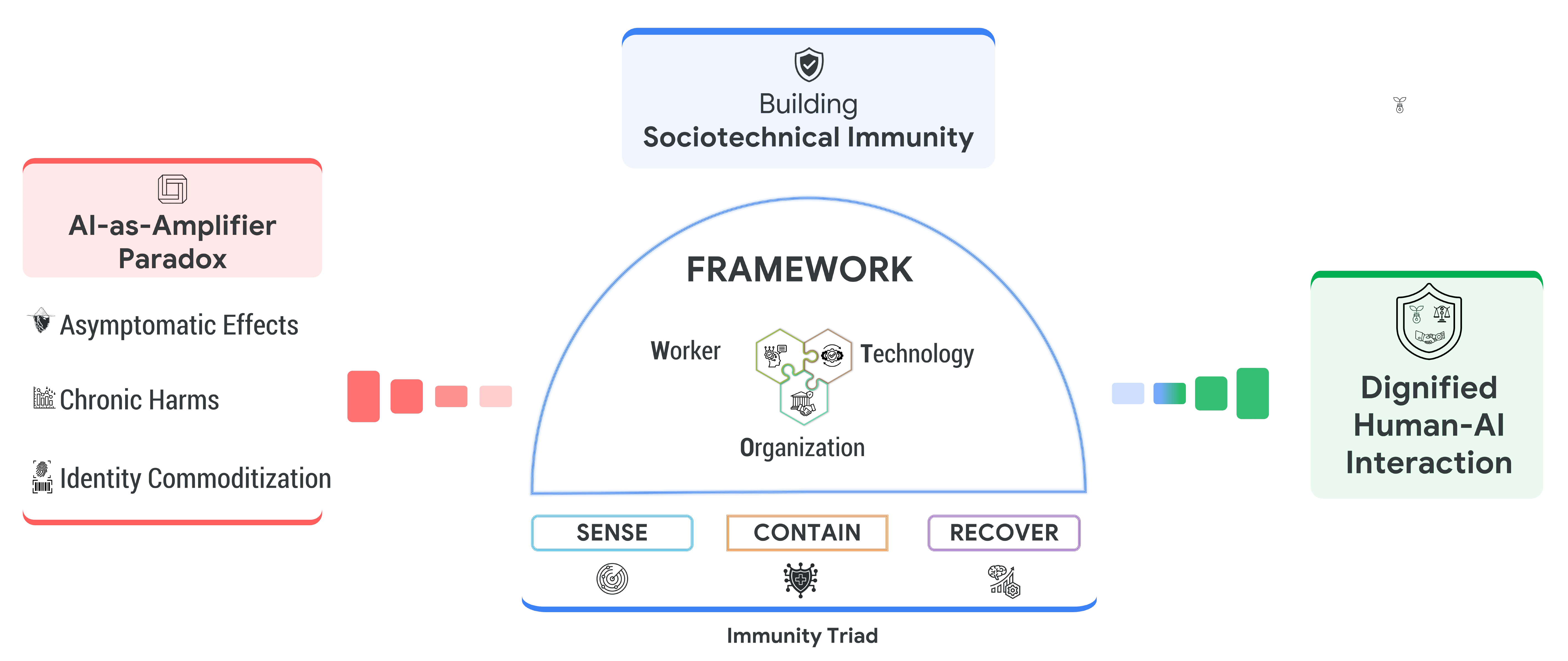}
     \vspace{-0.5em}
  \caption{Visual narrative of the erosion-to-dignity journey. The AI-as-Amplifier Paradox (left) threatens human expertise. The framework (middle) addresses this by building sociotechnical immunity across Worker-Technology-Organization levels using the Immunity Triad (Sense, Contain, Recover). This process ultimately fosters Dignified Human-AI Interaction (right).} 
            \label{fig:frameworkImmunity}
    \Description[figure]{A conceptual diagram with three main components arranged horizontally:
        Left section (red box): “AI-as-Amplifier Paradox” listing three stages with fading red squares: Asymptomatic Effects, Chronic Harms, and Identity Commoditization.
        Center section (blue semicircle): “FRAMEWORK” containing three interconnected hexagons labeled Worker, Technology, and Organization, forming a triangle.
        Right section (green box): “Dignified Human-AI Interaction” with ascending green bars showing positive progression.
        Top (blue banner): “Building Sociotechnical Immunity” with shield icon.
        Bottom (blue banner): “Immunity Triad” showing three connected elements: SENSE (radar icon), CONTAIN (bug/virus icon), and RECOVER (brain/refresh icon).
        The diagram illustrates how the framework addresses AI's paradoxical effects through an immunity system that protects dignified human-AI interaction across worker, technology, and organizational levels.
        
        }
    \vspace{-10pt}
\end{figure*}

\subsection{Framework Effectiveness Assessment}

\subsubsection{\textbf{Methods \& Analysis:}} \edit{We ran two participatory workshops to assess the Dignified Human-AI Interaction Framework's effectiveness in healthcare and software engineering. The first convened 7 participants (3 RadOncs, 2 physicists, 1 dosimetrist, 1 administrator; all new to our study to prevent developer-evaluator conflation and strengthen validity) to co-construct scenarios around RadPlan, testing \textit{within-domain} robustness.} The second convened 10 participants (5 developers, 3 UX researchers, 2 engineering managers) experienced in AI-assisted programming, demonstrating preliminary \textit{cross-domain} transfer. Both groups employed \textbf{participant-driven scenario-based design}~\cite{ehsan2022seamful,rosson2009scenario} to craft ecologically-situated scenarios, then systematically applied the framework's guidelines and questions. Participants critically evaluated feasibility and affordances post-application. In total, we analyzed 12 hours of data using methods detailed in Section~\ref{sec:data_collection_analysis}. Below, we present participant-generated walkthroughs and effectiveness findings.


\edit{
\subsubsection{\textbf{Framework Walkthrough (within-domain): RadPlan}}

\textit{Scenario:} Six months after Hospital Alpha deployed RadPlan, initial efficiency gains gave way to subtle concerns. Dr. Perez, a senior radiation oncologist, noticed that her junior dosimetrist, Ben, had stopped producing alternate plans and was often rubber-stamping RadPlan's first suggestion. Alicia herself felt her “sixth sense” for spotting weak plans was fading. Meanwhile, Dr. Chen, the department head, heard that while throughput had increased, many staff felt their work was becoming less meaningful.

\textbf{How the framework guided action:} \textbf{Baseline (pre-intervention):} The workshop began by mapping their starting point. Initially, RadPlan made planning faster, and clinicians trusted the tool. However, junior staff members gradually stopped creating manual alternates and allowed RadPlan to handle routine steps. This drift went unnoticed; no metrics flagged concerns, yet clinicians were approving plans on autopilot.

\noindent \textbf{Early Warning Signals:} Using the framework's questions, the team then identified the red flags they'd missed:
    \begin{itemize} [nolistsep,noitemsep,left=0.5em]
        \item \textit{Worker} (W3 Early Warning): Clinicians notice themselves approving plans without manual checks, sensing their vigilance slipping. Dr. Perez admitted, “I was clicking through plans like email.”
        \item \textit{Technology} (T3 Reliance Calibration): The RadPlan interface shows many single-click “instant” approvals (the red-dot indicator) on borderline cases. Ben recognized this pattern in his own logs.
        \item \textit{Organization} (O3 Human Authority \& Liability): Missing peer rationales in the plan record; approvals without documented justification. “We were all trusting the machine,” Dr. Chen observed.
    \end{itemize}

\noindent 
\textbf{Sense → Contain → Recover:} The group then walked through \textit{the immunity triad in action}:
    \begin{itemize} [nolistsep,noitemsep,left=0.5em]
        \item \textit{Sense:} Instant one-click approval of a borderline plan (red-dot) | \textit{Contain:} invoke an AI-off peer review | \textit{Recover:} record an override and prompt another clinician for rationale. “That red dot becomes a pause button,” Ben explained.
        \item \textit{Sense:} Plan approved with no peer rationale note | \textit{Contain:} block completion until a second clinician provides justification | \textit{Recover:} the reviewer's name and written rationale are saved with the plan, creating accountability.
    \end{itemize}

\noindent 
\textbf{Interventions:} The team designed targeted responses, each addressing a specific vulnerability:
    \begin{itemize} [nolistsep,noitemsep,left=0.5em]
    \item Establish a living do-not-automate list for complex cases requiring human judgment: \textit{preserve critical expertise} (O1 Humane Automation)
    \item Monthly AI-off alternate-planning drills: \textit{preserve skill} (O2 Saved Time). The team would dedicate saved time to skill maintenance.
    \item 'Red-dot' instant approval alerts: \textit{interrupt autopilot} (W3 Early Warning). Visual cues to break the automation trance.
    \item Require a second clinician to review and annotate high-risk plans: \textit{ensure human accountability} (T3 Reliance Calibration).
    \end{itemize}

\noindent 
\textbf{Envisioned outcomes:} After walking through these steps, the team reported that these interventions could lead to renewed ownership over planning: RadPlan continued to speed routine work, but clinicians once again held control of complex decisions. “I feel like a doctor again, not a rubber stamper,” Dr. Perez concluded.

Participants felt that the framework helped them sense quiet drift in routine planning, contain it with simple in-line safeguards and second-reader rationale, and recover through logged traces and drills. The pragmatic nature of the framework resonated with them with Dr. Chen reflecting, “This helps us survive the AI revolution today while building resilience for tomorrow.” It offered a tractable way to preserve skills and clinical agency across Human-AI settings.

} 

\subsubsection{\textbf{Framework Walkthrough (cross-domain): AI‑Assisted Billing Microservice}}
\textit{Scenario:} CodeCraft's platform team integrated "AutoCoder" to accelerate work on a new \textit{billing microservice}. Three months in, throughput soared, but a near-miss surfaced: an AI-generated tax-calculation snippet passed tests yet rounded pennies the wrong way. Evan (junior engineer) had merged it without review; Priya (senior) caught the error late while auditing logs. The team worried that unchecked reliance on AutoCoder could let small rounding errors quietly erode revenue over time. 

\textbf{How the framework guided action:} 
\textbf{Baseline (pre-intervention):} The workshop participants first mapped out how things had been working. Initially, AutoCoder produced scaffolding code that passed standard tests, and developers merged it quickly. AI had become their trusted first-draft writer: why question something that consistently delivers? Complex finance rules (tax formulas, compliance checks) were rarely handled manually. Under this baseline, output metrics were high, but developer vigilance had quietly relaxed. “We were living in a false paradise,” Priya noted, “the numbers looked great on paper.”
    
\noindent \textbf{Early Warning Signals:} Using the framework's questions, the team then identified the red flags they'd missed:
    \begin{itemize} [nolistsep,noitemsep,left=0.5em]
        \item \textit{Worker} (W2 Skill Atrophy): When pressed, the group realized that Priya's discovery of the rounding bug wasn't luck—it signaled emerging complacency across the team. “If our senior dev almost missed this, what are the juniors missing?,” Evan asked.
        \item \textit{Technology} (T3 Reliance Calibration): Digging into repository logs revealed an uncomfortable truth: there were many zero-edit commits (AI merges with no human edits). Code flowed straight from AutoCoder to production.
        \item \textit{Organization} (O3 Human Authority \& Liability): Most damning of all, critical billing code changes were going live without a named human reviewer or written explanation in the logs. “Who's accountable if this goes wrong?” The silence was telling.
    \end{itemize}
\noindent \textbf{Sense → Contain → Recover:} The group then practiced applying \textit{the immunity triad in action}: 
    \begin{itemize} [nolistsep,noitemsep,left=0.5em]
        \item \textit{Sense:} Zero-edit AutoCoder commit | \textit{Contain:} enforce an AI-off paired review | \textit{Recover:} log the override event. 
        “Like a circuit breaker,” Priya explained, “it stops the automatic flow and forces human engagement.”
        \item \textit{Sense:} Missing edge-case test (e.g., no negative-amount handling) | \textit{Contain:} require immediate manual implementation of test | \textit{Recover:} record new test code and author. This ensures accountability and learning.
    \end{itemize}
\noindent \textbf{Interventions:} With problems identified, the team designed their response. Each intervention addressed a specific vulnerability that they had uncovered. 
    \begin{itemize} [nolistsep,noitemsep,left=0.5em]
        \item Mark critical finance logic as human-only (“Do-Not-Automate” list): \textit{preserve judgment} (O1 Humane Automation). “Money math stays in human hands,” became their mantra.
        \item Enforce AI-off paired reviews on flagged commits: \textit{ensure oversight} (O3 Human Authority \& Liability). Two sets of human eyes on anything touching revenue.
        \item Display “Why is this safe?” uncertainty prompts in the IDE: \textit{surface limitations} (T1 Leverage Seams). Make AutoCoder's blind spots visible at the moment of decision.
        \item Conduct monthly manual coding drills (e.g., re-implement a tax rule from scratch): \textit{preserve skill} (W2 Skill Atrophy; O2 Saved Time). “Use it or lose it,” Priya reminded the group.
    \end{itemize}

\noindent \textbf{Envisioned outcomes:} The session concluded with the team envisioning their new reality. They would rewrite the tax snippet code manually, add thorough rounding tests, and enable override logging. With these steps, AutoCoder could still handle routine code while humans guarded precision-critical logic. “It's not about rejecting the AI,” Evan reflected, “it's about knowing when to take the wheel.”

The framework enabled participants to notice when their reliance on AI became excessive, respond by activating human-only safeguards and calibrated reviews, and recover by referencing override logs and reclaiming time for skill practice. This approach offered a transferable pattern for erosion-resistant AI.

\vspace{-5pt}
\subsubsection{\textbf{Framework Evaluation Findings}}
Below we present our findings on the effectiveness of the framework, related challenges, and how we may address the challenges.

\textbf{Shared Vocabulary as Operational Infrastructure: } The framework enabled participants to transform abstract values like dignity and safety into tangible artifacts, creating a shared grammar that supported coordination, negotiation, and institutional memory. Participants described that “the framework made disagreement productive.”. A RadOnc articulated how the framework provided “practical language for what we were already feeling but couldn't articulate to management.” The shared vocabulary enabled workers to present a united voice, as articulated by one participant: “Before, we each thought we’re going crazy watching our skills rust. Now we’ve the collective language and data to prove it… and [the administration] can’t just gaslight us if they try to cut headcount or question our value.”

\edit{
\textbf{Pragmatic Resistance Through Professional Excellence:} Participants linked observable signals (red-dot instant approvals, zero-edit merges) to actionable triggers (AI-off review, rationale requirements) and traceable records (override logs, peer annotations). This created what a Physicist called “quality metrics that double as survival tools.” Participants valued the frameworks “dual-purpose design…satisfying institutional requirements while preserving human expertise without confrontation.” A developer observed: “The same logs that prove system reliability prove we can't be replaced.” A dosimetrist observed: “Every time I document why I rejected the AI's suggestion, I'm building a case file for my own necessity.” Participants valued the “pragmatism in the framework” (manger) that turned institutional monitoring into self-preservation, “repurposing metrics meant to measure them into mechanisms that protect them” (UX researcher).
}

\textbf{Sustaining Expertise Protective Practices Over Time: }While the framework jumpstarted new protective routines, participants across both domains expressed sustainability concerns. The proposed fix was lightweight instrumentation: embedding drill completion rates in existing QA dashboards (whether for treatment planning or code review), appointing rotating “craft champions,” and reviewing one override log per clinical round or sprint. A physicist noted: “It has to live in our existing workflows or it dies.” An engineer agreed: “The moment it becomes ceremony, we've lost.”

The assessment workshops revealed the framework's value as a practical infrastructure for dignified Human-AI work. Participants gained a shared language to turn private anxieties about expertise erosion into collective evidence and convert routine metrics into “quality survival tools.” These facets satisfied institutional reporting requirements while safeguarding human expertise without requiring open confrontations. Appendix~\ref{sec:Metrics_Eval_Framework_Appendix} contains best practices on metrics and evaluation plan. Future work should evaluate the framework in additional domains, as outlined in Sec. ~\ref{sec: limitations_futurework}.
\vspace{-5pt} 
\section{Discussion \& Implications}\label{section:discussion_implications}
Our findings illustrate a microcosm of a larger phenomenon in the modern workplace: AI technologies can simultaneously uplift and undermine human labor. This \textit{AI-as-Amplifier Paradox} forces us to confront critical questions about what we truly value in the future of work. This section advances the conversation. We discuss the dual-purpose affordances of the framework and share implications for design, research, and practice.

\subsection{Addressing the Missing Middle: The Framework's Dual Architecture for Knowledge Worker Protection}
\edit{

The Dignified Human-AI Interaction Framework addresses a critical gap in worker protection that emerged during our study. Despite their relative privilege, knowledge workers in contemporary AI-mediated workplaces face a counterintuitive form of precarity. They remain structurally vulnerable in what participants termed the ``\textbf{missing middle}'' (R10, D03, P07): it includes professionals like radiation oncologists, software engineers, and data scientists who earn too much to qualify for unemployment assistance (typically exceeding \$100K thresholds), face NDAs that prohibit discussing workplace conditions, and work in states where unionization may be difficult or outlawed. Our \textbf{framework operates in the space between accommodation and resistance}, which postcolonial scholars recognize as \textit{tactical navigation from marginalized positions}~\cite{scott1985weapons,spivak2023can}. It addresses the structural trap by offering tactical tools that support survival under current constraints while laying the infrastructural groundwork for future collective resistance.


\vspace{-5pt}
\subsubsection{Dual-Purpose Design: Quality Standards as Worker Protection Infrastructure}
The framework's value proposition lies in its dual-purpose architecture: every mechanism simultaneously serves professional excellence and organizational goals while building worker power. This duality emerged from participants' recognition that visible resistance meant professional suicide. As detailed in ~\ref{sec:derive_framework}, the Sense-Contain-Recover triad intentionally serves institutional priorities, documentation, quality control, and workflow efficiency while enabling workers to preserve agency, sustain expertise, and build collective visibility.
Social Transparency exemplified this approach. While appearing to improve AI adoption, its 4W documentation (who, what, when, why) actually created forensic evidence of AI limitations. As one participant observed: “ST just didn’t help us calibrate reliance but helped us bring receipts when upper management tried to shove more AI into the workflow. We caught 23 dose constraint violations the AI missed last month. That's evidence.” (R12). This team used ST 4Ws to prevent a staff reduction, transforming quality assurance documentation into what Tang et al. call “data for advocacy," evidence for collective bargaining~\cite{tang2023back}.

\vspace{-5pt}
\subsubsection{Threading the Pragmatic Needle Between Survival and Transformation}
Taken together, the framework's components enable knowledge workers to navigate asymptomatic harms without direct confrontation. The framework transforms professional excellence, traditionally an individualizing force, into a collective protection infrastructure. Participants co-opted managerial goals to serve their own ends. This created what Khovanskaya and Sengers call “uneasy alliances”~\cite{khovanskaya2019uneasyalliance} between institutional data collection and worker advocacy. Workers generated evidence that surfaced AI's limitations while preserving the core elements of their craft.
This approach extends the organizing strategies developed for other worker categories, not because knowledge workers deserve special treatment, but because their structural position requires different tools. Gig workers successfully made precarity visible through public campaigns; knowledge workers bound by NDAs cannot. Factory workers leverage geographic concentration; distributed professionals cannot. The framework learns from these movements while providing mechanisms calibrated to professional constraints, adding new tools to the collective toolkit of knowledge workers’ resistance.

The framework is neither positioned as a complete answer nor premised on AI's inevitability. It is a pragmatic tactical response: equipping workers to “deal with the AI that's already here today” (P02). It operates through professional norms, using the system's own logics (quality, compliance) as vehicles for protection. This enables knowledge workers to start now, using professional development to protect their expertise while building infrastructure for future organizing. 
\vspace{-5pt}
} 

\subsection{Research \& Design Implications}

\textit{In terms of research implications,} our work reveals a critical blind spot in Human-Centered AI research: an overwhelming focus on enhancement metrics and neglect of erosion dynamics.  Current evaluation frameworks focus on efficiency and task performance, which are important but insufficient. These metrics capture what AI adds but ignore what it quietly subtracts. The AI-as-Amplifier Paradox demands an expanded evaluative lens that encompasses both enhancement and erosion. Human-Centered AI must ask not only “does AI help performance?” but also “does AI sustain workers over time?” We join calls to make \textbf{dignity} an explicit success criterion for AI in the workplace~\cite{clark2020artificial,kawakami2023sensing,stahl2021artificial,das2024sensible,roemmich2023emotion}. This shift raises the question of whether an AI system not only delivers output but also preserves (or even enhances) the user’s long-term agency, skills, and professional purpose. We require metrics that track long-term expertise retention, professional agency, and skill resilience.  Longitudinal studies, such as ours, can address these needs.

The field must embrace \textbf{erosion-resistant AI technology} as a design goal, alongside enhancement. This means building systems that preserve human capabilities even while automating tasks. In this case study, Social Transparency demonstrated a counterbalance to deskilling and overreliance by exposing AI's blind spots and foregrounding peer contributions. However, erosion resistance runs deeper than interface tweaks and challenges assumptions about automation. Rather than pursuing maximum efficiency through automation,  \textit{humane points of relevant automation:} complex areas that benefit from automation but  do not displace human roles. Our framework's "do-not-automate lists" operationalize this principle, but the field needs systematic methods for identifying these dignity-preserving human-only tasks. \textit{The argument here is not about limiting AI's potential} but about designing AI that amplifies dignity in work rather than merely work output. The distinction matters: a system that makes workers 50\% more productive today but 30\% less capable tomorrow fails the promise of human-centered design.

\textit{In terms of design implications,} the goal is to embed \textbf{sociotechnical immunity} from inception, not as an add-on later. In this case study, \textit{Social Transparency} (revealing who did what, when, and why) helped distribute credit between humans and AI, supporting contestability and calibrating reliance.  \textit{Seamful AI design}~\cite{ehsan2022seamful} can surface and leverage seams (mismatches, gaps, limitations) to redirect attention where human judgment adds most value, transforming potential weaknesses into collaboration points. Beyond these, designers can add \textit{cognitive forcing functions} to introduce checks at risk points, \textit{uncertainty communication} to align trust with competence, \textit{provenance trails} to create institutional memory from inputs and overrides, and \textit{AI-off drills} to maintain core skills despite automation.
These features can help builders resist the temptation to optimize everything for efficiency, instead balancing the benefits of automation with the deliberate preservation of human expertise. 


\subsection{The Economic Case for Dignified Human-AI Interaction: Preventing Catastrophic Costs}\label{sec:business_case_dignifiedHumanAI}

A Dignified Human-AI Interaction ethos serves everyone. 
Even profit-focused enterprises benefit from preventing catastrophic system failures. Preserving human expertise is a risk control and revenue protection strategy. 
Over-automated pipelines create single points of failure; when rare faults, distribution shifts, or adversarial inputs hit, teams without live skills cannot diagnose or stop the bleed. \textit{This is why we need humane, not unmitigated, automation.} History shows that every “foolproof” automated system eventually encounters the fool it was not proofed against: e.g., Knight Capital lost \$460 million in 45 minutes from algorithmic trading~\cite{SEC2013KnightCapital}; Boeing's MCAS killed 346 people~\cite{DOTOIG2020MAX}; Zillow Offers lost \$500 million due to AI mispricing~\cite{troncoso2023algorithm}. Regulators want human oversight: the EU AI Act~\cite{panezi2024article} requires continuous oversight for high-risk systems, and the NIST AI Risk Management Framework~\cite{ai2023artificial} centers oversight. Hollowing out entry-level roles to “let the AI do it” starves the talent pipeline, inflates recovery costs, and deepens vendor lock-in. “Better AI later” is not a plan; it is an unpriced option on a moving target: markets, data, and rules change faster than models mature. Companies that bet against human expertise are now desperately hiring the same experts to repair what their AI confidently broke~\cite{alsibai2025desperate}. Thus, even with a profit-driven mindset, a dignified Human-AI approach is pragmatic: sense asymptomatic drift (e.g., instant-approval metrics), contain automation creep (e.g., AI-off reviews, Seamful AI), and recover capability (e.g., reinvesting AI-based efficiency into AI-off drills and peer learning). The result is higher uptime, faster incident recovery, lower regulatory exposure, and a durable advantage that pure automation cannot match.
\vspace{-5pt} 
\section{Limitations and Future Directions} \label{sec: limitations_futurework}
This work marks the first articulation of how AI amplifies both utility and erosion over the long term, conceptualizing asymptomatic effects and chronic harms, and offering a framework for sociotechnical immunity as a response. Given this foundational contribution, our insights should be scoped accordingly. The framework underwent cross-domain verification and a multi-site deployment, yet findings came from a single domain. \edit{Future work can explore these dynamics in multiple domains (finance, cybersecurity, professional training) and populations (e.g., students during AI literacy acquisition) with varied regulatory environments and AI tools to refine immunity mechanisms.} Our longitudinal qualitative approach captured the lived experiences of skill erosion. Future mixed-methods studies can leverage our framework to quantify degradation rates, recovery timelines, and immunity thresholds across populations. Additionally, our case involved teams within single organizations; examining contexts where AI vendors and end users span different institutions would illuminate additional power dynamics.

Looking ahead, this work draws on Agre’s design philosophy of Critical Technical Practice: “at least for the foreseeable future, [we] will require a split identity…one foot planted in the craft work of design and the other foot planted in the reflexive work of critique.”~\cite{agre1997toward}. We have planted “one foot” in the craft of empirical design (deploying probes, crafting the framework). Now, we seek to learn from and with the HCI and AI communities to plant the other foot in the self-reflective realm of critique, interrogating AI's deeper implications for the future of workers.
\vspace{-5pt}
 
\section{Conclusions}
AI reshapes workflows and redefines workers. Our longitudinal study in radiation oncology showed that the AI-as-Amplifier Paradox manifests through a progression from asymptomatic effects to chronic harms to identity commoditization. However, this erosion is not inevitable. Through interventions such as Social Transparency and our multi-level framework for dignified human-AI interaction, we demonstrated how to build sociotechnical immunity: the capacity to detect, contain, and recover from AI's hidden harms before they ossify. The shift from the ``future of work'' to the ``future of workers'' demands that human dignity and skill preservation be treated as first-class requirements alongside efficiency. As AI proliferates, the question is not whether it will transform work but whether that transformation will honor or hollow out human expertise. Our framework offers concrete tools for stakeholders to navigate this terrain. Developed with workers rather than for them, it operates through professional norms, using institutional logics of quality and compliance as vehicles for protection. This dual-purpose design addresses knowledge workers as the ``missing middle'': professionals too privileged for conventional collective action yet structurally vulnerable to AI's erosive effects. By centering workers, not just workflows, we can ensure that the future of work does not undercut the future of workers.


\begin{acks}
With our deepest gratitude, we acknowledge the time that our participants generously invested in this project. Without their input, this project would not have been possible. We also want to thank the organizations, the sites for the case studies, for their cooperation. At multiple stages of the project over the years, we are appreciative of the insightful discussions with Michael Muller, Drew Hope, Tim Showalter, Bill Noyes, Peter Hobban, Jean Wright, Parham Alaei, and Julie Shade. 

\end{acks}

\bibliographystyle{ACM-Reference-Format}
\bibliography{referencesCleaned,upol_refs}


\begin{thebibliography}{106}


\ifx \showCODEN    \undefined \def \showCODEN     #1{\unskip}     \fi
\ifx \showISBNx    \undefined \def \showISBNx     #1{\unskip}     \fi
\ifx \showISBNxiii \undefined \def \showISBNxiii  #1{\unskip}     \fi
\ifx \showISSN     \undefined \def \showISSN      #1{\unskip}     \fi
\ifx \showLCCN     \undefined \def \showLCCN      #1{\unskip}     \fi
\ifx \shownote     \undefined \def \shownote      #1{#1}          \fi
\ifx \showarticletitle \undefined \def \showarticletitle #1{#1}   \fi
\ifx \showURL      \undefined \def \showURL       {\relax}        \fi
\providecommand\bibfield[2]{#2}
\providecommand\bibinfo[2]{#2}
\providecommand\natexlab[1]{#1}
\providecommand\showeprint[2][]{arXiv:#2}

\bibitem[ACR(2023)]%
        {ACR2023factsheet}
 \bibinfo{year}{2023}\natexlab{}.
\newblock \bibinfo{title}{ACR Fact Sheet | American College of Radiology}.
\newblock
\urldef\tempurl%
\url{https://www.acr.org/About-ACR/Fact-Sheet}
\showURL{%
\tempurl}


\bibitem[Acemoglu and Autor(2011)]%
        {acemoglu2011skills}
\bibfield{author}{\bibinfo{person}{Daron Acemoglu} {and} \bibinfo{person}{David Autor}.} \bibinfo{year}{2011}\natexlab{}.
\newblock \showarticletitle{Skills, tasks and technologies: Implications for employment and earnings}.
\newblock In \bibinfo{booktitle}{\emph{Handbook of labor economics}}. Vol.~\bibinfo{volume}{4}. \bibinfo{publisher}{Elsevier}, \bibinfo{pages}{1043--1171}.
\newblock


\bibitem[Acemoglu and Restrepo(2019)]%
        {acemoglu2019_automation_new_tasks}
\bibfield{author}{\bibinfo{person}{Daron Acemoglu} {and} \bibinfo{person}{Pascual Restrepo}.} \bibinfo{year}{2019}\natexlab{}.
\newblock \showarticletitle{Automation and New Tasks: How Technology Displaces and Reinstates Labor}.
\newblock \bibinfo{journal}{\emph{Journal of Economic Perspectives}} \bibinfo{volume}{33}, \bibinfo{number}{2} (\bibinfo{year}{2019}), \bibinfo{pages}{3--30}.
\newblock
\href{https://doi.org/10.1257/jep.33.2.3}{doi:\nolinkurl{10.1257/jep.33.2.3}}


\bibitem[Agre(1997)]%
        {agre1997toward}
\bibfield{author}{\bibinfo{person}{P Agre}.} \bibinfo{year}{1997}\natexlab{}.
\newblock \showarticletitle{Toward a critical technical practice: Lessons learned in trying to reform AI in Bowker}.
\newblock \bibinfo{journal}{\emph{Social science, technical systems, and cooperative work: Beyond the Great Divide}} (\bibinfo{year}{1997}).
\newblock


\bibitem[Ahmed(2022)]%
        {ahmed2022futureofwork}
\bibfield{author}{\bibinfo{person}{Alex~A. Ahmed}.} \bibinfo{year}{2022}\natexlab{}.
\newblock \showarticletitle{Who Owns the Future of Work?}. In \bibinfo{booktitle}{\emph{Extended Abstracts of the 2022 CHI Conference on Human Factors in Computing Systems}} (New Orleans, LA, USA) \emph{(\bibinfo{series}{CHI EA '22})}. \bibinfo{publisher}{Association for Computing Machinery}, \bibinfo{address}{New York, NY, USA}, Article \bibinfo{articleno}{3}, \bibinfo{numpages}{6}~pages.
\newblock
\showISBNx{9781450391566}
\href{https://doi.org/10.1145/3491101.3516386}{doi:\nolinkurl{10.1145/3491101.3516386}}


\bibitem[AI(2023)]%
        {ai2023artificial}
\bibfield{author}{\bibinfo{person}{NIST AI}.} \bibinfo{year}{2023}\natexlab{}.
\newblock \showarticletitle{Artificial intelligence risk management framework (AI RMF 1.0)}.
\newblock \bibinfo{journal}{\emph{URL: https://nvlpubs. nist. gov/nistpubs/ai/nist. ai}} (\bibinfo{year}{2023}), \bibinfo{pages}{100--1}.
\newblock


\bibitem[Akridge et~al\mbox{.}(2025)]%
        {akridge2025workersbustransit}
\bibfield{author}{\bibinfo{person}{Hunter Akridge}, \bibinfo{person}{Alice~Xiaodi Tang}, \bibinfo{person}{Nikolas Martelaro}, {and} \bibinfo{person}{Sarah~E. Fox}.} \bibinfo{year}{2025}\natexlab{}.
\newblock \showarticletitle{Punctuated and Prolonged: A Workers' Inquiry into Infrastructural Failures in Bus Transit}.
\newblock \bibinfo{journal}{\emph{Proc. ACM Hum.-Comput. Interact.}} \bibinfo{volume}{9}, \bibinfo{number}{2}, Article \bibinfo{articleno}{CSCW016} (\bibinfo{date}{May} \bibinfo{year}{2025}), \bibinfo{numpages}{25}~pages.
\newblock
\href{https://doi.org/10.1145/3710914}{doi:\nolinkurl{10.1145/3710914}}


\bibitem[Al-Sibai(2025)]%
        {alsibai2025desperate}
\bibfield{author}{\bibinfo{person}{Noor Al-Sibai}.} \bibinfo{year}{2025}\natexlab{}.
\newblock \bibinfo{booktitle}{\emph{Desperate Companies Now Hiring Humans to Fix What {AI} Botched}}.
\newblock Futurism.
\newblock
\urldef\tempurl%
\url{https://futurism.com/companies-hiring-humans-fix-ai}
\showURL{%
\tempurl}


\bibitem[Amershi et~al\mbox{.}(2019)]%
        {amershi2019guidelines}
\bibfield{author}{\bibinfo{person}{Saleema Amershi}, \bibinfo{person}{Dan Weld}, \bibinfo{person}{Mihaela Vorvoreanu}, \bibinfo{person}{Adam Fourney}, \bibinfo{person}{Besmira Nushi}, \bibinfo{person}{Penny Collisson}, \bibinfo{person}{Jina Suh}, \bibinfo{person}{Shamsi Iqbal}, \bibinfo{person}{Paul~N Bennett}, \bibinfo{person}{Kori Inkpen}, {et~al\mbox{.}}} \bibinfo{year}{2019}\natexlab{}.
\newblock \showarticletitle{Guidelines for human-AI interaction}. In \bibinfo{booktitle}{\emph{Proceedings of the 2019 chi conference on human factors in computing systems}}. \bibinfo{pages}{1--13}.
\newblock


\bibitem[Aquino et~al\mbox{.}(2022)]%
        {aquino2022utopia}
\bibfield{author}{\bibinfo{person}{Yves Saint~James Aquino}, \bibinfo{person}{Wendy~A. Rogers}, \bibinfo{person}{Annette~J. Braunack-Mayer}, \bibinfo{person}{Helen Frazer}, \bibinfo{person}{Khin~Than Win}, \bibinfo{person}{Nehmat Houssami}, \bibinfo{person}{Chris Degeling}, \bibinfo{person}{Christopher Semsarian}, {and} \bibinfo{person}{Stacy~M. Carter}.} \bibinfo{year}{2022}\natexlab{}.
\newblock \showarticletitle{Utopia versus dystopia: Professional perspectives on the impact of healthcare artificial intelligence on clinical roles and skills}.
\newblock \bibinfo{journal}{\emph{International journal of medical informatics}}  \bibinfo{volume}{169} (\bibinfo{year}{2022}), \bibinfo{pages}{104903}.
\newblock
\urldef\tempurl%
\url{https://api.semanticscholar.org/CorpusID:249615948}
\showURL{%
\tempurl}


\bibitem[Arvai et~al\mbox{.}(2025)]%
        {losingcontrol2025}
\bibfield{author}{\bibinfo{person}{Nora Arvai}, \bibinfo{person}{Gell{\'e}rt Katonai}, {and} \bibinfo{person}{Bertalan Mesko}.} \bibinfo{year}{2025}\natexlab{}.
\newblock \showarticletitle{Health Care Professionals' Concerns About Medical {AI} and Psychological Barriers and Strategies for Successful Implementation: Scoping Review}.
\newblock \bibinfo{journal}{\emph{J Med Internet Res}}  \bibinfo{volume}{27} (\bibinfo{date}{April} \bibinfo{year}{2025}), \bibinfo{pages}{e66986}.
\newblock


\bibitem[Autor(2015)]%
        {autor2015_why_jobs}
\bibfield{author}{\bibinfo{person}{David~H. Autor}.} \bibinfo{year}{2015}\natexlab{}.
\newblock \showarticletitle{Why Are There Still So Many Jobs? The History and Future of Workplace Automation}.
\newblock \bibinfo{journal}{\emph{Journal of Economic Perspectives}} \bibinfo{volume}{29}, \bibinfo{number}{3} (\bibinfo{year}{2015}), \bibinfo{pages}{3--30}.
\newblock
\href{https://doi.org/10.1257/jep.29.3.3}{doi:\nolinkurl{10.1257/jep.29.3.3}}


\bibitem[Baiocco et~al\mbox{.}(2022)]%
        {baiocco2022algorithmic}
\bibfield{author}{\bibinfo{person}{Sara Baiocco}, \bibinfo{person}{Enrique Fern{\'a}ndez-Mac{\'\i}as}, \bibinfo{person}{Uma Rani}, {and} \bibinfo{person}{Annarosa Pesole}.} \bibinfo{year}{2022}\natexlab{}.
\newblock \bibinfo{booktitle}{\emph{The Algorithmic Management of Work and its Implications in Different Contexts}}.
\newblock \bibinfo{type}{JRC Working Papers Series on Labour, Education and Technology} 2022-02. \bibinfo{institution}{Joint Research Centre, European Commission}.
\newblock


\bibitem[Braganza et~al\mbox{.}(2021)]%
        {braganza2021psych}
\bibfield{author}{\bibinfo{person}{Ashley Braganza}, \bibinfo{person}{Weifeng Chen}, \bibinfo{person}{Ana Canhoto}, {and} \bibinfo{person}{Serap Sap}.} \bibinfo{year}{2021}\natexlab{}.
\newblock \showarticletitle{Productive employment and decent work: The impact of AI adoption on psychological contracts, job engagement and employee trust}.
\newblock \bibinfo{journal}{\emph{Journal of Business Research}}  \bibinfo{volume}{131} (\bibinfo{year}{2021}), \bibinfo{pages}{485--494}.
\newblock
\showISSN{0148-2963}
\href{https://doi.org/10.1016/j.jbusres.2020.08.018}{doi:\nolinkurl{10.1016/j.jbusres.2020.08.018}}


\bibitem[Brynjolfsson et~al\mbox{.}(2025)]%
        {Brynjolfsson2025customerservice}
\bibfield{author}{\bibinfo{person}{Erik Brynjolfsson}, \bibinfo{person}{Danielle Li}, {and} \bibinfo{person}{Lindsey Raymond}.} \bibinfo{year}{2025}\natexlab{}.
\newblock \showarticletitle{Generative AI at Work*}.
\newblock \bibinfo{journal}{\emph{The Quarterly Journal of Economics}} \bibinfo{volume}{140}, \bibinfo{number}{2} (\bibinfo{date}{02} \bibinfo{year}{2025}), \bibinfo{pages}{889--942}.
\newblock
\showISSN{0033-5533}
\showeprint{https://academic.oup.com/qje/article-pdf/140/2/889/61701561/qjae044.pdf}
\href{https://doi.org/10.1093/qje/qjae044}{doi:\nolinkurl{10.1093/qje/qjae044}}


\bibitem[Budzy{\'{n}} et~al\mbox{.}(2025)]%
        {endoscopy2025deskilling}
\bibfield{author}{\bibinfo{person}{Krzysztof Budzy{\'{n}}}, \bibinfo{person}{Marcin Roma{\'{n}}czyk}, \bibinfo{person}{Diana Kitala}, \bibinfo{person}{Pawe{\l} Ko{\l}odziej}, \bibinfo{person}{Marek Bugajski}, \bibinfo{person}{Hans~O. Adami}, \bibinfo{person}{Johannes Blom}, \bibinfo{person}{Marek Buszkiewicz}, \bibinfo{person}{Natalie Halvorsen}, \bibinfo{person}{Cesare Hassan}, \bibinfo{person}{Tomasz Roma{\'{n}}czyk}, \bibinfo{person}{{\O}yvind Holme}, \bibinfo{person}{Krzysztof Jarus}, \bibinfo{person}{Shona Fielding}, \bibinfo{person}{Melina Kunar}, \bibinfo{person}{Maria Pellise}, \bibinfo{person}{Nastazja Pilonis}, \bibinfo{person}{Micha{\l}~Filip Kami{\'{n}}ski}, \bibinfo{person}{Mette Kalager}, \bibinfo{person}{Michael Bretthauer}, {and} \bibinfo{person}{Yuichi Mori}.} \bibinfo{year}{2025}\natexlab{}.
\newblock \showarticletitle{Endoscopist deskilling risk after exposure to artificial intelligence in colonoscopy: a multicentre, observational study}.
\newblock \bibinfo{journal}{\emph{The Lancet Gastroenterology {\&} Hepatology}} \bibinfo{volume}{10}, \bibinfo{number}{10} (\bibinfo{date}{01 Oct} \bibinfo{year}{2025}), \bibinfo{pages}{896--903}.
\newblock
\showISSN{2468-1253}
\href{https://doi.org/10.1016/S2468-1253(25)00133-5}{doi:\nolinkurl{10.1016/S2468-1253(25)00133-5}}


\bibitem[Butler et~al\mbox{.}(2024a)]%
        {butler2024coding}
\bibfield{author}{\bibinfo{person}{Jenna Butler}, \bibinfo{person}{Jina Suh}, \bibinfo{person}{Sankeerti Haniyur}, {and} \bibinfo{person}{Constance Hadley}.} \bibinfo{year}{2024}\natexlab{a}.
\newblock \bibinfo{title}{Dear Diary: A randomized controlled trial of Generative AI coding tools in the workplace}.
\newblock
\showeprint[arxiv]{2410.18334}~[cs.SE]
\urldef\tempurl%
\url{https://arxiv.org/abs/2410.18334}
\showURL{%
\tempurl}


\bibitem[Butler et~al\mbox{.}(2024b)]%
        {butler2024microsoftfowreport}
\bibfield{author}{\bibinfo{person}{Jenna Butler}, \bibinfo{person}{Mihaela Vorvoreanu}, \bibinfo{person}{Rebecca Janssen}, \bibinfo{person}{Abigail Sellen}, \bibinfo{person}{Nicole Immorlica}, \bibinfo{person}{Adam Troy}, \bibinfo{person}{Advait Sarkar}, \bibinfo{person}{Alex Farach}, \bibinfo{person}{Alex Chouldechova}, \bibinfo{person}{Alexandra Olteanu}, \bibinfo{person}{Alexia Cambon}, \bibinfo{person}{Arjun Radhakrishna}, \bibinfo{person}{Asta Roseway}, \bibinfo{person}{Ben Zorn}, \bibinfo{person}{Brent Hecht}, \bibinfo{person}{Daniel~G. Goldstein}, \bibinfo{person}{Dhruv Joshi}, \bibinfo{person}{Ed Cutrell}, \bibinfo{person}{Emre Kiciman}, \bibinfo{person}{Gonzalo Ramos}, \bibinfo{person}{Gustavo Soares}, \bibinfo{person}{Hanna Wallach}, \bibinfo{person}{Ian Drosos}, \bibinfo{person}{Jack Williams}, \bibinfo{person}{Jacki O'Neill}, \bibinfo{person}{Jake Hofman}, \bibinfo{person}{Jaime Teevan}, \bibinfo{person}{Javier Hernandez}, \bibinfo{person}{Jennifer~Wortman Vaughan}, \bibinfo{person}{Jina Suh},
  \bibinfo{person}{John Tang}, \bibinfo{person}{Justin Edwards}, \bibinfo{person}{Kalika Bali}, \bibinfo{person}{Kori Inkpen}, \bibinfo{person}{Krishna Madhavan}, \bibinfo{person}{Laylah Bulman}, \bibinfo{person}{Leon Reicherts}, \bibinfo{person}{Lev Tankelevitch}, \bibinfo{person}{Longqi Yang}, \bibinfo{person}{Martez Mott}, \bibinfo{person}{Millicent Ochieng}, \bibinfo{person}{Mercy Muchai}, \bibinfo{person}{Nancy Baym}, \bibinfo{person}{Najeeb Abdulhamid}, \bibinfo{person}{Nicolai Marquardt}, \bibinfo{person}{Ken Hinckley}, \bibinfo{person}{Michael Bentley}, \bibinfo{person}{Dave Brown}, \bibinfo{person}{Hugo Romat}, \bibinfo{person}{Nathalie Henry~Riche}, \bibinfo{person}{Samuel Maina}, \bibinfo{person}{Shamsi Iqbal}, \bibinfo{person}{Siân Lindley}, \bibinfo{person}{Stephanie Nyairo}, \bibinfo{person}{Su~Lin Blodgett}, \bibinfo{person}{Sumit Gulwani}, \bibinfo{person}{Sunayana Sitaram}, {and} \bibinfo{person}{Vu Le}.} \bibinfo{year}{2024}\natexlab{b}.
\newblock \bibinfo{booktitle}{\emph{Microsoft New Future of Work Report 2024}}.
\newblock \bibinfo{type}{{T}echnical {R}eport} MSR-TR-2024-56. \bibinfo{institution}{Microsoft}.
\newblock
\urldef\tempurl%
\url{https://www.microsoft.com/en-us/research/publication/microsoft-new-future-of-work-report-2024/}
\showURL{%
\tempurl}


\bibitem[Cazzaniga et~al\mbox{.}(2024)]%
        {imf2024_genai_future_of_work}
\bibfield{author}{\bibinfo{person}{Mauro Cazzaniga}, \bibinfo{person}{Florence Jaumotte}, \bibinfo{person}{Longji Li}, \bibinfo{person}{Giovanni Melina}, \bibinfo{person}{Augustus~J. Panton}, \bibinfo{person}{Carlo Pizzinelli}, \bibinfo{person}{Emma~J. Rockall}, {and} \bibinfo{person}{Marina~Mendes Tavares}.} \bibinfo{year}{2024}\natexlab{}.
\newblock \bibinfo{booktitle}{\emph{Gen-AI: Artificial Intelligence and the Future of Work}}.
\newblock \bibinfo{type}{IMF Staff Discussion Note} SDN/2024/001. \bibinfo{institution}{International Monetary Fund}, \bibinfo{address}{Washington, DC}.
\newblock
\href{https://doi.org/10.5089/9798400262548.006}{doi:\nolinkurl{10.5089/9798400262548.006}}


\bibitem[Chang et~al\mbox{.}(2023)]%
        {chang2023provider}
\bibfield{author}{\bibinfo{person}{Leslie Chang}, \bibinfo{person}{Sara~Rachel Alcorn}, {and} \bibinfo{person}{Jean~L Wright}.} \bibinfo{year}{2023}\natexlab{}.
\newblock \bibinfo{title}{Provider perspectives on radiation oncology quality practices in peer review.}
\newblock


\bibitem[Chen et~al\mbox{.}(2022)]%
        {chen2022displacement}
\bibfield{author}{\bibinfo{person}{Ni Chen}, \bibinfo{person}{Zhi Li}, {and} \bibinfo{person}{Bo Tang}.} \bibinfo{year}{2022}\natexlab{}.
\newblock \showarticletitle{Can digital skill protect against job displacement risk caused by artificial intelligence? Empirical evidence from 701 detailed occupations}.
\newblock \bibinfo{journal}{\emph{PLOS ONE}} \bibinfo{volume}{17}, \bibinfo{number}{11} (\bibinfo{date}{11} \bibinfo{year}{2022}), \bibinfo{pages}{1--13}.
\newblock
\href{https://doi.org/10.1371/journal.pone.0277280}{doi:\nolinkurl{10.1371/journal.pone.0277280}}


\bibitem[Chen et~al\mbox{.}(2021)]%
        {chen2021radiographers}
\bibfield{author}{\bibinfo{person}{Yaru Chen}, \bibinfo{person}{Charitini Stavropoulou}, \bibinfo{person}{Radhika Narasinkan}, \bibinfo{person}{Adrian Baker}, {and} \bibinfo{person}{Harry Scarbrough}.} \bibinfo{year}{2021}\natexlab{}.
\newblock \showarticletitle{Professionals' responses to the introduction of AI innovations in radiology and their implications for future adoption: a qualitative study}.
\newblock \bibinfo{journal}{\emph{BMC Health Services Research}} \bibinfo{volume}{21}, \bibinfo{number}{1} (\bibinfo{date}{14 Aug} \bibinfo{year}{2021}), \bibinfo{pages}{813}.
\newblock
\showISSN{1472-6963}
\href{https://doi.org/10.1186/s12913-021-06861-y}{doi:\nolinkurl{10.1186/s12913-021-06861-y}}


\bibitem[Chen and Chan(2024)]%
        {chen2024ghostwriters}
\bibfield{author}{\bibinfo{person}{Zenan Chen} {and} \bibinfo{person}{Jason Chan}.} \bibinfo{year}{2024}\natexlab{}.
\newblock \showarticletitle{Large Language Model in Creative Work: The Role of Collaboration Modality and User Expertise}.
\newblock \bibinfo{journal}{\emph{Manage. Sci.}} \bibinfo{volume}{70}, \bibinfo{number}{12} (\bibinfo{date}{Dec.} \bibinfo{year}{2024}), \bibinfo{pages}{9101–9117}.
\newblock
\showISSN{0025-1909}
\href{https://doi.org/10.1287/mnsc.2023.03014}{doi:\nolinkurl{10.1287/mnsc.2023.03014}}


\bibitem[Cheon(2023)]%
        {cheon2023bigtech}
\bibfield{author}{\bibinfo{person}{EunJeong Cheon}.} \bibinfo{year}{2023}\natexlab{}.
\newblock \showarticletitle{Powerful Futures: How a Big Tech Company Envisions Humans and Technologies in the Workplace of the Future}.
\newblock \bibinfo{journal}{\emph{Proc. ACM Hum.-Comput. Interact.}} \bibinfo{volume}{7}, \bibinfo{number}{CSCW2}, Article \bibinfo{articleno}{312} (\bibinfo{date}{Oct.} \bibinfo{year}{2023}), \bibinfo{numpages}{35}~pages.
\newblock
\href{https://doi.org/10.1145/3610103}{doi:\nolinkurl{10.1145/3610103}}


\bibitem[Chowdhary et~al\mbox{.}(2023)]%
        {chowdhary2023can}
\bibfield{author}{\bibinfo{person}{Shreya Chowdhary}, \bibinfo{person}{Anna Kawakami}, \bibinfo{person}{Mary~L Gray}, \bibinfo{person}{Jina Suh}, \bibinfo{person}{Alexandra Olteanu}, {and} \bibinfo{person}{Koustuv Saha}.} \bibinfo{year}{2023}\natexlab{}.
\newblock \showarticletitle{Can Workers Meaningfully Consent to Workplace Wellbeing Technologies?}. In \bibinfo{booktitle}{\emph{Proceedings of the 2023 ACM Conference on Fairness, Accountability, and Transparency}}. \bibinfo{pages}{569--582}.
\newblock


\bibitem[Clark and Gevorkyan(2020)]%
        {clark2020artificial}
\bibfield{author}{\bibinfo{person}{Charles~MA Clark} {and} \bibinfo{person}{Aleksandr~V Gevorkyan}.} \bibinfo{year}{2020}\natexlab{}.
\newblock \showarticletitle{Artificial intelligence and human flourishing}.
\newblock \bibinfo{journal}{\emph{American Journal of Economics and Sociology}} \bibinfo{volume}{79}, \bibinfo{number}{4} (\bibinfo{year}{2020}), \bibinfo{pages}{1307--1344}.
\newblock


\bibitem[Crowston and Bolici(2025)]%
        {crowston2025upskilling}
\bibfield{author}{\bibinfo{person}{Kevin Crowston} {and} \bibinfo{person}{Francesco Bolici}.} \bibinfo{year}{2025}\natexlab{}.
\newblock \showarticletitle{Deskilling and upskilling with AI systems}.
\newblock \bibinfo{journal}{\emph{Information Research an international electronic journal}} \bibinfo{volume}{30}, \bibinfo{number}{iConf} (\bibinfo{date}{March} \bibinfo{year}{2025}), \bibinfo{pages}{1009–1023}.
\newblock
\href{https://doi.org/10.47989/ir30iConf47143}{doi:\nolinkurl{10.47989/ir30iConf47143}}


\bibitem[Cui et~al\mbox{.}(2024)]%
        {cui2024coding}
\bibfield{author}{\bibinfo{person}{Zheyuan~Kevin Cui}, \bibinfo{person}{Mert Demirer}, \bibinfo{person}{Sonia Jaffe}, \bibinfo{person}{Leon Musolff}, \bibinfo{person}{Sida Peng}, {and} \bibinfo{person}{Tobias Salz}.} \bibinfo{year}{2024}\natexlab{}.
\newblock \showarticletitle{The effects of generative ai on high skilled work: Evidence from three field experiments with software developers}.
\newblock \bibinfo{journal}{\emph{Available at SSRN 4945566}} (\bibinfo{year}{2024}).
\newblock


\bibitem[Das~Swain et~al\mbox{.}(2024)]%
        {das2024sensible}
\bibfield{author}{\bibinfo{person}{Vedant Das~Swain}, \bibinfo{person}{Lan Gao}, \bibinfo{person}{Abhirup Mondal}, \bibinfo{person}{Gregory~D Abowd}, {and} \bibinfo{person}{Munmun De~Choudhury}.} \bibinfo{year}{2024}\natexlab{}.
\newblock \showarticletitle{Sensible and sensitive AI for worker wellbeing: factors that inform adoption and resistance for information workers}. In \bibinfo{booktitle}{\emph{Proceedings of the 2024 CHI Conference on Human Factors in Computing Systems}}. \bibinfo{pages}{1--30}.
\newblock


\bibitem[Das~Swain et~al\mbox{.}(2023)]%
        {das2023algorithmic}
\bibfield{author}{\bibinfo{person}{Vedant Das~Swain}, \bibinfo{person}{Lan Gao}, \bibinfo{person}{William~A Wood}, \bibinfo{person}{Srikruthi~C Matli}, \bibinfo{person}{Gregory~D Abowd}, {and} \bibinfo{person}{Munmun De~Choudhury}.} \bibinfo{year}{2023}\natexlab{}.
\newblock \showarticletitle{Algorithmic power or punishment: Information worker perspectives on passive sensing enabled ai phenotyping of performance and wellbeing}. In \bibinfo{booktitle}{\emph{Proceedings of the 2023 CHI Conference on Human Factors in Computing Systems}}. \bibinfo{pages}{1--17}.
\newblock


\bibitem[Das~Swain and Saha(2024)]%
        {das2024teacher}
\bibfield{author}{\bibinfo{person}{Vedant Das~Swain} {and} \bibinfo{person}{Koustuv Saha}.} \bibinfo{year}{2024}\natexlab{}.
\newblock \showarticletitle{Teacher, trainer, counsel, spy: how generative AI can bridge or widen the gaps in worker-centric digital phenotyping of wellbeing}. In \bibinfo{booktitle}{\emph{Proceedings of the 3rd Annual Meeting of the Symposium on Human-Computer Interaction for Work}}. \bibinfo{pages}{1--13}.
\newblock


\bibitem[Das~Swain et~al\mbox{.}(2025)]%
        {das2025ai}
\bibfield{author}{\bibinfo{person}{Vedant Das~Swain}, \bibinfo{person}{Qiuyue"~Joy" Zhong}, \bibinfo{person}{Jash~Rajesh Parekh}, \bibinfo{person}{Yechan Jeon}, \bibinfo{person}{Roy Zimmermann}, \bibinfo{person}{Mary~P Czerwinski}, \bibinfo{person}{Jina Suh}, \bibinfo{person}{Varun Mishra}, \bibinfo{person}{Koustuv Saha}, {and} \bibinfo{person}{Javier Hernandez}.} \bibinfo{year}{2025}\natexlab{}.
\newblock \showarticletitle{Ai on my shoulder: Supporting emotional labor in front-office roles with an llm-based empathetic coworker}. In \bibinfo{booktitle}{\emph{Proceedings of the 2025 CHI Conference on Human Factors in Computing Systems}}. \bibinfo{pages}{1--29}.
\newblock


\bibitem[De-Arteaga et~al\mbox{.}(2020)]%
        {De-Arteaga2020}
\bibfield{author}{\bibinfo{person}{Maria De-Arteaga}, \bibinfo{person}{Riccardo Fogliato}, {and} \bibinfo{person}{Alexandra Chouldechova}.} \bibinfo{year}{2020}\natexlab{}.
\newblock \showarticletitle{A Case for Humans-in-the-Loop: Decisions in the Presence of Erroneous Algorithmic Scores}. In \bibinfo{booktitle}{\emph{Proceedings of the 2020 CHI Conference on Human Factors in Computing Systems}} (Honolulu, HI, USA) \emph{(\bibinfo{series}{CHI '20})}. \bibinfo{publisher}{Association for Computing Machinery}, \bibinfo{address}{New York, NY, USA}, \bibinfo{pages}{1–12}.
\newblock
\showISBNx{9781450367080}
\href{https://doi.org/10.1145/3313831.3376638}{doi:\nolinkurl{10.1145/3313831.3376638}}


\bibitem[Ehsan et~al\mbox{.}(2021)]%
        {ehsan2021expanding}
\bibfield{author}{\bibinfo{person}{Upol Ehsan}, \bibinfo{person}{Q~Vera Liao}, \bibinfo{person}{Michael Muller}, \bibinfo{person}{Mark~O Riedl}, {and} \bibinfo{person}{Justin~D Weisz}.} \bibinfo{year}{2021}\natexlab{}.
\newblock \showarticletitle{Expanding explainability: Towards social transparency in ai systems}. In \bibinfo{booktitle}{\emph{Proceedings of the 2021 CHI Conference on Human Factors in Computing Systems}}. \bibinfo{pages}{1--19}.
\newblock


\bibitem[Ehsan et~al\mbox{.}(2022)]%
        {ehsan2022seamful}
\bibfield{author}{\bibinfo{person}{Upol Ehsan}, \bibinfo{person}{Q~Vera Liao}, \bibinfo{person}{Samir Passi}, \bibinfo{person}{Mark~O Riedl}, {and} \bibinfo{person}{Hal Daume~III}.} \bibinfo{year}{2022}\natexlab{}.
\newblock \showarticletitle{Seamful XAI: Operationalizing Seamful Design in Explainable AI}.
\newblock \bibinfo{journal}{\emph{arXiv preprint arXiv:2211.06753}} (\bibinfo{year}{2022}).
\newblock


\bibitem[Ehsan and Riedl(2020)]%
        {ehsan2020human}
\bibfield{author}{\bibinfo{person}{Upol Ehsan} {and} \bibinfo{person}{Mark~O Riedl}.} \bibinfo{year}{2020}\natexlab{}.
\newblock \showarticletitle{Human-centered Explainable AI: Towards a Reflective Sociotechnical Approach}, In \bibinfo{booktitle}{International Conference on Human-Computer Interaction}.
\newblock \bibinfo{journal}{\emph{arXiv preprint arXiv:2002.01092}}, \bibinfo{pages}{449--466}.
\newblock


\bibitem[Ehsan et~al\mbox{.}(2023)]%
        {ehsan2023charting}
\bibfield{author}{\bibinfo{person}{Upol Ehsan}, \bibinfo{person}{Koustuv Saha}, \bibinfo{person}{Munmun De~Choudhury}, {and} \bibinfo{person}{Mark~O Riedl}.} \bibinfo{year}{2023}\natexlab{}.
\newblock \showarticletitle{Charting the Sociotechnical Gap in Explainable AI: A Framework to Address the Gap in XAI}.
\newblock \bibinfo{journal}{\emph{Proceedings of the ACM on Human-Computer Interaction}} \bibinfo{volume}{7}, \bibinfo{number}{CSCW1} (\bibinfo{year}{2023}), \bibinfo{pages}{1--32}.
\newblock


\bibitem[Ferrario et~al\mbox{.}(2024)]%
        {ferrario2024addressing}
\bibfield{author}{\bibinfo{person}{Andrea Ferrario}, \bibinfo{person}{Alberto Termine}, {and} \bibinfo{person}{Alessandro Facchini}.} \bibinfo{year}{2024}\natexlab{}.
\newblock \showarticletitle{Addressing social misattributions of large language models: An HCXAI-based approach}.
\newblock \bibinfo{journal}{\emph{arXiv preprint arXiv:2403.17873}} (\bibinfo{year}{2024}).
\newblock


\bibitem[Filippucci et~al\mbox{.}(2024)]%
        {oecd2024_ai_productivity}
\bibfield{author}{\bibinfo{person}{Francesco Filippucci}, \bibinfo{person}{Peter Gal}, \bibinfo{person}{Cecilia Jona-Lasinio}, \bibinfo{person}{Alvaro Leandro}, {and} \bibinfo{person}{Giuseppe Nicoletti}.} \bibinfo{year}{2024}\natexlab{}.
\newblock \bibinfo{booktitle}{\emph{The impact of Artificial Intelligence on productivity, distribution and growth: Key mechanisms, initial evidence and policy challenges}}.
\newblock \bibinfo{type}{{T}echnical {R}eport} No. 15. \bibinfo{institution}{OECD Publishing}, \bibinfo{address}{Paris}.
\newblock
\href{https://doi.org/10.1787/8d900037-en}{doi:\nolinkurl{10.1787/8d900037-en}}


\bibitem[Fox et~al\mbox{.}(2020)]%
        {fox2020workerdesign}
\bibfield{author}{\bibinfo{person}{Sarah~E. Fox}, \bibinfo{person}{Vera Khovanskaya}, \bibinfo{person}{Clara Crivellaro}, \bibinfo{person}{Niloufar Salehi}, \bibinfo{person}{Lynn Dombrowski}, \bibinfo{person}{Chinmay Kulkarni}, \bibinfo{person}{Lilly Irani}, {and} \bibinfo{person}{Jodi Forlizzi}.} \bibinfo{year}{2020}\natexlab{}.
\newblock \showarticletitle{Worker-Centered Design: Expanding HCI Methods for Supporting Labor}. In \bibinfo{booktitle}{\emph{Extended Abstracts of the 2020 CHI Conference on Human Factors in Computing Systems}} (Honolulu, HI, USA) \emph{(\bibinfo{series}{CHI EA '20})}. \bibinfo{publisher}{Association for Computing Machinery}, \bibinfo{address}{New York, NY, USA}, \bibinfo{pages}{1–8}.
\newblock
\showISBNx{9781450368193}
\href{https://doi.org/10.1145/3334480.3375157}{doi:\nolinkurl{10.1145/3334480.3375157}}


\bibitem[Fragiadakis et~al\mbox{.}(2024)]%
        {fragiadakis2024evaluating}
\bibfield{author}{\bibinfo{person}{George Fragiadakis}, \bibinfo{person}{Christos Diou}, \bibinfo{person}{George Kousiouris}, {and} \bibinfo{person}{Mara Nikolaidou}.} \bibinfo{year}{2024}\natexlab{}.
\newblock \showarticletitle{Evaluating human-ai collaboration: A review and methodological framework}.
\newblock \bibinfo{journal}{\emph{arXiv preprint arXiv:2407.19098}} (\bibinfo{year}{2024}).
\newblock


\bibitem[Greenbaum(1996)]%
        {greenbaum1996backtolabor}
\bibfield{author}{\bibinfo{person}{Joan Greenbaum}.} \bibinfo{year}{1996}\natexlab{}.
\newblock \showarticletitle{Back to labor: returning to labor process discussions in the study of work}. In \bibinfo{booktitle}{\emph{Proceedings of the 1996 ACM Conference on Computer Supported Cooperative Work}} (Boston, Massachusetts, USA) \emph{(\bibinfo{series}{CSCW '96})}. \bibinfo{publisher}{Association for Computing Machinery}, \bibinfo{address}{New York, NY, USA}, \bibinfo{pages}{229–237}.
\newblock
\showISBNx{0897917650}
\href{https://doi.org/10.1145/240080.240259}{doi:\nolinkurl{10.1145/240080.240259}}


\bibitem[Gupta et~al\mbox{.}(2018)]%
        {gupta2018futureofworkers}
\bibfield{author}{\bibinfo{person}{Sarita Gupta}, \bibinfo{person}{Stephen Lerner}, {and} \bibinfo{person}{Joseph McCartin}.} \bibinfo{year}{2018}\natexlab{}.
\newblock \bibinfo{title}{It’s Not the 'Future of Work,' It’s the Future of Workers That’s in Doubt - The American Prospect}.
\newblock
\urldef\tempurl%
\url{https://prospect.org/2018/08/31/future-work-future-workers-doubt/}
\showURL{%
\tempurl}
\newblock
\shownote{[Online; accessed 2025-11-15]}.


\bibitem[Hayes and Downie(2025)]%
        {hayes2025fowinnovation}
\bibfield{author}{\bibinfo{person}{Molly Hayes} {and} \bibinfo{person}{Amanda Downie}.} \bibinfo{year}{2025}\natexlab{}.
\newblock \bibinfo{title}{AI and the Future of Work | IBM}.
\newblock
\urldef\tempurl%
\url{https://www.ibm.com/think/insights/ai-and-the-future-of-work}
\showURL{%
\tempurl}
\newblock
\shownote{[Online; accessed 2025-09-01]}.


\bibitem[Holzinger et~al\mbox{.}(2025)]%
        {holzinger2024oversight}
\bibfield{author}{\bibinfo{person}{Andreas Holzinger}, \bibinfo{person}{Kurt Zatloukal}, {and} \bibinfo{person}{Heimo Müller}.} \bibinfo{year}{2025}\natexlab{}.
\newblock \showarticletitle{Is human oversight to AI systems still possible?}
\newblock \bibinfo{journal}{\emph{New Biotechnology}}  \bibinfo{volume}{85} (\bibinfo{year}{2025}), \bibinfo{pages}{59--62}.
\newblock
\showISSN{1871-6784}
\href{https://doi.org/10.1016/j.nbt.2024.12.003}{doi:\nolinkurl{10.1016/j.nbt.2024.12.003}}


\bibitem[Hui et~al\mbox{.}(2024)]%
        {hui2024freelance}
\bibfield{author}{\bibinfo{person}{Xiang Hui}, \bibinfo{person}{Oren Reshef}, {and} \bibinfo{person}{Luofeng Zhou}.} \bibinfo{year}{2024}\natexlab{}.
\newblock \showarticletitle{The Short-Term Effects of Generative Artificial Intelligence on Employment: Evidence from an Online Labor Market}.
\newblock \bibinfo{journal}{\emph{Organization Science}} \bibinfo{volume}{35}, \bibinfo{number}{6} (\bibinfo{year}{2024}), \bibinfo{pages}{1977--1989}.
\newblock
\showeprint{https://doi.org/10.1287/orsc.2023.18441}
\href{https://doi.org/10.1287/orsc.2023.18441}{doi:\nolinkurl{10.1287/orsc.2023.18441}}


\bibitem[Jaffe et~al\mbox{.}(2024)]%
        {jaffe2024generative}
\bibfield{author}{\bibinfo{person}{Sonia Jaffe}, \bibinfo{person}{Neha~Parikh Shah}, \bibinfo{person}{Jenna Butler}, \bibinfo{person}{Alex Farach}, \bibinfo{person}{Alexia Cambon}, \bibinfo{person}{Brent Hecht}, \bibinfo{person}{Michael Schwarz}, {and} \bibinfo{person}{Jaime Teevan}.} \bibinfo{year}{2024}\natexlab{}.
\newblock \bibinfo{booktitle}{\emph{Generative AI in Real-World Workplaces}}.
\newblock \bibinfo{type}{{T}echnical {R}eport} MSR-TR-2024-29. \bibinfo{institution}{Microsoft}.
\newblock
\urldef\tempurl%
\url{https://www.microsoft.com/en-us/research/publication/generative-ai-in-real-world-workplaces/}
\showURL{%
\tempurl}


\bibitem[Karataş and Yüce(2024)]%
        {kratas2024teacher}
\bibfield{author}{\bibinfo{person}{Fatih Karataş} {and} \bibinfo{person}{Erkan Yüce}.} \bibinfo{year}{2024}\natexlab{}.
\newblock \showarticletitle{AI and the Future of Teaching: Preservice Teachers’ Reflections on the Use of Artificial Intelligence in Open and Distributed Learning}.
\newblock \bibinfo{journal}{\emph{The International Review of Research in Open and Distributed Learning}} \bibinfo{volume}{25}, \bibinfo{number}{3} (\bibinfo{date}{Aug.} \bibinfo{year}{2024}), \bibinfo{pages}{304–325}.
\newblock
\href{https://doi.org/10.19173/irrodl.v25i3.7785}{doi:\nolinkurl{10.19173/irrodl.v25i3.7785}}


\bibitem[Kawakami et~al\mbox{.}(2023)]%
        {kawakami2023sensing}
\bibfield{author}{\bibinfo{person}{Anna Kawakami}, \bibinfo{person}{Shreya Chowdhary}, \bibinfo{person}{Shamsi~T Iqbal}, \bibinfo{person}{Q~Vera Liao}, \bibinfo{person}{Alexandra Olteanu}, \bibinfo{person}{Jina Suh}, {and} \bibinfo{person}{Koustuv Saha}.} \bibinfo{year}{2023}\natexlab{}.
\newblock \showarticletitle{Sensing wellbeing in the workplace, why and for whom? envisioning impacts with organizational stakeholders}.
\newblock \bibinfo{journal}{\emph{Proceedings of the ACM on Human-Computer Interaction}} \bibinfo{volume}{7}, \bibinfo{number}{CSCW2} (\bibinfo{year}{2023}), \bibinfo{pages}{1--33}.
\newblock


\bibitem[Khovanskaya and Sengers(2019)]%
        {khovanskaya2019uneasyalliance}
\bibfield{author}{\bibinfo{person}{Vera Khovanskaya} {and} \bibinfo{person}{Phoebe Sengers}.} \bibinfo{year}{2019}\natexlab{}.
\newblock \showarticletitle{Data Rhetoric and Uneasy Alliances: Data Advocacy in US Labor History}. In \bibinfo{booktitle}{\emph{Proceedings of the 2019 on Designing Interactive Systems Conference}} (San Diego, CA, USA) \emph{(\bibinfo{series}{DIS '19})}. \bibinfo{publisher}{Association for Computing Machinery}, \bibinfo{address}{New York, NY, USA}, \bibinfo{pages}{1391–1403}.
\newblock
\showISBNx{9781450358507}
\href{https://doi.org/10.1145/3322276.3323691}{doi:\nolinkurl{10.1145/3322276.3323691}}


\bibitem[Lan(2024)]%
        {lan2024teacher}
\bibfield{author}{\bibinfo{person}{Yanzhen Lan}.} \bibinfo{year}{2024}\natexlab{}.
\newblock \showarticletitle{Through tensions to identity-based motivations: Exploring teacher professional identity in Artificial Intelligence-enhanced teacher training}.
\newblock \bibinfo{journal}{\emph{Teaching and Teacher Education}}  \bibinfo{volume}{151} (\bibinfo{year}{2024}), \bibinfo{pages}{104736}.
\newblock
\showISSN{0742-051X}
\href{https://doi.org/10.1016/j.tate.2024.104736}{doi:\nolinkurl{10.1016/j.tate.2024.104736}}


\bibitem[Langer et~al\mbox{.}(2024)]%
        {langer2024oversight}
\bibfield{author}{\bibinfo{person}{Markus Langer}, \bibinfo{person}{Kevin Baum}, {and} \bibinfo{person}{Nadine Schlicker}.} \bibinfo{year}{2024}\natexlab{}.
\newblock \showarticletitle{Effective Human Oversight of AI-Based Systems: A Signal Detection Perspective on the Detection of Inaccurate and Unfair Outputs}.
\newblock \bibinfo{journal}{\emph{Minds and Machines}} \bibinfo{volume}{35}, \bibinfo{number}{1} (\bibinfo{date}{05 Nov} \bibinfo{year}{2024}), \bibinfo{pages}{1}.
\newblock
\showISSN{1572-8641}
\href{https://doi.org/10.1007/s11023-024-09701-0}{doi:\nolinkurl{10.1007/s11023-024-09701-0}}


\bibitem[Lee et~al\mbox{.}(2025)]%
        {lee2025criticalthinking}
\bibfield{author}{\bibinfo{person}{Hao-Ping~(Hank) Lee}, \bibinfo{person}{Advait Sarkar}, \bibinfo{person}{Lev Tankelevitch}, \bibinfo{person}{Ian Drosos}, \bibinfo{person}{Sean Rintel}, \bibinfo{person}{Richard Banks}, {and} \bibinfo{person}{Nicholas Wilson}.} \bibinfo{year}{2025}\natexlab{}.
\newblock \showarticletitle{The Impact of Generative AI on Critical Thinking: Self-Reported Reductions in Cognitive Effort and Confidence Effects From a Survey of Knowledge Workers}. In \bibinfo{booktitle}{\emph{Proceedings of the 2025 CHI Conference on Human Factors in Computing Systems}} \emph{(\bibinfo{series}{CHI '25})}. \bibinfo{publisher}{Association for Computing Machinery}, \bibinfo{address}{New York, NY, USA}, Article \bibinfo{articleno}{1121}, \bibinfo{numpages}{22}~pages.
\newblock
\showISBNx{9798400713941}
\href{https://doi.org/10.1145/3706598.3713778}{doi:\nolinkurl{10.1145/3706598.3713778}}


\bibitem[Lobel(2018)]%
        {lobel2018ndas}
\bibfield{author}{\bibinfo{person}{Orly Lobel}.} \bibinfo{year}{2018}\natexlab{}.
\newblock \showarticletitle{NDAs are out of control. Here’s What needs to change}.
\newblock \bibinfo{journal}{\emph{Harvard Business Review}}  \bibinfo{volume}{30} (\bibinfo{year}{2018}).
\newblock


\bibitem[Lombi and Rossero(2024)]%
        {lombi2024radiologist}
\bibfield{author}{\bibinfo{person}{Linda Lombi} {and} \bibinfo{person}{Eleonora Rossero}.} \bibinfo{year}{2024}\natexlab{}.
\newblock \showarticletitle{How artificial intelligence is reshaping the autonomy and boundary work of radiologists. A qualitative study}.
\newblock \bibinfo{journal}{\emph{Sociology of Health \& Illness}} \bibinfo{volume}{46}, \bibinfo{number}{2} (\bibinfo{year}{2024}), \bibinfo{pages}{200--218}.
\newblock
\showeprint{https://onlinelibrary.wiley.com/doi/pdf/10.1111/1467-9566.13702}
\href{https://doi.org/10.1111/1467-9566.13702}{doi:\nolinkurl{10.1111/1467-9566.13702}}


\bibitem[Lushnikova et~al\mbox{.}(2025)]%
        {lushnikova2025cscwfutureofwork}
\bibfield{author}{\bibinfo{person}{Alina Lushnikova}, \bibinfo{person}{Michael Muller}, \bibinfo{person}{Shaowen Bardzell}, \bibinfo{person}{Toby Jia-Jun Li}, {and} \bibinfo{person}{Saiph Savage}.} \bibinfo{year}{2025}\natexlab{}.
\newblock \showarticletitle{CSCW Contributions to Critical Futures of Work}. In \bibinfo{booktitle}{\emph{Companion Publication of the 2025 Conference on Computer-Supported Cooperative Work and Social Computing}} \emph{(\bibinfo{series}{CSCW Companion '25})}. \bibinfo{publisher}{Association for Computing Machinery}, \bibinfo{address}{New York, NY, USA}, \bibinfo{pages}{161–167}.
\newblock
\showISBNx{9798400714801}
\href{https://doi.org/10.1145/3715070.3748296}{doi:\nolinkurl{10.1145/3715070.3748296}}


\bibitem[Macnamara et~al\mbox{.}(2024)]%
        {macnamara2024skilldecay}
\bibfield{author}{\bibinfo{person}{Brooke~N. Macnamara}, \bibinfo{person}{Ibrahim Berber}, \bibinfo{person}{M.~Cenk {\c{C}}avu{\c{s}}o{\u{g}}lu}, \bibinfo{person}{Elizabeth~A. Krupinski}, \bibinfo{person}{Naren Nallapareddy}, \bibinfo{person}{Noelle~E. Nelson}, \bibinfo{person}{Philip~J. Smith}, \bibinfo{person}{Amy~L. Wilson-Delfosse}, {and} \bibinfo{person}{Soumya Ray}.} \bibinfo{year}{2024}\natexlab{}.
\newblock \showarticletitle{Does using artificial intelligence assistance accelerate skill decay and hinder skill development without performers' awareness?}
\newblock \bibinfo{journal}{\emph{Cognitive Research: Principles and Implications}} \bibinfo{volume}{9}, \bibinfo{number}{1} (\bibinfo{date}{12 Jul} \bibinfo{year}{2024}), \bibinfo{pages}{46}.
\newblock
\showISSN{2365-7464}
\href{https://doi.org/10.1186/s41235-024-00572-8}{doi:\nolinkurl{10.1186/s41235-024-00572-8}}


\bibitem[Maier and Seligman(1976)]%
        {maier1976learnedhelplessness}
\bibfield{author}{\bibinfo{person}{Steven~F Maier} {and} \bibinfo{person}{Martin~E Seligman}.} \bibinfo{year}{1976}\natexlab{}.
\newblock \showarticletitle{Learned helplessness: theory and evidence.}
\newblock \bibinfo{journal}{\emph{Journal of experimental psychology: general}} \bibinfo{volume}{105}, \bibinfo{number}{1} (\bibinfo{year}{1976}), \bibinfo{pages}{3}.
\newblock


\bibitem[Miceli et~al\mbox{.}(2025)]%
        {miceli2025epistemicauthorityai}
\bibfield{author}{\bibinfo{person}{Milagros Miceli}, \bibinfo{person}{Adio-Adet Dinika}, \bibinfo{person}{Krystal Kauffman}, \bibinfo{person}{Camilla Salim~Wagner}, \bibinfo{person}{Laurenz Sachenbacher}, \bibinfo{person}{Alex Hanna}, {and} \bibinfo{person}{Timnit Gebru}.} \bibinfo{year}{2025}\natexlab{}.
\newblock \showarticletitle{Methodological Considerations for Centering Workers’ Epistemic Authority in AI Research}.
\newblock \bibinfo{journal}{\emph{Proceedings of the AAAI/ACM Conference on AI, Ethics, and Society}} \bibinfo{volume}{8}, \bibinfo{number}{2} (\bibinfo{date}{Oct.} \bibinfo{year}{2025}), \bibinfo{pages}{1698--1710}.
\newblock
\href{https://doi.org/10.1609/aies.v8i2.36667}{doi:\nolinkurl{10.1609/aies.v8i2.36667}}


\bibitem[Otis et~al\mbox{.}(2023)]%
        {otis2023entrepreneurs}
\bibfield{author}{\bibinfo{person}{Nicholas~G Otis}, \bibinfo{person}{Rowan~P Clarke}, \bibinfo{person}{Solene Delecourt}, \bibinfo{person}{David Holtz}, {and} \bibinfo{person}{Rembrand Koning}.} \bibinfo{year}{2023}\natexlab{}.
\newblock \bibinfo{title}{The Uneven Impact of Generative AI on Entrepreneurial Performance}.
\newblock
\href{https://doi.org/10.31219/osf.io/hdjpk}{doi:\nolinkurl{10.31219/osf.io/hdjpk}}


\bibitem[Ou et~al\mbox{.}(2025)]%
        {ou2025psychological}
\bibfield{author}{\bibinfo{person}{Min Ou}, \bibinfo{person}{Hope Koch}, {and} \bibinfo{person}{Qin Weng}.} \bibinfo{year}{2025}\natexlab{}.
\newblock \showarticletitle{Algorithmic Oversight of Knowledge Workers: A Research Agenda on Technology, Control, and Psychological Contracts}. In \bibinfo{booktitle}{\emph{Proceedings of the 2025 Computers and People Research Conference}}. ACM, \bibinfo{pages}{1--10}.
\newblock
\href{https://doi.org/10.1145/3716489.3728454}{doi:\nolinkurl{10.1145/3716489.3728454}}


\bibitem[Panezi(2024)]%
        {panezi2024article}
\bibfield{author}{\bibinfo{person}{Argyri Panezi}.} \bibinfo{year}{2024}\natexlab{}.
\newblock \showarticletitle{Article 14 Human Oversight}.
\newblock \bibinfo{journal}{\emph{The EU Artificial Intelligence (AI) Act: A Commentary}} (\bibinfo{year}{2024}).
\newblock


\bibitem[Passi(2025)]%
        {passi2025_oversightproblem}
\bibfield{author}{\bibinfo{person}{Samir Passi}.} \bibinfo{year}{2025}\natexlab{}.
\newblock \showarticletitle{Agentic AI has a Human Oversight Problem}.
\newblock \bibinfo{journal}{\emph{Available at SSRN 5529058}} (\bibinfo{year}{2025}), \bibinfo{numpages}{30}~pages.
\newblock
\href{https://doi.org/10.2139/ssrn.5529058}{doi:\nolinkurl{10.2139/ssrn.5529058}}


\bibitem[Passi et~al\mbox{.}(2024)]%
        {passi2024appropriate}
\bibfield{author}{\bibinfo{person}{Samir Passi}, \bibinfo{person}{Shipi Dhanorkar}, {and} \bibinfo{person}{Mihaela Vorvoreanu}.} \bibinfo{year}{2024}\natexlab{}.
\newblock \bibinfo{booktitle}{\emph{Appropriate reliance on Generative AI: Research synthesis}}.
\newblock \bibinfo{type}{{T}echnical {R}eport} MSR-TR-2024-7. \bibinfo{institution}{Microsoft}.
\newblock
\urldef\tempurl%
\url{https://www.microsoft.com/en-us/research/publication/appropriate-reliance-on-generative-ai-research-synthesis/}
\showURL{%
\tempurl}


\bibitem[Passi et~al\mbox{.}(2025)]%
        {passi2025overreliancechapter}
\bibfield{author}{\bibinfo{person}{Samir Passi}, \bibinfo{person}{Shipi Dhanorkar}, {and} \bibinfo{person}{Mihaela Vorvoreanu}.} \bibinfo{year}{2025}\natexlab{}.
\newblock \bibinfo{booktitle}{\emph{Addressing Overreliance on AI}}.
\newblock \bibinfo{publisher}{Springer Nature Singapore}, \bibinfo{address}{Singapore}, \bibinfo{pages}{1--34}.
\newblock
\showISBNx{978-981-97-8440-0}
\href{https://doi.org/10.1007/978-981-97-8440-0_98-1}{doi:\nolinkurl{10.1007/978-981-97-8440-0_98-1}}


\bibitem[Passi and Vorvoreanu(2022)]%
        {passi2022overreliance}
\bibfield{author}{\bibinfo{person}{Samir Passi} {and} \bibinfo{person}{Mihaela Vorvoreanu}.} \bibinfo{year}{2022}\natexlab{}.
\newblock \bibinfo{booktitle}{\emph{Overreliance on AI: Literature Review}}.
\newblock \bibinfo{type}{{T}echnical {R}eport} MSR-TR-2022-12. \bibinfo{institution}{Microsoft}.
\newblock
\urldef\tempurl%
\url{https://www.microsoft.com/en-us/research/publication/overreliance-on-ai-literature-review/}
\showURL{%
\tempurl}


\bibitem[Prather et~al\mbox{.}(2023)]%
        {prather2023noviceprogrammer}
\bibfield{author}{\bibinfo{person}{James Prather}, \bibinfo{person}{Brent~N. Reeves}, \bibinfo{person}{Paul Denny}, \bibinfo{person}{Brett~A. Becker}, \bibinfo{person}{Juho Leinonen}, \bibinfo{person}{Andrew Luxton-Reilly}, \bibinfo{person}{Garrett Powell}, \bibinfo{person}{James Finnie-Ansley}, {and} \bibinfo{person}{Eddie~Antonio Santos}.} \bibinfo{year}{2023}\natexlab{}.
\newblock \showarticletitle{“It’s Weird That it Knows What I Want”: Usability and Interactions with Copilot for Novice Programmers}.
\newblock \bibinfo{journal}{\emph{ACM Trans. Comput.-Hum. Interact.}} \bibinfo{volume}{31}, \bibinfo{number}{1}, Article \bibinfo{articleno}{4} (\bibinfo{date}{Nov.} \bibinfo{year}{2023}), \bibinfo{numpages}{31}~pages.
\newblock
\showISSN{1073-0516}
\href{https://doi.org/10.1145/3617367}{doi:\nolinkurl{10.1145/3617367}}


\bibitem[Rawashdeh(2023)]%
        {rawashdeh2023displacement}
\bibfield{author}{\bibinfo{person}{Awni Rawashdeh}.} \bibinfo{year}{2023}\natexlab{}.
\newblock \showarticletitle{The consequences of artificial intelligence: an investigation into the impact of AI on job displacement in accounting}.
\newblock \bibinfo{journal}{\emph{Journal of Science and Technology Policy Management}} \bibinfo{volume}{16}, \bibinfo{number}{3} (\bibinfo{date}{11} \bibinfo{year}{2023}), \bibinfo{pages}{506--535}.
\newblock
\showISSN{2053-4620}
\showeprint{https://www.emerald.com/jstpm/article-pdf/16/3/506/9659776/jstpm-02-2023-0030.pdf}
\href{https://doi.org/10.1108/JSTPM-02-2023-0030}{doi:\nolinkurl{10.1108/JSTPM-02-2023-0030}}


\bibitem[Reicherts et~al\mbox{.}(2025)]%
        {reicherts2025helpmethink}
\bibfield{author}{\bibinfo{person}{Leon Reicherts}, \bibinfo{person}{Zelun~Tony Zhang}, \bibinfo{person}{Elisabeth von Oswald}, \bibinfo{person}{Yuanting Liu}, \bibinfo{person}{Yvonne Rogers}, {and} \bibinfo{person}{Mariam Hassib}.} \bibinfo{year}{2025}\natexlab{}.
\newblock \showarticletitle{AI, Help Me Think—but for Myself: Assisting People in Complex Decision-Making by Providing Different Kinds of Cognitive Support}. In \bibinfo{booktitle}{\emph{Proceedings of the 2025 CHI Conference on Human Factors in Computing Systems}} \emph{(\bibinfo{series}{CHI '25})}. \bibinfo{publisher}{Association for Computing Machinery}, \bibinfo{address}{New York, NY, USA}, Article \bibinfo{articleno}{255}, \bibinfo{numpages}{19}~pages.
\newblock
\showISBNx{9798400713941}
\href{https://doi.org/10.1145/3706598.3713295}{doi:\nolinkurl{10.1145/3706598.3713295}}


\bibitem[Renkema and Tursunbayeva(2024)]%
        {Renkema2024futureacademics}
\bibfield{author}{\bibinfo{person}{Maarten Renkema} {and} \bibinfo{person}{Aizhan Tursunbayeva}.} \bibinfo{year}{2024}\natexlab{}.
\newblock \showarticletitle{The future of work of academics in the age of Artificial Intelligence: State-of-the-art and a research roadmap}.
\newblock \bibinfo{journal}{\emph{Futures}}  \bibinfo{volume}{163} (\bibinfo{year}{2024}), \bibinfo{pages}{103453}.
\newblock
\showISSN{0016-3287}
\href{https://doi.org/10.1016/j.futures.2024.103453}{doi:\nolinkurl{10.1016/j.futures.2024.103453}}


\bibitem[Ridley(2025)]%
        {ridley2025human}
\bibfield{author}{\bibinfo{person}{Michael Ridley}.} \bibinfo{year}{2025}\natexlab{}.
\newblock \showarticletitle{Human-centered explainable artificial intelligence: An Annual Review of Information Science and Technology (ARIST) paper}.
\newblock \bibinfo{journal}{\emph{Journal of the Association for Information Science and Technology}} \bibinfo{volume}{76}, \bibinfo{number}{1} (\bibinfo{year}{2025}), \bibinfo{pages}{98--120}.
\newblock


\bibitem[Roemmich et~al\mbox{.}(2023)]%
        {roemmich2023emotion}
\bibfield{author}{\bibinfo{person}{Kat Roemmich}, \bibinfo{person}{Florian Schaub}, {and} \bibinfo{person}{Nazanin Andalibi}.} \bibinfo{year}{2023}\natexlab{}.
\newblock \showarticletitle{Emotion AI at work: Implications for workplace surveillance, emotional labor, and emotional privacy}. In \bibinfo{booktitle}{\emph{Proceedings of the 2023 CHI conference on human factors in computing systems}}. \bibinfo{pages}{1--20}.
\newblock


\bibitem[Rose et~al\mbox{.}({[n.\,d.]})]%
        {rosehistory}
\bibfield{author}{\bibinfo{person}{Christopher Rose}, \bibinfo{person}{Allen~S Lichter}, {and} \bibinfo{person}{FASCO FASTRO}.} \bibinfo{year}{[n.\,d.]}\natexlab{}.
\newblock \showarticletitle{History of Radiation Oncology in the United States}.
\newblock  (\bibinfo{year}{[n.\,d.]}).
\newblock


\bibitem[Rosson and Carroll(2009)]%
        {rosson2009scenario}
\bibfield{author}{\bibinfo{person}{Mary~Beth Rosson} {and} \bibinfo{person}{John~M Carroll}.} \bibinfo{year}{2009}\natexlab{}.
\newblock \showarticletitle{Scenario based design}.
\newblock \bibinfo{journal}{\emph{Human-computer interaction. boca raton, FL}} (\bibinfo{year}{2009}), \bibinfo{pages}{145--162}.
\newblock


\bibitem[Scarbrough et~al\mbox{.}(2025)]%
        {scarbrough2025sensemaking}
\bibfield{author}{\bibinfo{person}{Harry Scarbrough}, \bibinfo{person}{Yaru Chen}, {and} \bibinfo{person}{Gerardo Patriotta}.} \bibinfo{year}{2025}\natexlab{}.
\newblock \showarticletitle{The AI of the Beholder: Intra-Professional Sensemaking of an Epistemic Technology}.
\newblock \bibinfo{journal}{\emph{Journal of Management Studies}} \bibinfo{volume}{62}, \bibinfo{number}{5} (\bibinfo{year}{2025}), \bibinfo{pages}{1885--1913}.
\newblock
\showeprint{https://onlinelibrary.wiley.com/doi/pdf/10.1111/joms.13065}
\href{https://doi.org/10.1111/joms.13065}{doi:\nolinkurl{10.1111/joms.13065}}


\bibitem[Schultz et~al\mbox{.}(2021)]%
        {schultz2021qualitative}
\bibfield{author}{\bibinfo{person}{Olivia~A Schultz}, \bibinfo{person}{Robert~S Hight}, \bibinfo{person}{Stanley Gutiontov}, \bibinfo{person}{Ravi Chandra}, \bibinfo{person}{Jeanne Farnan}, {and} \bibinfo{person}{Daniel~W Golden}.} \bibinfo{year}{2021}\natexlab{}.
\newblock \showarticletitle{Qualitative study of interprofessional collaboration in radiation oncology clinics: Is there a need for further education?}
\newblock \bibinfo{journal}{\emph{International Journal of Radiation Oncology* Biology* Physics}} \bibinfo{volume}{109}, \bibinfo{number}{3} (\bibinfo{year}{2021}), \bibinfo{pages}{661--669}.
\newblock


\bibitem[Schulze-Cleven and Vachon(2023)]%
        {cleven2023laborfutureofwork}
\bibfield{author}{\bibinfo{person}{Tobias Schulze-Cleven} {and} \bibinfo{person}{Todd~E. Vachon}.} \bibinfo{year}{2023}\natexlab{}.
\newblock \showarticletitle{Building the Future of Work Today - A Labor Studies Perspective}.
\newblock \bibinfo{journal}{\emph{Labor Studies Journal}} \bibinfo{volume}{48}, \bibinfo{number}{3} (\bibinfo{year}{2023}), \bibinfo{pages}{225--233}.
\newblock
\showeprint{https://doi.org/10.1177/0160449X231180375}
\href{https://doi.org/10.1177/0160449X231180375}{doi:\nolinkurl{10.1177/0160449X231180375}}


\bibitem[Scott(1985)]%
        {scott1985weapons}
\bibfield{author}{\bibinfo{person}{James~C Scott}.} \bibinfo{year}{1985}\natexlab{}.
\newblock \bibinfo{booktitle}{\emph{Weapons of the weak: Everyday forms of peasant resistance}}.
\newblock \bibinfo{publisher}{yale university Press}.
\newblock


\bibitem[Scott(1990)]%
        {scott1990domination}
\bibfield{author}{\bibinfo{person}{James~C Scott}.} \bibinfo{year}{1990}\natexlab{}.
\newblock \bibinfo{booktitle}{\emph{Domination and the arts of resistance: Hidden transcripts}}.
\newblock \bibinfo{publisher}{Yale university press}.
\newblock


\bibitem[{Securities and Exchange Commission}(2013)]%
        {SEC2013KnightCapital}
\bibfield{author}{\bibinfo{person}{{Securities and Exchange Commission}}.} \bibinfo{year}{2013}\natexlab{}.
\newblock \bibinfo{booktitle}{\emph{In the Matter of Knight Capital Americas LLC}}.
\newblock \bibinfo{type}{Administrative Proceeding} File No. 3-15570, Securities Exchange Act Release No. 34-70694. \bibinfo{institution}{U.S. Securities and Exchange Commission}, \bibinfo{address}{Washington, D.C.} \bibinfo{pages}{1--18} pages.
\newblock
\urldef\tempurl%
\url{https://www.sec.gov/files/litigation/admin/2013/34-70694.pdf}
\showURL{%
\tempurl}


\bibitem[Shneiderman(2020)]%
        {shneiderman2020human}
\bibfield{author}{\bibinfo{person}{Ben Shneiderman}.} \bibinfo{year}{2020}\natexlab{}.
\newblock \showarticletitle{Human-centered artificial intelligence: Reliable, safe \& trustworthy}.
\newblock \bibinfo{journal}{\emph{International Journal of Human--Computer Interaction}} \bibinfo{volume}{36}, \bibinfo{number}{6} (\bibinfo{year}{2020}), \bibinfo{pages}{495--504}.
\newblock


\bibitem[Shukla et~al\mbox{.}(2025)]%
        {shukla2025deskilling}
\bibfield{author}{\bibinfo{person}{Prakash Shukla}, \bibinfo{person}{Phuong Bui}, \bibinfo{person}{Sean~S Levy}, \bibinfo{person}{Max Kowalski}, \bibinfo{person}{Ali Baigelenov}, {and} \bibinfo{person}{Paul Parsons}.} \bibinfo{year}{2025}\natexlab{}.
\newblock \showarticletitle{De-skilling, Cognitive Offloading, and Misplaced Responsibilities: Potential Ironies of AI-Assisted Design}. In \bibinfo{booktitle}{\emph{Proceedings of the Extended Abstracts of the CHI Conference on Human Factors in Computing Systems}} \emph{(\bibinfo{series}{CHI EA '25})}. \bibinfo{publisher}{Association for Computing Machinery}, \bibinfo{address}{New York, NY, USA}, Article \bibinfo{articleno}{171}, \bibinfo{numpages}{7}~pages.
\newblock
\showISBNx{9798400713958}
\href{https://doi.org/10.1145/3706599.3719931}{doi:\nolinkurl{10.1145/3706599.3719931}}


\bibitem[Simkute et~al\mbox{.}(2025a)]%
        {simkute2025calculator}
\bibfield{author}{\bibinfo{person}{Auste Simkute}, \bibinfo{person}{Viktor Kewenig}, \bibinfo{person}{Abigail Sellen}, \bibinfo{person}{Sean Rintel}, {and} \bibinfo{person}{Lev Tankelevitch}.} \bibinfo{year}{2025}\natexlab{a}.
\newblock \bibinfo{title}{The New Calculator? Practices, Norms, and Implications of Generative AI in Higher Education}.
\newblock
\showeprint[arxiv]{2501.08864}~[cs.HC]
\urldef\tempurl%
\url{https://arxiv.org/abs/2501.08864}
\showURL{%
\tempurl}


\bibitem[Simkute et~al\mbox{.}(2025b)]%
        {simkute2025ironies}
\bibfield{author}{\bibinfo{person}{Auste Simkute}, \bibinfo{person}{Lev Tankelevitch}, \bibinfo{person}{Viktor Kewenig}, \bibinfo{person}{Ava~Elizabeth Scott}, \bibinfo{person}{Abigail Sellen}, {and} \bibinfo{person}{Sean Rintel}.} \bibinfo{year}{2025}\natexlab{b}.
\newblock \showarticletitle{Ironies of Generative AI: Understanding and Mitigating Productivity Loss in Human-AI Interaction}.
\newblock \bibinfo{journal}{\emph{International Journal of Human–Computer Interaction}} \bibinfo{volume}{41}, \bibinfo{number}{5} (\bibinfo{year}{2025}), \bibinfo{pages}{2898--2919}.
\newblock
\showeprint{https://doi.org/10.1080/10447318.2024.2405782}
\href{https://doi.org/10.1080/10447318.2024.2405782}{doi:\nolinkurl{10.1080/10447318.2024.2405782}}


\bibitem[Somers and Sloan(2023)]%
        {somers2023generative}
\bibfield{author}{\bibinfo{person}{Meredith Somers} {and} \bibinfo{person}{MIT Sloan}.} \bibinfo{year}{2023}\natexlab{}.
\newblock \showarticletitle{How generative AI can boost highly skilled workers’ productivity}.
\newblock \bibinfo{journal}{\emph{Ideas That Matter}} (\bibinfo{year}{2023}).
\newblock


\bibitem[Soroko et~al\mbox{.}(2023)]%
        {soroko2023social}
\bibfield{author}{\bibinfo{person}{Daria Soroko}, \bibinfo{person}{Gian-Luca Savino}, \bibinfo{person}{Nicholas Gray}, {and} \bibinfo{person}{Johannes Sch{\"o}ning}.} \bibinfo{year}{2023}\natexlab{}.
\newblock \showarticletitle{Social Transparency in Network Monitoring and Security Systems}. In \bibinfo{booktitle}{\emph{Proceedings of the 22nd International Conference on Mobile and Ubiquitous Multimedia}}. \bibinfo{pages}{37--53}.
\newblock


\bibitem[Spektor et~al\mbox{.}(2023)]%
        {spektor2023designingwellbeing}
\bibfield{author}{\bibinfo{person}{Franchesca Spektor}, \bibinfo{person}{Sarah~E Fox}, \bibinfo{person}{Ezra Awumey}, \bibinfo{person}{Christine~A. Riordan}, \bibinfo{person}{Hye~Jin Rho}, \bibinfo{person}{Chinmay Kulkarni}, \bibinfo{person}{Marlen Martinez-Lopez}, \bibinfo{person}{Betsy Stringam}, \bibinfo{person}{Ben Begleiter}, {and} \bibinfo{person}{Jodi Forlizzi}.} \bibinfo{year}{2023}\natexlab{}.
\newblock \showarticletitle{Designing for Wellbeing: Worker-Generated Ideas on Adapting Algorithmic Management in the Hospitality Industry}. In \bibinfo{booktitle}{\emph{Proceedings of the 2023 ACM Designing Interactive Systems Conference}} (Pittsburgh, PA, USA) \emph{(\bibinfo{series}{DIS '23})}. \bibinfo{publisher}{Association for Computing Machinery}, \bibinfo{address}{New York, NY, USA}, \bibinfo{pages}{623–637}.
\newblock
\showISBNx{9781450398930}
\href{https://doi.org/10.1145/3563657.3596018}{doi:\nolinkurl{10.1145/3563657.3596018}}


\bibitem[Spivak(2023)]%
        {spivak2023can}
\bibfield{author}{\bibinfo{person}{Gayatri~Chakravorty Spivak}.} \bibinfo{year}{2023}\natexlab{}.
\newblock \showarticletitle{Can the subaltern speak?}
\newblock In \bibinfo{booktitle}{\emph{Imperialism}}. \bibinfo{publisher}{Routledge}, \bibinfo{pages}{171--219}.
\newblock


\bibitem[Stahl et~al\mbox{.}(2021)]%
        {stahl2021artificial}
\bibfield{author}{\bibinfo{person}{Bernd~Carsten Stahl}, \bibinfo{person}{Andreas Andreou}, \bibinfo{person}{Philip Brey}, \bibinfo{person}{Tally Hatzakis}, \bibinfo{person}{Alexey Kirichenko}, \bibinfo{person}{Kevin Macnish}, \bibinfo{person}{S~Laulh{\'e} Shaelou}, \bibinfo{person}{Andrew Patel}, \bibinfo{person}{Mark Ryan}, {and} \bibinfo{person}{David Wright}.} \bibinfo{year}{2021}\natexlab{}.
\newblock \showarticletitle{Artificial intelligence for human flourishing--Beyond principles for machine learning}.
\newblock \bibinfo{journal}{\emph{Journal of Business Research}}  \bibinfo{volume}{124} (\bibinfo{year}{2021}), \bibinfo{pages}{374--388}.
\newblock


\bibitem[Stogiannos et~al\mbox{.}(2025)]%
        {survey2025anxiety}
\bibfield{author}{\bibinfo{person}{Nikolaos Stogiannos}, \bibinfo{person}{Gemma Walsh}, \bibinfo{person}{Benard Ohene-Botwe}, \bibinfo{person}{Kevin McHugh}, \bibinfo{person}{Ben Potts}, \bibinfo{person}{Winnie Tam}, \bibinfo{person}{Chris O'Sullivan}, \bibinfo{person}{Anton~Sheahan Quinsten}, \bibinfo{person}{Christopher Gibson}, \bibinfo{person}{Rodrigo~Garcia Gorga}, \bibinfo{person}{David Sipos}, \bibinfo{person}{Elona Dybeli}, \bibinfo{person}{Moreno Zanardo}, \bibinfo{person}{Cl{\'a}udia S{\'a}~dos Reis}, \bibinfo{person}{Nejc Mekis}, \bibinfo{person}{Carst Buissink}, \bibinfo{person}{Andrew England}, \bibinfo{person}{Charlotte Beardmore}, \bibinfo{person}{Altino Cunha}, \bibinfo{person}{Amanda Goodall}, \bibinfo{person}{Mark John-Matthews, Janice Stand~McEntee}, \bibinfo{person}{Yiannis Kyratsis}, {and} \bibinfo{person}{Christina Malamateniou}.} \bibinfo{year}{2025}\natexlab{}.
\newblock \showarticletitle{R-AI-diographers: a European survey on perceived impact of AI on professional identity, careers, and radiographers' roles}.
\newblock \bibinfo{journal}{\emph{Insights into Imaging}} \bibinfo{volume}{16}, \bibinfo{number}{1} (\bibinfo{date}{17 Feb} \bibinfo{year}{2025}), \bibinfo{pages}{43}.
\newblock
\showISSN{1869-4101}
\href{https://doi.org/10.1186/s13244-025-01918-6}{doi:\nolinkurl{10.1186/s13244-025-01918-6}}


\bibitem[Sum(2025)]%
        {sum2025futureoflabor}
\bibfield{author}{\bibinfo{person}{Cella~M. Sum}.} \bibinfo{year}{2025}\natexlab{}.
\newblock \showarticletitle{From the Future of Work to the Future of Labor: Centering Worker Resistance in the Age of AI and Automation}. In \bibinfo{booktitle}{\emph{Companion Publication of the 2025 Conference on Computer-Supported Cooperative Work and Social Computing}} \emph{(\bibinfo{series}{CSCW Companion '25})}. \bibinfo{publisher}{Association for Computing Machinery}, \bibinfo{address}{New York, NY, USA}, \bibinfo{pages}{19–21}.
\newblock
\showISBNx{9798400714801}
\href{https://doi.org/10.1145/3715070.3747335}{doi:\nolinkurl{10.1145/3715070.3747335}}


\bibitem[Sum et~al\mbox{.}(2025)]%
        {sum2025workerresistsurveillance}
\bibfield{author}{\bibinfo{person}{Cella~M. Sum}, \bibinfo{person}{Caroline Shi}, {and} \bibinfo{person}{Sarah~E. Fox}.} \bibinfo{year}{2025}\natexlab{}.
\newblock \showarticletitle{'It's Always a Losing Game': How Workers Understand and Resist Surveillance Technologies on the Job}.
\newblock \bibinfo{journal}{\emph{Proc. ACM Hum.-Comput. Interact.}} \bibinfo{volume}{9}, \bibinfo{number}{2}, Article \bibinfo{articleno}{CSCW004} (\bibinfo{date}{May} \bibinfo{year}{2025}), \bibinfo{numpages}{32}~pages.
\newblock
\href{https://doi.org/10.1145/3710902}{doi:\nolinkurl{10.1145/3710902}}


\bibitem[Tang et~al\mbox{.}(2023a)]%
        {tang2023backtolabor}
\bibfield{author}{\bibinfo{person}{Joice Tang}, \bibinfo{person}{McKane Andrus}, \bibinfo{person}{Samuel So}, \bibinfo{person}{Udayan Tandon}, \bibinfo{person}{Andr\'{e}s Monroy-Hern\'{a}ndez}, \bibinfo{person}{Vera Khovanskaya}, \bibinfo{person}{Sean~A. Munson}, \bibinfo{person}{Mark Zachry}, {and} \bibinfo{person}{Sucheta Ghoshal}.} \bibinfo{year}{2023}\natexlab{a}.
\newblock \showarticletitle{Back to “ Back to Labor”: Revisiting Political Economies of Computer-Supported Cooperative Work}. In \bibinfo{booktitle}{\emph{Companion Publication of the 2023 Conference on Computer Supported Cooperative Work and Social Computing}} (Minneapolis, MN, USA) \emph{(\bibinfo{series}{CSCW '23 Companion})}. \bibinfo{publisher}{Association for Computing Machinery}, \bibinfo{address}{New York, NY, USA}, \bibinfo{pages}{522–526}.
\newblock
\showISBNx{9798400701290}
\href{https://doi.org/10.1145/3584931.3611285}{doi:\nolinkurl{10.1145/3584931.3611285}}


\bibitem[Tang et~al\mbox{.}(2023b)]%
        {tang2023back}
\bibfield{author}{\bibinfo{person}{Joice Tang}, \bibinfo{person}{McKane Andrus}, \bibinfo{person}{Samuel So}, \bibinfo{person}{Udayan Tandon}, \bibinfo{person}{Andr{\'e}s Monroy-Hern{\'a}ndez}, \bibinfo{person}{Vera Khovanskaya}, \bibinfo{person}{Sean~A Munson}, \bibinfo{person}{Mark Zachry}, {and} \bibinfo{person}{Sucheta Ghoshal}.} \bibinfo{year}{2023}\natexlab{b}.
\newblock \showarticletitle{Back to “Back to Labor”: Revisiting Political Economies of Computer-Supported Cooperative Work}. In \bibinfo{booktitle}{\emph{Companion Publication of the 2023 Conference on Computer Supported Cooperative Work and Social Computing}}. \bibinfo{pages}{522--526}.
\newblock


\bibitem[Tasci(2025)]%
        {tasci2025aiknowledgeworkers}
\bibfield{author}{\bibinfo{person}{Murat Tasci}.} \bibinfo{year}{2025}\natexlab{}.
\newblock \bibinfo{booktitle}{\emph{AI and Knowledge Worker Displacement: Unprecedented Shifts in Unemployment Trends}}.
\newblock \bibinfo{type}{Economic Research Note}. \bibinfo{institution}{JPMorgan Chase}.
\newblock
\newblock
\shownote{Senior U.S. Economist Analysis}.


\bibitem[Toner-Rodgers(2024)]%
        {rodgers2024researchers}
\bibfield{author}{\bibinfo{person}{Aidan Toner-Rodgers}.} \bibinfo{year}{2024}\natexlab{}.
\newblock \bibinfo{booktitle}{\emph{Artificial Intelligence, Scientific Discovery, and Product Innovation}}.
\newblock \bibinfo{type}{Papers} 2412.17866. \bibinfo{institution}{arXiv.org}.
\newblock
\href{https://doi.org/None}{doi:\nolinkurl{None}}


\bibitem[Troncoso et~al\mbox{.}(2023)]%
        {troncoso2023algorithm}
\bibfield{author}{\bibinfo{person}{Isamar Troncoso}, \bibinfo{person}{Runshan Fu}, \bibinfo{person}{Nikhil Malik}, {and} \bibinfo{person}{Davide Proserpio}.} \bibinfo{year}{2023}\natexlab{}.
\newblock \showarticletitle{Algorithm failures and consumers’ response: Evidence from Zillow}.
\newblock \bibinfo{journal}{\emph{Available at SSRN 4520172}} (\bibinfo{year}{2023}).
\newblock


\bibitem[Turner et~al\mbox{.}(2022)]%
        {turner2022integrating}
\bibfield{author}{\bibinfo{person}{Sandra~L Turner}, \bibinfo{person}{Stephanie Tesson}, \bibinfo{person}{Phyllis Butow}, \bibinfo{person}{Burcu Vachan}, \bibinfo{person}{Ming-Ka Chan}, {and} \bibinfo{person}{Timothy Shaw}.} \bibinfo{year}{2022}\natexlab{}.
\newblock \showarticletitle{Integrating leadership development into radiation oncology training: A qualitative analysis of resident interviews}.
\newblock \bibinfo{journal}{\emph{International Journal of Radiation Oncology* Biology* Physics}} \bibinfo{volume}{113}, \bibinfo{number}{1} (\bibinfo{year}{2022}), \bibinfo{pages}{26--36}.
\newblock


\bibitem[{U.S. Department of Transportation, Office of Inspector General}(2020)]%
        {DOTOIG2020MAX}
\bibfield{author}{\bibinfo{person}{{U.S. Department of Transportation, Office of Inspector General}}.} \bibinfo{year}{2020}\natexlab{}.
\newblock \bibinfo{title}{Timeline of Activities Leading to Return to Service of the Boeing 737 MAX and Related FAA Oversight}.
\newblock
\urldef\tempurl%
\url{https://www.oig.dot.gov/library-item/38834}
\showURL{%
\tempurl}
\newblock
\shownote{Federal review materials identify MCAS as a significant contributing factor in both accidents}.


\bibitem[Webb(2019)]%
        {webb2019impact}
\bibfield{author}{\bibinfo{person}{Michael Webb}.} \bibinfo{year}{2019}\natexlab{}.
\newblock \showarticletitle{The impact of artificial intelligence on the labor market}.
\newblock \bibinfo{journal}{\emph{Available at SSRN 3482150}} (\bibinfo{year}{2019}).
\newblock


\bibitem[Wong(2021)]%
        {wong2021softresistance}
\bibfield{author}{\bibinfo{person}{Richmond~Y. Wong}.} \bibinfo{year}{2021}\natexlab{}.
\newblock \showarticletitle{Tactics of Soft Resistance in User Experience Professionals' Values Work}.
\newblock \bibinfo{journal}{\emph{Proc. ACM Hum.-Comput. Interact.}} \bibinfo{volume}{5}, \bibinfo{number}{CSCW2}, Article \bibinfo{articleno}{355} (\bibinfo{date}{Oct.} \bibinfo{year}{2021}), \bibinfo{numpages}{28}~pages.
\newblock
\href{https://doi.org/10.1145/3479499}{doi:\nolinkurl{10.1145/3479499}}


\bibitem[WorkLab(2024)]%
        {MSWorklab2024AIchangingwork}
\bibfield{author}{\bibinfo{person}{Micrososft WorkLab}.} \bibinfo{year}{2024}\natexlab{}.
\newblock \bibinfo{title}{AI is already changing work—Microsoft included}.
\newblock
\urldef\tempurl%
\url{https://www.microsoft.com/en-us/worklab/ai-is-already-changing-work-microsoft-included}
\showURL{%
\tempurl}
\newblock
\shownote{[Online; accessed 2025-09-01]}.


\bibitem[{World Economic Forum}(2025)]%
        {wef2025_future_of_jobs}
\bibfield{author}{\bibinfo{person}{{World Economic Forum}}.} \bibinfo{year}{2025}\natexlab{}.
\newblock \bibinfo{title}{The Future of Jobs Report 2023}.
\newblock
\urldef\tempurl%
\url{https://www.weforum.org/publications/the-future-of-jobs-report-2025/}
\showURL{%
\tempurl}


\bibitem[Yeverechyahu et~al\mbox{.}(2025)]%
        {yeverechyahu2025coding}
\bibfield{author}{\bibinfo{person}{Doron Yeverechyahu}, \bibinfo{person}{Raveesh Mayya}, {and} \bibinfo{person}{Gal Oestreicher-Singer}.} \bibinfo{year}{2025}\natexlab{}.
\newblock \bibinfo{title}{The Impact of Large Language Models on Open-source Innovation: Evidence from GitHub Copilot}.
\newblock
\showeprint[arxiv]{2409.08379}~[cs.SE]
\urldef\tempurl%
\url{https://arxiv.org/abs/2409.08379}
\showURL{%
\tempurl}


\bibitem[Zhang et~al\mbox{.}(2022)]%
        {zhang2022gigwellbeing}
\bibfield{author}{\bibinfo{person}{Angie Zhang}, \bibinfo{person}{Alexander Boltz}, \bibinfo{person}{Chun~Wei Wang}, {and} \bibinfo{person}{Min~Kyung Lee}.} \bibinfo{year}{2022}\natexlab{}.
\newblock \showarticletitle{Algorithmic Management Reimagined For Workers and By Workers: Centering Worker Well-Being in Gig Work}. In \bibinfo{booktitle}{\emph{Proceedings of the 2022 CHI Conference on Human Factors in Computing Systems}} (New Orleans, LA, USA) \emph{(\bibinfo{series}{CHI '22})}. \bibinfo{publisher}{Association for Computing Machinery}, \bibinfo{address}{New York, NY, USA}, Article \bibinfo{articleno}{14}, \bibinfo{numpages}{20}~pages.
\newblock
\showISBNx{9781450391573}
\href{https://doi.org/10.1145/3491102.3501866}{doi:\nolinkurl{10.1145/3491102.3501866}}


\bibitem[Zhou and Lee(2024)]%
        {zhou2024artists}
\bibfield{author}{\bibinfo{person}{Eric Zhou} {and} \bibinfo{person}{Dokyun Lee}.} \bibinfo{year}{2024}\natexlab{}.
\newblock \showarticletitle{Generative artificial intelligence, human creativity, and art}.
\newblock \bibinfo{journal}{\emph{PNAS Nexus}} \bibinfo{volume}{3}, \bibinfo{number}{3} (\bibinfo{date}{03} \bibinfo{year}{2024}), \bibinfo{pages}{pgae052}.
\newblock
\showeprint{https://academic.oup.com/pnasnexus/article-pdf/3/3/pgae052/57464715/pgae052.pdf}
\href{https://doi.org/10.1093/pnasnexus/pgae052}{doi:\nolinkurl{10.1093/pnasnexus/pgae052}}


\end{thebibliography}

\newpage
\appendix
\section{Appendix}
\edit{
\subsection{Further details on Workshops and Interviews } \label{sec: appendix_methods_further_details}
\subsubsection{\textbf{Workshop Timeline and Activities (WS1–WS5)}}

We conducted five participatory workshops (WS1–WS5) across the 12-month RadPlan deployment and spanned all
four roles (RadOncs, Physicists, Dosimetrists, and Hospital Administrators). Workshops were staged in early, mid, and late phases to trace the shift from first-wave optimism to asymptomatic effects, chronic harms, and identity commoditization, and to co-construct the Dignified Human-AI Interaction Framework.

\paragraph{\textbf{Workshop 1(month 2).}}  
WS1 occurred after RadPlan's initial rollout (near the second month) when conventional metrics indicated clear success (for example, faster planning cycles and improved dose-volume histograms). Participants collaboratively mapped how they currently planned cases with and without RadPlan, annotated typical decision points, and surfaced early intuitions about “where the AI helps” versus “where it might confuse.” They created simple stick-figure drawings of “a day with RadPlan,” marking where AI changed the sequence, which parts felt easier, and what kinds of judgment still felt uniquely human. Participants grouped these drawings into themes that later informed the early panels of the composite comic in ~\autoref{fig:comicStrips}, capturing the wave of optimism and focus on efficiency gains. Isolated concerns about “shortcuts creeping in” emerged here but remained largely based on aggregate success measures.

\paragraph{\textbf{Workshop 2 (month 3).}}  
By roughly the third month of use, participants began articulating “intuition rust” and unease about accepting RadPlan’s first suggestion without fully understanding why it was good. In response, the research and clinical teams prepared low-fidelity mock-ups of a Social Transparency (ST) layer based on the 4W template (who did what, when, and why) shown in ~\autoref{fig:ST_Radplan}, using synthetic cases to satisfy governance constraints. WS2 centered on critiquing and refining these mock-ups. Participants annotated which events should appear in the 4W log, how peer rationales should be displayed, and where transparency cues should sit so as not to overwhelm the treatment planning screen. These activities directly informed the implemented ST overlay illustrated in ~\autoref{fig:ST_Radplan}. A second drawing exercise invited participants to sketch “how I rely on RadPlan now versus before,” seeding visual
motifs of vigilance drift that later appeared in ~\autoref{fig:comicStrips}.

\paragraph{\textbf{Workshop 3 (Month 5)}}.  
After ST had been deployed and used in practice for several weeks, WS3 became the turning point described in~\autoref{sec:derive_framework}. Participants reviewed 4W traces alongside their own earlier drawings. Small groups annotated segments in the logs where approvals felt too rapid, rationales felt thin, or human edits clustered in particular phases of the workflow. Each annotated segment was tagged as an asymptomatic effect (e.g., subtle vigilance drift or fewer alternate plans) or as evidence that such effects were hardening into chronic harms (e.g., demonstrable skill loss or narrowing autonomy over time). Participants then drafted “questions we wish someone had asked before we deployed RadPlan,” focusing on craft, skill maintenance, and decision latitude. These exercises consolidated the expertise erosion
cascade that structures the Findings and produced the first rough clusters of Worker, Technology, and Organization concerns that underlie the later framework.

\paragraph{\textbf{Workshop 4 (month 8)}}  
WS4 took place as chronic harms became more visible and identity concerns started to surface more explicitly. Building on WS3 outputs and interim analysis, participants revisited composite drawings and selected ST traces and sorted them into the emerging Sense–Contain–Recover triad across Worker, Technology, and Organization levels. Participants used this structure to brainstorm concrete routines that could strengthen sociotechnical immunity without derailing existing workflows. Examples included periodic AI-off alternate-planning drills, recording brief rationales for high-risk overrides, and maintaining living “do-not-automate” lists for fragile tasks. Participants also commented on which draft framework questions felt actionable inside their existing meetings and quality review structures, and which felt too
burdensome.

\paragraph{\textbf{Workshop 5 (month 11)}}  
WS5 occurred near the end of the 12-month study, when identity commoditization and worries about the “future of the profession” had become salient. Participants used the now-matured framework and its starter-pack questions to walk through composite RadPlan scenarios and representative ST traces. The focus was on refining wording and granularity so that Worker, Technology, and Organization questions could be embedded in existing routines such as clinical rounds and planning conferences rather than requiring entirely new ceremonies. Particular attention was given to Organizational prompts around Humane Automation and “saved time,” and to Worker-level questions that helped junior and senior staff notice early warning signals of skill atrophy. 

Throughout WS1–WS5, drawing exercises and collaborative panel sequencing gradually produced the visual expertise erosion arc describing the AI-as-Amplifier paradox in~\autoref{fig:ParadoxCurve}, which participants endorsed as a faithful depiction of how RadPlan initially amplified efficiency while quietly eroding expertise and identity over time.

\subsubsection{\textbf{Semi-structured Interviews}}

The 24 semi-structured interviews ran in parallel with the workshops across the 12-month deployment. We conducted 8 interviews in months 2–3, 10 in months 5–7, and 6 in months 10–11. This distribution allowed participants to reflect on baseline practice before RadPlan, their evolving experiences as the system became routine, and, for those who used it, the impact of the ST overlay. The interview guide remained stable but was applied flexibly. Each session covered five core areas:

\begin{itemize} [nolistsep,noitemsep,left=0.5em]
    \item  \textit{Role and workflow.} Participants described their role (RadOnc, Physicist, Dosimetrist, Administrator) and walked through how they conducted treatment planning before RadPlan and how RadPlan fit into their current workflow.
    \item \textit{Changes in practice and skill.} Participants reflected on whether their planning behavior had changed since RadPlan deployment, including the number of alternate plans they prepared, how often they double-checked AI suggestions, and how “sharp” their intuition felt. Interviewers probed for early signs of asymptomatic effects, such as approving plans more quickly or skipping exploratory work, and for any deliberate strategies to keep skills from rusting.
    \item \textit{AI limitations and oversight.} Participants described situations where RadPlan behaved unexpectedly, produced unsatisfying suggestions, or was difficult to interpret, and how they detected and handled those cases. This block illuminated informal verification routines and the extent to which participants relied on peers, checklists, or their own judgment to monitor AI behavior.
    \item \textit{Social Transparency (when applicable)}. For participants interviewed after ST deployment, the guide included additional prompts about the 4W log. Interviewers asked how seeing who did what, when, and why affected trust, disagreement, and collaboration, and whether ST changed how participants balanced AI output against human judgment. Earlier interviews, conducted before ST was introduced, naturally did not include these prompts.
    \item \textit{Identity, values, and suggestions.} Finally, participants reflected on how AI integration had shifted their sense of expertise, autonomy, and meaning at work, and suggested changes that could preserve skills and dignity while retaining legitimate efficiency gains.
\end{itemize}

Interviews lasted 45–60 minutes, were audio-recorded with consent, and were transcribed with pseudonyms and removal of identifying clinical details. The structure above matches the themes reported in the Findings and is intended to be used as an outline rather than as a fixed script.

\subsubsection{\textbf{Data Privacy and Governance}}

Hospital Alpha’s data privacy and governance policies, along with IRB requirements, prohibit releasing raw artifacts that could expose proprietary interfaces or identifiable clinical traces. For instance,~\autoref{fig:ST_Radplan} uses a reconstructed interface with synthetic data to illustrate RadPlan’s core functionality rather than screenshots from production systems with patient data. The same constraints apply to workshop artifacts that referenced internal systems or specific cases.



\subsection{Metrics and Evaluation Plan} \label{sec:Metrics_Eval_Framework_Appendix}

To ensure this framework facilitates dignified Human-AI interactions, we propose tracking specific metrics and a phased evaluation plan. Key metrics include \textit{appropriate reliance} (alignment between AI accuracy and human trust, avoiding both over-trust and under-use), \textit{de-skilling indicators} (e.g., manual skill checks, error rates without AI), and dignity signals (self-reported autonomy, engagement, and willingness to critique the AI). The evaluation plan begins with \textbf{pilot deployments}, comparing teams using the framework to those without it: tracking skill retention and decision quality over time. Organizations should use feedback loops to refine questions and practices based on metric outcomes.

For adoption, we recommend an \textbf{iterative and reflexive rollout}: start small (e.g., one department), let champions (early adopters, team advocates, and power users) tailor the starter questions, assess outcomes after 3–6 months, and adapt and scale gradually. To test cross-domain usability, apply the framework in varied industries (e.g., healthcare, finance, engineering) and adapt with minimal effort. Above all, keep the framework practical: simple enough for routine use (e.g., check-ins, design reviews, training), yet robust enough to surface deep issues.
}

\end{document}
\endinput